\documentclass[
    superscriptaddress,
    prb,
    preprintnumbers,
    twocolumn,	%
    final,		%
    nobalancelastpage,
    nonbibnotes,
    showpacs,
    fleqn,
    a4paper,
    showkeys
]{revtex4-1}

\usepackage[utf8]{inputenc}
\usepackage{graphicx}
\usepackage{amsmath}
\usepackage{mathtools}
\usepackage{amsfonts}
\usepackage{color}
\usepackage{epstopdf}
\usepackage{subfigure}
\usepackage[
	colorlinks,
	linkcolor=blue,
	anchorcolor=blue,
	citecolor=blue,
	urlcolor=blue,
	menucolor=blue,
	filecolor=blue,
	pagecolor=blue
]{hyperref}

\newcommand{\avrgs}[1]{{\overline{#1}}}
\newcommand{\pinis}{pln}
\newcommand{\benis}{ben}

\newcommand{\dw}[1]{#1${}^\circ$}
\newcommand{\pinex}[2]{#1_{\rm \pinis{#2}}}
\newcommand{\benex}[2]{#1_{\rm \benis{#2}}}
\newcommand{\pin}[1]{#1_{\rm \pinis}}
\newcommand{\ben}[1]{#1_{\rm \benis}}
\newcommand{\pinsup}[1]{#1^{(\rm \pinis)}}
\newcommand{\bensup}[1]{#1^{(\rm \benis)}}

\newcommand{\bensupex}[2]{#1^{(\rm \benis{#2})}}
\newcommand{\up}[1]{#1_{\rm up}}
\newcommand{\dn}[1]{#1_{\rm dn}}

\DeclareMathOperator\erf{erf}

\newif\iftwocolumn
\twocolumntrue

\begin{document}

%\allowdisplaybreaks

\preprint{}

\title{Identification of the defect distribution at ferroelectric domain walls from the evolution of nonlinear dielectric response during aging process}

\author{Pavel Mokrý}
\email{mokry@ipp.cas.cz}
\affiliation{Regional Center for Special Optics and Optoelectronic Systems (TOPTEC), Institute of Plasma Physics, Academy of Sciences of the Czech Republic, Sobotecká 1660, CZ-51101 Turnov, Czech Republic}
%\affiliation{Faculty of Mechatronics, Informatics, and Interdisciplinary Studies, Technical University of Liberec, Studentská 2, CZ-46117 Liberec, Czech Republic}

\author{Tomáš Sluka}
%\affiliation{University of Geneva, DPMC, CH-1211 Geneva 4, Switzerland}
\affiliation{Ceramics Laboratory, Swiss Federal Institute of Technology (EPFL), CH-1015 Lausanne, Switzerland}


\date{\today}

\begin{abstract}
Motion of ferroelectric domain walls greatly contributes to the macroscopic dielectric and piezoelectric response of ferroelectric materials. 
The domain wall motion through the ferroelectric material is however hindered by pinning on crystal defects which substantially reduces these contributions.
Here, using thermodynamic models based on the Landau-Ginzburg-Devonshire theory, we find a relation between microscopic reversible motion of non-ferroelastic \dw{180} domain walls interacting with a periodic array of pinning centers and the nonlinear macroscopic permittivity.  
We show, that the reversible motion of domain walls can be split into two basic modes.  
First, the bending of a domain wall between pinning centers, and, second, the uniform movement of domain wall plane.  
We show that their respective contributions may change when the distribution of pinning centers is rearranged during the material aging.  
We demonstrate that it is possible to indicate which mechanism of the domain wall motion is affected during material aging. This allows to judge whether the defects only homogeneously accumulate at domain walls or prefer to align in certain directions inside the domain wall plane. We suggest that this information can be obtained using simple macroscopic dielectric measurements and a proper analysis of the nonlinear response.
Our results may therefore serve as a simple and useful tool to obtain details on domain wall pinning in an aging process.
\end{abstract} 

\maketitle

%\makenomenclature

%====
%
%====
\section{Introduction}
\label{sec:Intro}

Domain wall motion has been found responsible for the great contribution to dielectric and piezoelectric response of ferroelectric ceramic~\cite{arlt_dielectric_1985,li_extrinsic_1991,xu_domain_2001,jones_domain_2012}, single crystals~\cite{braun_creep_2005,kleemann_universal_2007}, and thin films~\cite{kim_evaluation_2003,gharb_dielectric_2005,bassiri-gharb_domain_2007}.
It is therefore expected that there exists a direct correlation between the fundamental parameters of domain wall motion and the parameters of enhanced macroscopic dielectric response of the studied ferroelectric polydomain systems.
In small applied electric fields, the domain walls perform reversible motions, which can be identified by anhysteretic dielectric response with \textit{nonlinear} field dependence of permittivity.
With an increase in the amplitude of applied electric field, the dielectric response becomes hysteretic and follows the so called Rayleigh law~\cite{li_extrinsic_1991}.
In order to explain phenomenological origins of the observed domain wall (extrinsic) contributions, several theoretical concepts of pinning mechanism have been proposed: elastic interaction between domain walls and ceramic grains~\cite{arlt_force_1991}, defect orientation~\cite{robels_domain_1993}, presence of electrode-adjacent passive layers~\cite{kopal_displacements_1999,bratkovsky_very_2001,mokry_size_2004}, domain wall pinning due to random bonds and random fields in disordered systems~\cite{nattermann_interface_1990}, or statistical effects of domain wall depinning~\cite{boser_statistical_1987}.

%----

The greatest challenge in characterization of ferroelectric samples is an unambiguous determination of physical mechanism standing behind the macroscopically observed extrinsic enhancement in dielectric response.
Techniques frequently used for a dielectric characterization of polydomain ferroelectrics are based on the measurement of frequency dependence of complex permittivity~\cite{braun_creep_2005,kleemann_universal_2007} and on the measurement of Rayleigh-like \textit{linear} field dependence of permittivity in the sub-switching fields~\cite{gharb_dielectric_2005,zhu_influence_2011,marincel_influence_2014}.
Unfortunately, it is known that high-field electric cycling often substantially influences the delicate pinning condition on the ferroelectric domain walls~\cite{morozov_hardening-softening_2008,glaum_-aging_2012}.
It means that Rayleigh-type dielectric measurement in sub-switching regime may not be a suitable dielectric characterization method for studying the aging processes.

%----

Recently, an alternative characterization method~\cite{mokry_method_2008}, which is based on the analysis of \textit{nonlinear} field dependence of macroscopic permittivity in weak electric fields:
\begin{equation}
	\label{eq:01:EpsF:Taylor}
	\varepsilon_f(E) = \varepsilon_F + \gamma \,E^2,
\end{equation}
where $\varepsilon_F$ is the small-signal permittivity and $\gamma$ is the dielectric nonlinearity constant, has been used to study the aging of polydomain ferroelectrics~\cite{mokry_evidence_2009}. 
It was observed that the time-dependent parameters $\varepsilon_F(t)$ and $\gamma(t)$ measured during the aging of [111]-oriented tetragonal lead zirconate titanate films (PZT) follow the relationship 
\begin{equation}
	\label{eq:01:GammaVsEps:ben}
	\sqrt{\gamma(t)}\propto \varepsilon_F(t) - \varepsilon_c,
\end{equation}
where $\varepsilon_c$ is the time-independent value of the permittivity of crystal lattice. 
This observation brought an evidence that the macroscopic dielectric response in the studied PZT samples was controlled by reversible bending movements of the \dw{180} domain walls.
The method based on the analysis of weak-field nonlinear dielectric measurements presented in Ref.~\cite{mokry_evidence_2009} was however developed with the assumption of the infinitely strong interaction of the domain wall with the pining center.
This represents a severe limitation for its applicability as it is evident that the interaction between real domain walls and pinning centers has a finite strength. 

%----

The aforementioned issues have motivated the work presented below, where we develop a general thermodynamic model for macroscopic nonlinear dielectric response produced by reversible motion of pinned non-ferroelastic \dw{180} domain walls in ferroelectrics.
In contrast to the model presented earlier~\cite{mokry_evidence_2009}, we consider %
a general situation of finite strength interaction between \dw{180} domain walls and pinning centers. 
In the first step, we show that the general motion of non-ferroelastic \dw{180} domain walls can be decomposed into two modes:
(i) oscillation of planar walls around the pinning center, where the wall keeps its electroneutrality (This mode will be referred to as \textit{planar mode}, PM.), and
(ii) bending movements of domain walls between pinning centers, which are characterized by the appearance of uncompensated bound charge due to the discontinuous normal component of polarization across the domain wall thickness. (This mode will be referred to as \textit{bending mode}, BM.)
We calculate simple formulae for the field dependence of nonlinear extrinsic permittivity controlled by the both considered modes.
The both contributions to the extrinsic permittivity are expressed as functions of dielectric parameters of the crystal lattice and representative distances between pinning centers.

%----

Now, for further understanding, it is  important to note that identification of the two types of contributions (with reasonable certainty) is not expected to be possible experimentally from the measurement of an actual value of permittivity, including its nonlinear component, at a given time. It is possible only from the observation of their characteristic changes during processes associated with the rearrangement of pinning centers on the domain wall. Such rearrangements take place during processes like material aging or fatigue.
This complication is associated with the necessity to decompose the dielectric response into the lattice (intrinsic) and domain wall (extrinsic) contributions.
It was shown in Ref.~\cite{mokry_evidence_2009} that the decomposition is feasible in polydomain ferroelectric films because the intrinsic contribution to the dielectric response is constant in time while the extrinsic contribution changes substantially when pinning centers rearrange in time.
However, the particular procedure of decomposition, which is presented in Ref.~\cite{mokry_evidence_2009}, is applicable only to a special case, when the BM dominates the reversible domain wall motion. 
In order to eliminate this restraint, a general procedure of decomposition, which is applicable to a general aging process, is introduced in this work.

%----

Therefore, we demonstrate that, once the time changes are observed, it is possible to indicate which mode of the domain wall motion is affected by material aging.
This allows to judge how the pinning centers rearrange at the domain walls, namely, it is possible to distinguish whether they homogeneously accumulate at domain walls and/or align in certain directions in the domain wall plane.
In order to support our suggestion, we have performed a series of numerical simulations of aging experiments in polydomain BaTiO${}_3$ (BT) and PbTiO${}_3$ (PT) in tetragonal phase. 
Our results may therefore serve as a simple and useful tool to obtain details on domain wall pinning from conventional dielectric measurements.

%====
%
%====
\section{Macroscopic dielectric response in films with \dw{180} domain walls}
\label{sec:MacroDielResp}

%---- Figure 1
\begin{figure}[t]
	\begin{center}
		\includegraphics[width=0.48\textwidth]{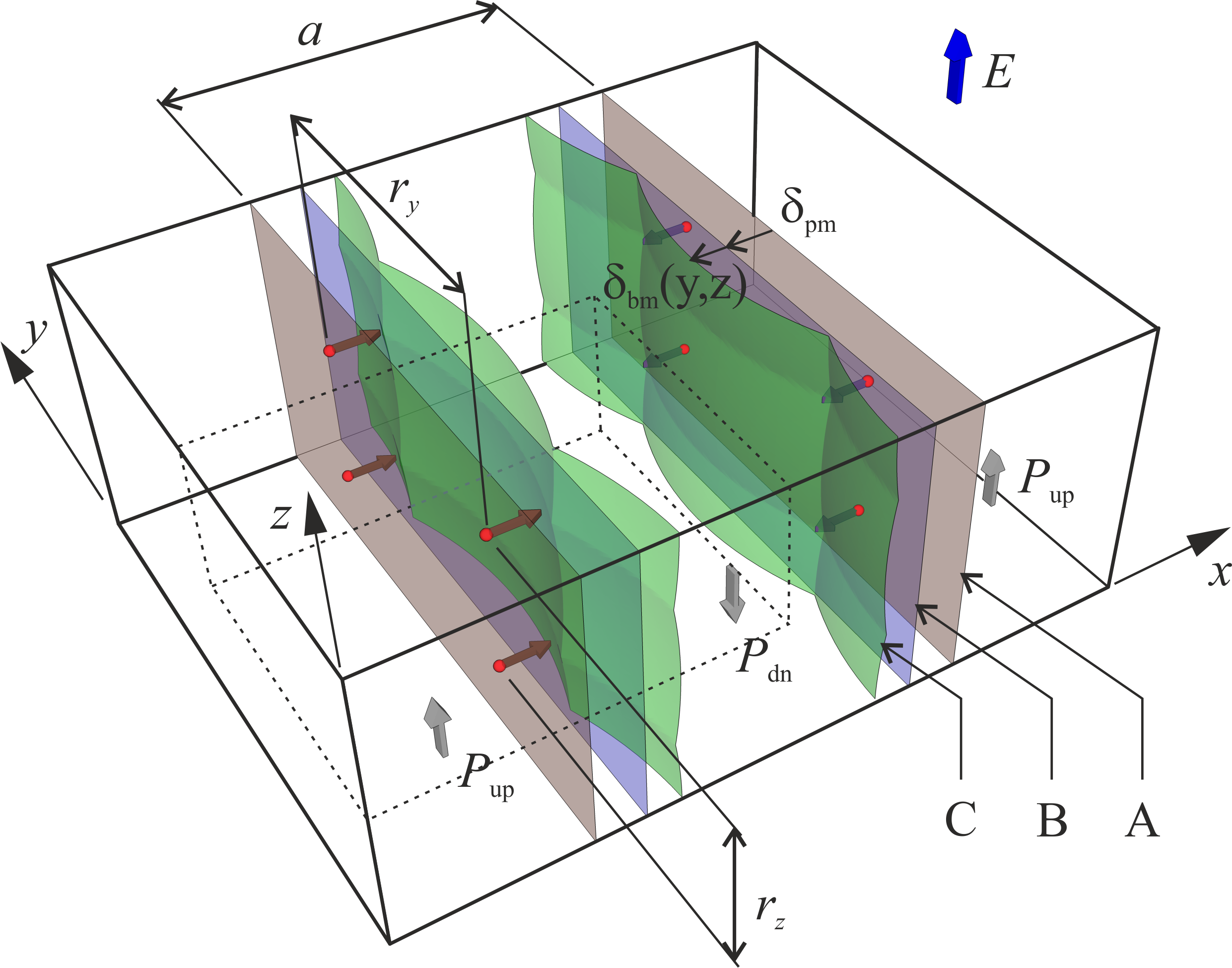}
	\end{center}
	\caption{Reversible movements of \dw{180} ferroelectric domain walls pinned by crystal lattice defects. Domain walls in the absence of the external electric field (Position A) are pinned by pinning centers indicated by red spheres. When the external electric field $E$ is applied to the system, the equilibrium position of the domain wall (Position C) is given by the superposition of two mechanism: uniform displacement $\pin{\delta}$ of the wall from the pinning center (Position B) and the spatially dependent displacement $\ben{\delta}(y,z)$ due to domain wall bending.}
	\label{fig:DWPinning}
\end{figure}
%---- Figure 1
%
In this section, we analyze the nonlinear macroscopic dielectric response of the polydomain ferroelectric system shown in Fig.~\ref{fig:DWPinning}. 
We consider a ferroelectric sample with crystallographic axes oriented along the $x$, $y$ and $z$ axes of the coordinate system.
Non-ferroelastic \dw{180} domain walls perpendicular to the $x$ axis separate anti-parallel domains with the vectors of polarization oriented along, $\up{P}$, and against, $\dn{P}$, the $z$ axis. 
In the absence of external electric field, the average distance between domain walls is denoted by the symbol $a$.
While the domain walls in real materials are pinned by randomly distributed lattice defects (pinning centers), we are limited to assume a regular periodic array of pinning centers in our model.
We consider that the largest reduction of the extrinsic dielectric response of the ferroelectric system is produced by the pinning centers, which are localized in representative distances $r_y$ and $r_z$ in the directions of $y$ and $z$ axes, respectively. 

%---- 

When the external AC electric field with the amplitude $E$ is applied in the direction of $z$ axis, domain wall displacement $\delta_w$ produces a change in the volume fractions $v_{\rm up}$ and $v_{\rm dn}$ of domains with polarization $P_{\rm up}$ and $P_{\rm dn}$, respectively:
\begin{subequations}
\label{eq:II:vup-vdn}
\begin{align}	
	\label{eq:II:vup-vdn:vup}
	\up{v}&= \frac 12 + \frac 1{r_y r_z a} \int_{0}^{r_y}dy \int_{0}^{r_z} \delta_w(y,\,z)\,dz,\\
	\label{eq:II:vup-vdn:vdn}
	\dn{v}&= \frac 12 - \frac 1{r_y r_z a} \int_{0}^{r_y}dy \int_{0}^{r_z} \delta_w(y,\,z)\,dz.
\end{align}	
\end{subequations}
Since we have limited our study to \dw{180} ferroelectric domain structure, the symmetry of crystal lattice in the ferroelectric state yields the following expressions for polarizations $\up{P}$ and $\dn{P}$ in the form of Taylor series with respect to the applied field $E$:
\begin{subequations}
\label{eq:II:Pup-Pdn}
\begin{align}
	\label{eq:II:Pup-Pdn:Pup}
	\up{P}(E) &= P_0 + \varepsilon_c\,E - \beta_c\,E^2 + \gamma_c\,E^3 + \cdots,\\
	\label{eq:II:Pup-Pdn:Pdn}
	\dn{P}(E) &= -P_0 + \varepsilon_c\,E + \beta_c\,E^2 + \gamma_c\,E^3 - \cdots
\end{align}
\end{subequations}
where $P_0$ is the spontaneous polarization, $\varepsilon_c$ is the lattice permittivity along the ferroelectric axis, and symbols $\beta_c$ and $\gamma_c$ stand for bulk dielectric nonlinearity constants, respectively. 

%----

The macroscopic polarization response of the polydomain sample can be expressed as the volume-averaged polarization responses of anti-parallel domains:
\begin{equation}
	\label{eq:II:Pf-def}
	P_f(E) = \up{v}(E)\,\up{P}(E) + \dn{v}(E)\,\dn{P}(E).
\end{equation}
It is convenient to introduce the net spontaneous polarization of the polydomain film due to domain wall displacements:
\begin{equation}
	\label{eq:II:PN-def}
	P_N(E)=P_0\,\left[\up{v}(E)-\dn{v}(E)\right]
\end{equation}
Using Eq.~(\ref{eq:II:PN-def}) and considering $\up{v}(E)+\dn{v}(E)=1$, the macroscopic polarization of the polydomain film given by Eq.~(\ref{eq:II:Pf-def}) can be expressed in the form:
\begin{equation}
	\label{eq:II:Pf-is}
	P_f(E) = P_N(E) + \varepsilon_c E - \frac{\beta_c P_N(E)}{P_0}\,E^2+\gamma_c\,E^3 + \cdots.
\end{equation}

%====
%
%====
\section{Dynamics of pinned \dw{180} domain walls}
\label{sec:ReversibleModel}

Objective of this Section is to formulate the equations of state for the net spontaneous polarization $P_N$ and to express extrinsic contributions to the small-signal permittivity and dielectric nonlinearity constant of the ferroelectric system due to the reversible motion of pinned \dw{180} domain walls.
The value of $P_N$ at the external field $E$ is given by the equilibrium displacement $\delta_w$ of the domain wall through Eqs.~(\ref{eq:II:vup-vdn}) and (\ref{eq:II:PN-def}).
The equation of state for $P_N$ can be expressed using a standard procedure from the condition for the minimum of the thermodynamic potential $H=\int_V h\,dV$, where the function
\begin{equation}
    \label{eq:III:psi-def}
    h=
    \psi^{(e)}_{\rm bulk} + \psi_{\rm ela} + \psi_{\rm es} + 
    \psi_{\rm wall} + \psi_{\rm ele} + \psi_{\rm pin} + D_i\,\varphi_{,i},
\end{equation}
is  the volume density of the electric enthalpy.
The thermodynamic potential $h$ is a function of the ferroelectric part of polarization $P_i$, electric displacement $D_i$, elastic strain $e_{ij}$, and electrostatic potential $\varphi$.
The indexes $i,\,j,\,\dots n$ with values 1, 2, and 3  denote the directions of the $x_1=x$, $x_2=y$, and $x_3=z$ axes of the coordinate system, respectively. 
The symbol $(\cdot)_{,i}=\partial(\cdot)/\partial x_i$ stands for the partial derivative.
The electric enthalpy includes the bulk free energy density $\psi^{(e)}_{\rm bulk}=\alpha_{ij}\,P_iP_j + \alpha_{ijkl}\,P_iP_jP_kP_l+ \alpha_{ijklmn}\,P_iP_jP_kP_lP_mP_n$, the elastic $\psi_{\rm ela}=(1/2)\,c_{ijkl}\, e_{ij} e_{kl}$, where $c_{ijkl}$ is the elastic stiffness tensor, electrostriction $\psi_{\rm es}=-q_{ijkl}\, e_{ij} P_k P_l$, where $q_{ijkl}$ are the electrostriction coefficients, domain wall $\psi_{\rm wall}=(1/2)\, G_{ijkl}\, P_{i,j} P_{k,l}$, where $G_{ijkl}$ are the gradient energy coefficients, and electrostatic $\psi_{\rm ele}=[1/(2 \varepsilon_0 \varepsilon_B)] (D_i-P_i)^2$, where
$\varepsilon_0$ and $\varepsilon_B$ are permittivity of vacuum
and relative background permittivity, respectively. 
Formulae for particular contributions to the volume density of electric enthalpy $h$ are expressed in Appendix~\ref{apndx:ThermoModel}.
In this work, we introduce the additional term 
\iftwocolumn
\begin{multline}
    \label{eq:III:f:pin}
    \psi_{\text{pin}}\left(x,\,y,\,z,\,P_i\right)= 
        \frac \xi{\pi^{3/2}w^3}\,P_i^2\, \times \\
        \exp\left[{-\frac{\left(x-x_d\right)^2 + \left(y-y_d\right)^2 + \left(z-z_d\right)^2}{w^2}}\right],
\end{multline}
%Tady to se zobrazuje nejak divne
\else
\begin{equation}
    \label{eq:III:f:pin}
    \psi_{\text{pin}}\left(x,\,y,\,z,\,P_i\right)= 
        \frac \xi{\pi^{3/2}w^3}\,P_i^2\, 
        \exp\left[{-\frac{\left(x-x_d\right)^2 + \left(y-y_d\right)^2 + \left(z-z_d\right)^2}{w^2}}\right],
\end{equation}
\fi
in order to specify the interaction of the domain wall with a pinning center.
Symbols $x_d$, $y_d$, and $z_d$ stand for the $x$-, $y$-, and $z$-coordinate of the pinning center and symbols $\xi$ and $w$ stand for the finite strength and radius of domain wall-defect interaction, respectively.
It should be noted here that the term $\psi_{\text{pin}}$ introduces a defect as a nanoscale energy perturbation, but it produces long-range electrostatic and elastic interactions due to its coupling with all remaining energy components [see e.g. Eqs.~(\ref{eq:A:f}) in Appendix~\ref{apndx:ThermoModel}].
Finally, the last term $D_i\,\varphi_{,i}$ corresponds to the subtracted work of external electric sources.

%----

In order to develop an analytical thermodynamic model for the finite-strength interaction of the non-ferroelastic \dw{180} domain wall with a pinning center, we consider a simplified situation, where the polarization is oriented along $x_3=z$ axis and that the $P_3$ component of the polarization changes its value linearly from $-P_0$ to $P_0$ across the domain wall of thickness $a_w$.
The points with zero polarization in the middle of the domain wall comprise a curved surface, which is given by the equation  $x=x_d+\delta_w(y,\,z)$.
This situation can be described mathematically by considering the ansatz of the domain wall profile in the form of piecewise linear function:
\begin{subequations}
	\label{eq:III:Piw}
	\begin{align}
	\label{eq:III:Piw:12}
	\hspace{-3mm}
	P^{(w)}_1\left(x_w\right) &= P^{(w)}_2\left(x_w\right) = 0,\\
	\label{eq:III:Piw:3}
	\hspace{-3mm}
	P^{(w)}_3\left(x_w\right) &= 
	\begin{dcases*}
	-P_0 & for $x_w<-a_w/2$,\\
	2 x_w P_0 /a_w & for $\left|x_w\right|\leq a_w/2$,\\
	P_0 & for $a_w/2<x_w$,
	\end{dcases*}
	\end{align}
\end{subequations}
where symbol $x_w$ stands for the distance from the mid-point of the domain wall.
Then, we consider that the spatial distribution of polarization $P_i$ within the ferroelectric can be approximated by: 
\begin{equation}
	\label{eq:III:PinApprox:3}
	P_i(x,\,y,\,z)\approx P^{(w)}_i(x-x_d-\delta_w).
\end{equation}
The assumption of the piecewise linear profile of the polarization at domain walls will be later proven as an adequate model by the correspondence between the analytical and phase field simulation results.

%----

An essential concept in our analysis, which is indicated in Fig.~\ref{fig:DWPinning}, is the consideration that the reversible domain wall displacement $\delta_w$ in the $x$-direction is given by the sum of two contributions:
\begin{equation}
\label{eq:II:deltaw}
\delta_w(y,\,z) = \pin{\delta} + \ben{\delta}(y,\,z),
\end{equation}
where $\pin{\delta}$ is the uniform displacement of the domain wall due to its PM motion (indicated by B in Fig.~\ref{fig:DWPinning}), which is controlled by the local interaction of \dw{180} domain wall with the pinning center, and $\ben{\delta}(y,\,z)$ is the spatially dependent displacement of the domain wall due to its BM motion (indicated by C in Fig.~\ref{fig:DWPinning}).
When Eq.~(\ref{eq:II:deltaw}) is substituted into Eqs.~(\ref{eq:II:vup-vdn}) and when the linearity of the operation of integration in Eqs.~(\ref{eq:II:vup-vdn}) is considered, it is convenient to express the net spontaneous polarization given by Eq.~(\ref{eq:II:PN-def}) as a sum of three contributions:
\begin{equation}
	\label{eq:II:PN-is}
	P_N = P_{N,0} + \pin{P} + \ben{P},
\end{equation}
where $P_{N,0}$ is the net spontaneous polarization at zero applied field. Symbols $\pin{P}$ and $\ben{P}$ stand for the field dependent contributions due to the uniform reversible displacement of domain wall (PM) and due to the reversible domain wall bending (BM), respectively. 
In accordance with the definition of symbols $\pin{P}$ and $\ben{P}$, it is noted here that these two contributions are produced by $\pin{\delta}$ and $\ben{\delta}$, respectively, as it is indicated in Fig.~\ref{fig:DWPinning}.

%----

In order to express the equations of state for the quantities $\pin{P}$ and $\ben{P}$, the ansatz for the distribution of polarization within the domain wall given by Eq.~(\ref{eq:III:Piw}) is substituted into the volume density of electric enthalpy given by Eq.~(\ref{eq:III:psi-def}) and integrated over the volume of the sample.
Using a detailed analysis of particular contributions to the electric enthalpy $h$ given by Eq.~(\ref{eq:III:psi-def}) it is possible to show that, in the first approximation, the $\psi_{\rm pin}$ is the only contribution to $h$, which depends solely on $\pin{P}$.
The reason for this is that $\psi_{\rm pin}$ is the only contribution to $h$, which is sensitive to the uniform displacement $\pin{\delta}$ of the domain wall from the pinning center at $x_d$.
On the other hand, it is possible to show that the remaining contributions $\psi^{(e)}_{\rm bulk} + \psi_{\rm ela} + \psi_{\rm es} + \psi_{\rm wall} + \psi_{\rm ele}$ depend solely on $\ben{P}$.
The reason for this is that spatially non-uniform displacement $\ben{\delta}(y,\,z)$ of the domain wall, which is proportional to $\ben{P}$, is associated with an increase in the domain wall area and with the production of  depolarizing field due to the discontinuity of the normal component of polarization across the domain wall thickness.
As a result of the above arguments, it is clear that, in the first approximation,
the thermodynamic function $H$ can be expressed in the form:
\begin{equation}
	\label{eq:Hw:def}	
	H = H_{0} + ar_yr_z\,\left[\pin{h}\left(\pin{P}\right) + \ben{h}\left(\ben{P}\right) - P_N E\right],
\end{equation}	
where 
\begin{equation}
    \pin{h}\left(\pin{P}\right) = \frac{1}{a r_yr_z} \int_0^a dx \int_0^{r_y} dy \int_0^{r_z} 
        \psi_{\rm pin}\, dz
\end{equation}
and
\iftwocolumn
\begin{multline}
\ben{h}\left(\ben{P}\right) = \frac{1}{a r_yr_z} \int_0^a dx \int_0^{r_y} dy \int_0^{r_z} \\
\left(
\psi^{(e)}_{\rm bulk} + \psi_{\rm ela} + \psi_{\rm es} + \psi_{\rm wall} + \psi_{\rm ele}
\right)\, dz
\end{multline}
\else
\begin{equation}
\ben{h}\left(\ben{P}\right) = \frac{1}{a r_yr_z} \int_0^a dx \int_0^{r_y} dy \int_0^{r_z} 
\left(
\psi^{(e)}_{\rm bulk} + \psi_{\rm ela} + \psi_{\rm es} + \psi_{\rm wall} + \psi_{\rm ele}
\right)\, dz
\end{equation}
\fi
are the contributions to the macroscopic volume density of electric enthalpy due to PM and BM motions of the domain wall, respectively.
Symbol $H_0$ is the constant term in the expansion of the thermodynamic potential $H$, which is independent on the reversible displacements of the domain wall.

%----

The value of the first contribution $\pin{h}$ is controlled by the additional term in the volume density of electric enthalpy, which describes the interaction of domain wall with a pinning center.
The most significant terms of $\pin{h}$ with respect to $\pin{P}$ are calculated in Appendix~\ref{apndx:Model:pin} and equal
\begin{subequations}
\label{eq:hpin}	
\begin{equation}
\label{eq:hpin:is}	
\pin{h} = 
\frac{a\xi\,\Phi\left(a_w/w\right)}{r_yr_zw^2}\, \pin{P^2} - \frac{a^3\xi\, \Psi\left(a_w/w\right)}{48P_0^3r_yr_zw^4}\, \pin{P^4} + \cdots,
\end{equation}
where 
\begin{align}
\label{eq:hpin:Phi}
\Phi\left(\tau\right) &= \frac {1}{\tau^2}\left[
{\rm erf}\left(\tau/2\right)-\frac{\tau\,\exp\left(-\tau^2/4\right)}{\sqrt{\pi}}
\right],\\
\label{eq:hpin:Psi}
\Psi\left(\tau\right) &= 
\frac{\tau}{\sqrt{\pi}} \exp\left(-\tau^2/4\right)
\end{align}
\end{subequations}
are numerical functions that depend on the spatial distribution of polarization within the domain wall.

%----

The second contribution to the macroscopic electric enthalpy $\ben{h}$ is due to the BM motion of \dw{180} domain wall.
Its value is controlled by energy of the electric field produced by uncompensated bound charges due to the discontinuity in the normal component of polarization across the domain wall thickness and due to the increase in the area of the bent domain wall. 
The leading terms of $\ben{h}$ with respect to $\ben{P}$ are calculated in Appendix~\ref{apndx:Model:ben} and equal
\begin{equation}
\label{eq:Hw:def:hben}	
\ben{h} =  
\frac{a a_w }{8\,\varepsilon_c r_y^2}\, \ben{P^2} - \frac{0.17\, a^3 a_w}{\varepsilon_c P_0^2 r_y^4 }\, \ben{P^4} + \cdots.
\end{equation}

%----

The equilibrium response of the net spontaneous polarization is found from the minimum of the thermodynamic function $H$ considering the principle of superposition for the net spontaneous polarization given by Eq.~(\ref{eq:II:PN-is}). 
Substituting Eq.~(\ref{eq:II:PN-is}) into Eq.~(\ref{eq:Hw:def}), the equilibrium values of $\pin{P}$ and $\ben{P}$ are found by solving the system of equations of state:
\begin{subequations}
\label{eq:IV:EoS}
\begin{align}
	\label{eq:IV:EoS:hpin}
	\partial \pin{h}/\partial \pin{P} -E &= 0,\\
	\label{eq:IV:EoS:hben}
	\partial \ben{h}/\partial \ben{P} -E &= 0.
\end{align}
\end{subequations}

%----

Solution of Eq.~(\ref{eq:IV:EoS:hpin}) can be expressed in the form of Taylor series (see Appendix~\ref{apndx:Model:pin}):
\begin{subequations}
	\label{eq:IV:Ppin-Taylor}
\begin{equation}
	\label{eq:IV:Ppin-Taylor:Ppin}
	\pin{P}(E) = \pin{\varepsilon}\,E + \pin{\gamma}\,E^3 + \cdots,
\end{equation}
where 
\begin{align}
\label{eq:IV:Ppin-Taylor:EpsL}
\pin{\varepsilon} &= \frac{\pin{\mu}\, r_y r_z w^2}{2 a \xi\, \Phi\left(a_w/w\right)},\\
\label{eq:IV:Ppin-Taylor:Gamma}
\pin{\gamma} &= \frac{\pin{\nu}\, 
	r_y^3 r_z^3 w^4\,\Psi\left(a_w/w\right)
}{
192\, a P_0^2 \xi^3\, \Phi^4\left(a_w/w\right)
}
\end{align}
\end{subequations}
stand for the small-signal permittivity and dielectric nonlinearity constant of the extrinsic dielectric response due to the movement of planar pinned domain walls, respectively. 
Symbols $\pin{\mu}$ and $\pin{\nu}$ are dimensionless numerical refinement factors, which are computed in Appendix~\ref{apndx:Refinement}.

%----

Similarly, solution of Eq.~(\ref{eq:IV:EoS:hben}) yield the Taylor expansion (see Appendix~\ref{apndx:Model:ben}):
\begin{subequations}
	\label{eq:IV:Pben-Taylor}
\begin{equation}
	\label{eq:IV:Pben-Taylor:Pben}
	\ben{P}(E) = \ben{\varepsilon}\,E + \ben{\gamma}\,E^3 + \cdots,
\end{equation}
where
\begin{align}
	\label{eq:IV:Pben-Taylor:EpsL}
	\ben{\varepsilon} &= \frac{4\,\ben{\mu}\,\varepsilon_c r_y^2}{a a_w},\\
	\label{eq:IV:Pben-Taylor:Gamma}
	\ben{\gamma} &= \frac{
		172.8\,\ben{\nu}\,\varepsilon_c^3 r_y^4
	}{
		a a_w^3 P_0^2
	}
\end{align}
\end{subequations}
stand for the small-signal permittivity and dielectric nonlinearity constant of the extrinsic dielectric response due to the bending movement of domain walls, respectively.
Symbols $\ben{\mu}$ and $\ben{\nu}$ are dimensionless numerical refinement factors, which are computed in Appendix~\ref{apndx:Refinement}.

%----

The field-dependent net spontaneous polarization can be expressed in the form:
\begin{subequations}
	\label{eq:IV:SuperP}
\begin{equation}
	P_N = P_{N,0} + \varepsilon_L\, E + \gamma_w\, E^3 + \cdots,
\end{equation}
where 
\begin{align}
	\label{eq:IV:SuperP:EpsL}
	\varepsilon_L &= \pin{\varepsilon} + \ben{\varepsilon},\\
	\label{eq:IV:SuperP:GammaW}
	\gamma_w &= \pin{\gamma} + \ben{\gamma}
\end{align}
\end{subequations}
are the small-signal contribution to the extrinsic permittivity and the nonlinearity constant of the anhysteretic field dependence of nonlinear extrinsic permittivity, respectively.

%----

In the final step of our analysis, it is convenient to introduce the field-dependent extrinsic contribution to permittivity $\varepsilon_w(E)$ by the formula:
\begin{equation}
	\label{eq:IV:Epsw-def}
	P_N(E) = P_{N,0} + \varepsilon_w(E)\,E.
\end{equation}
It follows from Eq.~(\ref{eq:IV:SuperP}) that $\varepsilon_w(E)$ can be expressed in the form of Taylor series
\begin{equation}
	\label{eq:IV:Epsw-ser}
	\varepsilon_w(E)=\varepsilon_L + \gamma_w E^2 + \cdots.
\end{equation}
Since in dielectric experiments it is possible to directly measure only the field dependence of  macroscopic permittivity $\varepsilon_f(E)$, which is defined as
\begin{equation}
	\varepsilon_f(E) = \frac{P_f(E)-P_f(0)}{E},
\end{equation}
it follows from Eq.~(\ref{eq:II:Pf-is}) that
\begin{equation}
	\varepsilon_f(E)\approx \varepsilon_F + \gamma\, E^2 + \cdots,
\end{equation}
where
\begin{subequations}
	\label{eq:IV:DielMacro}
\begin{align}
	\label{eq:IV:DielMacro:EpsF}
	\varepsilon_F &= \varepsilon_c + \varepsilon_L,\\
	\label{eq:IV:DielMacro:Gamma}
	\gamma &= \gamma_c + \gamma_w
\end{align}
\end{subequations}
and symbols $\varepsilon_c$ and $\gamma_c$ stand for the lattice values of permittivity and dielectric nonlinearity constant [see Eqs.~(\ref{eq:II:Pup-Pdn})].

%----

Analytical results obtained above describe the general nonlinear macroscopic dielectric response of ferroelectric polydomain system controlled by the reversible motion of non-ferroelastic \dw{180} domain walls interacting with pinning centers. 
It should be noted here that the continuous spatial distribution of the polarization vector was considered in the analytical solution of the thermodynamic model. 
This may seem to be inconsistent with the well-known concept of domain wall pinning at atomic planes by Miller and Weinreich~\cite{miller_mechanism_1960}. 
However, the recent results based on the atomistic molecular dynamic simulations by Shin et al.~\cite{shin_nucleation_2007} indicated that the classical model by Miller and Weinreich is outdated and that the critical domain nuclei responsible for the side-wise domain wall motion have actually much lower activation energy. 
This should reasonably justify our use of continuous thermodynamic model even the phenomena we analyse are on the atomistic level.
In the next Section, we will show that these results can be used for the identification of  microscopic processes responsible for aging or fatigue in ferroelectric polydomain films.

%====
%
%====
\section{Nonlinear dielectric characterization during aging}
\label{sec:Aging}

It was shown in Sec.~\ref{sec:MacroDielResp} that the macroscopic dielectric response of polydomain ferroelectric systems is given by the superposition of intrinsic (lattice) and extrinsic (domain wall) contributions and that these contributions can be, in principle, decomposed in the nonlinear dielectric response of these systems~\cite{mokry_evidence_2009}. Such a decomposition is possible because the intrinsic parameters $\varepsilon_c$ and $\gamma_c$ are virtually time-independent while the extrinsic parameters $\varepsilon_L$ and $\gamma_w$ undergo the process of aging, which are frequently explained by defect migration. Several types of defects that may act as pinning centers have been discussed in the works on lead titanate~\cite{morozov_hardening-softening_2008,morozov_charge_2010} and bismuth ferrite~\cite{rojac_strong_2010,rojac_bifeo3_2014}. 
In addition, several mechanisms that represent a driving force for defect formation and migration have been suggested~\cite{chandrasekaran_defect_2013,morozov_hardening-softening_2008,genenko_mechanisms_2015}. 
In the terms of our model, some of these mechanisms may result in two different processes that may be responsible for the dielectric aging in polydomain ferroelectric: (i) uniform accumulation of pinning centers at the domain wall, which is expressed by an increase in the value of pinning center concentration $n=1/\left(r_yr_z\right)$, and, (ii) redistribution and alignment of pinning centers along certain directions on the domain wall, which is expressed by a change in the value of the anisotropy of the pinning center distribution $\eta=r_y/r_z$.
The key objective of this Section is to show that the nonlinear analysis of time changes in the correlation between the small-signal permittivity $\varepsilon_F$ and  dielectric nonlinearity constant $\gamma$ allows to obtain information about the preferred modes of domain wall motion (PM or BM) and, consequently, also some details about the microscopic process responsible for the dielectric aging.

%----

An important result presented in Sec.~\ref{sec:MacroDielResp} is that the  dielectric parameters of extrinsic dielectric response due to the PM motion of domain walls, i.e. $\pin{\varepsilon}$ and $\pin{\gamma}$, depend only on the concentration of pinning centers on the domain wall $n$, but do not depend on the anisotropy of the pinning centers distribution $\eta$ [see Eqs.~(\ref{eq:IV:Ppin-Taylor})]. On the contrary, the dielectric parameters due to BM motion of domain walls, i.e. $\ben{\varepsilon}$ and $\ben{\gamma}$,  depend on the both parameters $n$ and $\eta$ [see Eqs.~(\ref{eq:IV:Pben-Taylor})].
Let us consider for a moment that the anisotropy parameter $\eta$ is constant and the aging process is driven by uniform accumulation of pinning centers on the domain wall.
Under this assumption, the both contributions to the extrinsic small-signal permittivity due to PM and BM domain wall motion $\pin{\varepsilon}$ and $\ben{\varepsilon}$ change proportionally to $1/n$. 
On the contrary, there exist a difference in scaling the dielectric nonlinearity constants. In the case of PM domain wall motion, it is $\pin{\gamma}\propto 1/n^{3}$, whereas it is $\ben{\gamma}\propto 1/n^{2}$ in the case of BM domain wall motion.

%----

The above statement indicates that the nonlinear extrinsic dielectric response due to PM motion of domain walls is characterized by a typical correlation between $\pin{\gamma}$ and $\pin{\varepsilon}$, which can be expressed as:
\begin{equation}
	\label{eq:IV:GammaVsEpsL:pin}
	\pin{\gamma} = 
	\left[
	\frac{
		\pin{\nu}\, a^2\, \Psi\left(a_w/w\right)
	}{
		24\,\pin{\mu}^3\,w^2\, \Phi\left(a_w/w\right)
	}
	\right]\, 
    \frac{\varepsilon_c^3}{P_0^2}\,
	\left(\frac{\pin{\varepsilon}}{\varepsilon_c}\right)^3.
\end{equation}
This result suggests that when the correlation ${\gamma}\propto{\varepsilon}^3$ is identified in the extrinsic dielectric response during aging process, it is a rather unambiguous indication that the aging process is dominated by PM domain wall motion.
Since the dielectric response in the PM is insensitive to the anisotropy parameter $\eta$, it can be clearly concluded that the experimentally observed aging process is caused by the accumulation of the pinning centers on the domain wall. 

%----

Similarly, a correlation between $\ben{\gamma}$ and $\ben{\varepsilon}$, which satisfies:
\begin{equation}
	\label{eq:IV:GammaVsEpsL:ben}
	\ben{\gamma} = \left(
	\frac{
		10.8\,\ben{\nu}\,a
	}{
		\ben{\mu}^2\,a_w
	}
	\right)\, 
    \frac{\varepsilon_c^3}{P_0^2}\,
	\left(\frac{\ben{\varepsilon}}{\varepsilon_c}\right)^2,
\end{equation}
represents a clear indication that the nonlinear dielectric response is dominated by the BM motion of domain walls.
Identification of the microscopic origin of the aging process, however, is more difficult and requires more detailed analysis. 
First it should be noted that there exist a critical defect concentration $n_{\rm crit}$ which represents a threshold between the dominant modes in the dielectric response. 
For smaller values of pinning center concentration than $n_{\rm crit}$, the contribution of PM to extrinsic permittivity becomes negligible compared to BM and makes further analysis of the microscopic origin of the aging process impossible, since the aging in the BM can be caused by the both aforementioned processes. 
However, above this critical concentration the effects of PM and BM modes on the nonlinear dielectric response are comparable.
It means that when the ${\gamma}\propto{\varepsilon}^2$ is identified in the extrinsic dielectric response during the aging process in samples with supercritical concentration of pinning centers at the domain wall, it is an indication that the aging process is caused by the redistribution and alignment of pinning centers along certain directions in the domain wall plane.

%----

Thus, the results above indicate that the microscopic process, which is responsible for the dielectric aging, can be determined by means of the identification of the dominant mode of domain wall motion in the extrinsic dielectric response. 
The last task that has to be resolved is the decomposition of the intrinsic and extrinsic responses from the macroscopic dielectric response of the ferroelectric system.

%----

In order to identify the aging mechanism in a real ferroelectric system, we suggest the following procedure.
Let us introduce a general aging mechanism, which is characterized by a general correlation between the small-signal extrinsic permittivity $\varepsilon_L$ and the extrinsic nonlinearity constant $\gamma_w$ in the form
\begin{equation}
	\gamma_w=B\, \left(\varepsilon_L/\varepsilon_c\right)^p.
\end{equation}
It follows from Eqs.~(\ref{eq:IV:DielMacro}) that the coupling between the macroscopic small-signal permittivity $\varepsilon_F$ and the macroscopic nonlinearity constant $\gamma$ satisfies the relation:
\begin{equation}
	\label{eq:IV:Gamma:Is}
	\gamma\left(\varepsilon_F\right) = \gamma_c + B\,\left(\varepsilon_F/\varepsilon_c-1\right)^p.
\end{equation}
It is considered that during aging, the lattice parameters $\varepsilon_c$ and $\gamma_c$ remains constant and the evolution of the macroscopic dielectric response (i.e. parameters $\varepsilon_F$ and $\gamma$) is produced due to change in the extrinsic dielectric response parameters $\varepsilon_L$ and $\gamma_w$.

%----

Let us conduct an aging experiment and measure the time dependence of macroscopic values $\varepsilon_F$ and $\gamma$.
Let us define the symbols $\varepsilon_{\min}$ and $\gamma_{\min}$ for the minimum measured values of $\varepsilon_F$ and $\gamma$, respectively.
Let us define the parameter $A$
\begin{equation}
	\label{eq:IV:A:def}
	A\left(\varepsilon_F\right) = 
	\frac{
		\log\left[\gamma\left(\varepsilon_F\right)/\gamma_{\min}\right]
	}{
		\log\left[\varepsilon_F/\varepsilon_{\min}\right]
	}.
\end{equation}
Substituting Eq.~(\ref{eq:IV:Gamma:Is}) into Eq.~(\ref{eq:IV:A:def}), it can be calculated with the use of elementary calculus rules that 
\begin{equation}
	\label{eq:IV:A:lim}
	\lim_{\varepsilon_F\rightarrow\infty}A\left(\varepsilon_F\right) = p.
\end{equation}
It means that using the estimation of the limiting value of parameter $A$ for very large values of macroscopic small-signal permittivity from measured experimental data, it is possible to determine the dominant aging mechanism in the ferroelectric system.

%----

Unfortunately, it is rather difficult to achieve the condition given by Eq.~(\ref{eq:IV:A:lim}) in real experimental conditions. 
The reason is that using dielectric measurements, it is possible to study only intermediate stages of the aging process. 
Now, the key objective is to find a procedure for the determination of the  extrapolated value of $A$ for very large values of macroscopic small-signal permittivity $\varepsilon_F$. 

%----

One possibility is offered by expressing the Taylor expansion of $A$ given by Eq.~(\ref{eq:IV:A:def}) with respect to the inverse macroscopic small-signal permittivity $\varepsilon_F$ normalized to its minimal value  measured during aging experiment.
For integer numbers of $p$ in Eq.~(\ref{eq:IV:Gamma:Is}), one yields:
\iftwocolumn
\begin{multline}
\label{eq:IV:ATaylor:Is}
	A\left(\varepsilon_{\min}/\varepsilon_F\right) \approx 
		p - 
		\log\left(\frac{B\,\varepsilon_{\min}^p}{\gamma_{\min}\varepsilon_c^p}\right)
		\frac{1}{\log\left(\varepsilon_{\min}/\varepsilon_F\right)} + \\
		+ \left(\frac{p\,\varepsilon_c}{\varepsilon_{\min}}\right)
		\frac{\varepsilon_{\min}/\varepsilon_F}{\log\left(\varepsilon_{\min}/\varepsilon_F\right)}
		+ \cdots.
\end{multline}
\else
\begin{equation}
\label{eq:IV:ATaylor:Is}
	A\left(\varepsilon_{\min}/\varepsilon_F\right) \approx 
	p - \log\left(\frac{B\,\varepsilon_{\min}^p}{\gamma_{\min}\varepsilon_c^p}\right)
	\frac{1}{\log\left(\varepsilon_{\min}/\varepsilon_F\right)} + 
	\left(\frac{p\,\varepsilon_c}{\varepsilon_{\min}}\right)
	\frac{\varepsilon_{\min}/\varepsilon_F}{\log\left(\varepsilon_{\min}/\varepsilon_F\right)}
	+\cdots.
\end{equation}
\fi
In order to estimate (i) the value $p$, which identifies the mechanism of the aging, (ii) lattice permittivity $\varepsilon_c$, and (iii) parameter $B$, which contains information on the pinning condition on the domain wall, we suggests to compute the  parameter $A$ from the values of $\varepsilon_F$ and $\gamma$ measured during the aging experiment and to fit the values of parameter $A$ to the function of the form:
\begin{equation}
\label{eq:IV:ATaylor:Fit}
	A\left(\varepsilon_{\min}/\varepsilon_F\right) = 
	C_1 - 
	\frac{C_2}{\log\left(\varepsilon_{\min}/\varepsilon_F\right)} + 
	\frac{C_3\,\left(\varepsilon_{\min}/\varepsilon_F\right)}{\log\left(\varepsilon_{\min}/\varepsilon_F\right)},
\end{equation}
where $C_i$ are the fitting parameters.
Comparing the coefficients in the Taylor expansions in Eqs.~(\ref{eq:IV:ATaylor:Is}) and (\ref{eq:IV:ATaylor:Fit}), it is possible to roughly estimate the unknown parameters of the ferroelectric system:
\begin{subequations}
\label{eq:IV:AgingExtrapol}
\begin{align}
\label{eq:IV:AgingExtrapol:n}
p &= C_1,\\
\label{eq:IV:AgingExtrapol:EpsC}
\varepsilon_c &= \varepsilon_{\min}\,\left(C_3/C_1\right),\\
\label{eq:IV:AgingExtrapol:B}
B &= \gamma_{\min}\,e^{C_2}\,\left(C_3/C_1\right)^{C_1}.
\end{align}
\end{subequations}
Thus, it is seen that it is possible to decompose the intrinsic and extrinsic contributions to the dielectric response of ferroelectric polydomain samples and that it is possible to identify the dominant aging mechanism by proper analysis of the nonlinear dielectric response. 

%====
%
%====
\section{Phase-field model simulations of aging}
\label{sec:PFM}

The key objective of this Section is to provide independent arguments supporting the idea that results presented in the above Section can be used to study processes of aging in real ferroelectric systems. 
We have performed numerical simulations of two aging mechanisms using the standard thermodynamic model~\cite{zhirnov_contribution_1959,cao_theory_1991,hu_three-dimensional_1998}.
The process of aging has been modeled by means of a series of numerical simulations performed on a representative volume element (RVE) in a periodic structure of a ferroelectric material with a \dw{180} domain wall. 
The RVE is considered as a cuboid with dimensions $a$, $r_y$, and $r_z$ along $x$, $y$, and $z$ axes of the coordinate system as it is indicated by a dashed line in Fig.~\ref{fig:DWPinning}. 
In accordance with two aging mechanisms considered in the above Section, we have performed two types of numerical simulations. 
The PM aging mechanism can be modeled in the system, where the parameter $r_y$ equals a constant value $r_0$, which is smaller than the value of parameter $r_z$. 
The BM aging mechanism, where the dielectric response is controlled by BM movements of \dw{180} domain walls, can be modeled in the system with the constant value of parameter $r_z=r_0$ and decreasing value of parameter $r_y$.

%----

We have performed numerical simulations of two aforementioned aging mechanisms using the numerical method, which is described in Appendix~\ref{apndx:ThermoModel}.
In the numerical simulations, the parameter $r_0$ was set to be equal to 4\,nm. 
The effect of aging is introduced into the numerical simulations by means of a decrease in the representative distance between pinning centers in the particular direction. 
The parameter in the RVE, which is subjected to change during the simulation of aging, is denoted by a symbol $r$ and its value has been decreasing from 25\,nm to 4\,nm. 

%----

%---- Figure 3
\begin{figure}[t]
	\centering
	\subfigure[Bending mode (BM) aging]{
		\includegraphics[width=0.48\textwidth]{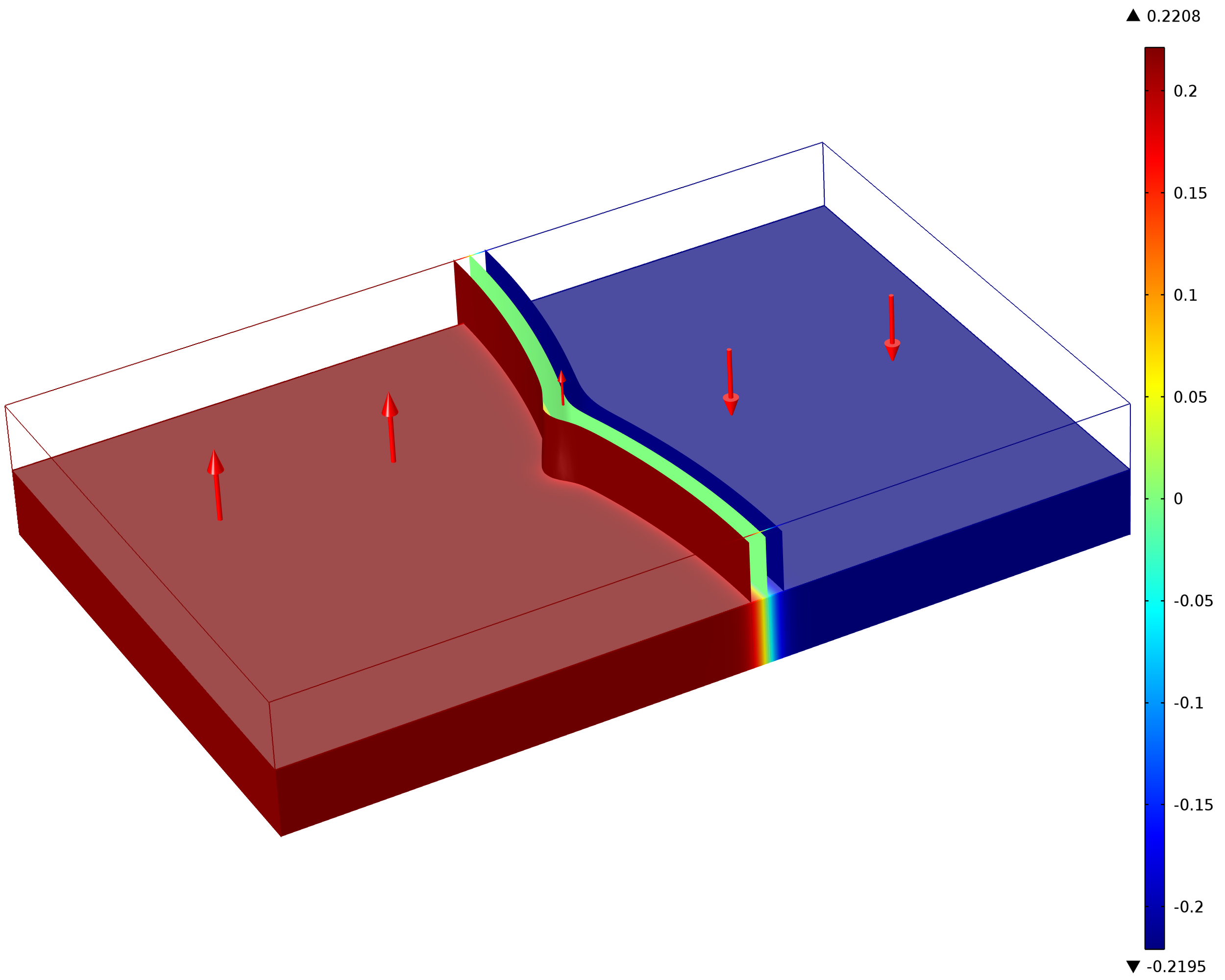}
		\label{fig:PFMres:a}
	}
	\subfigure[Planar mode (PM) aging]{
		\includegraphics[width=0.48\textwidth]{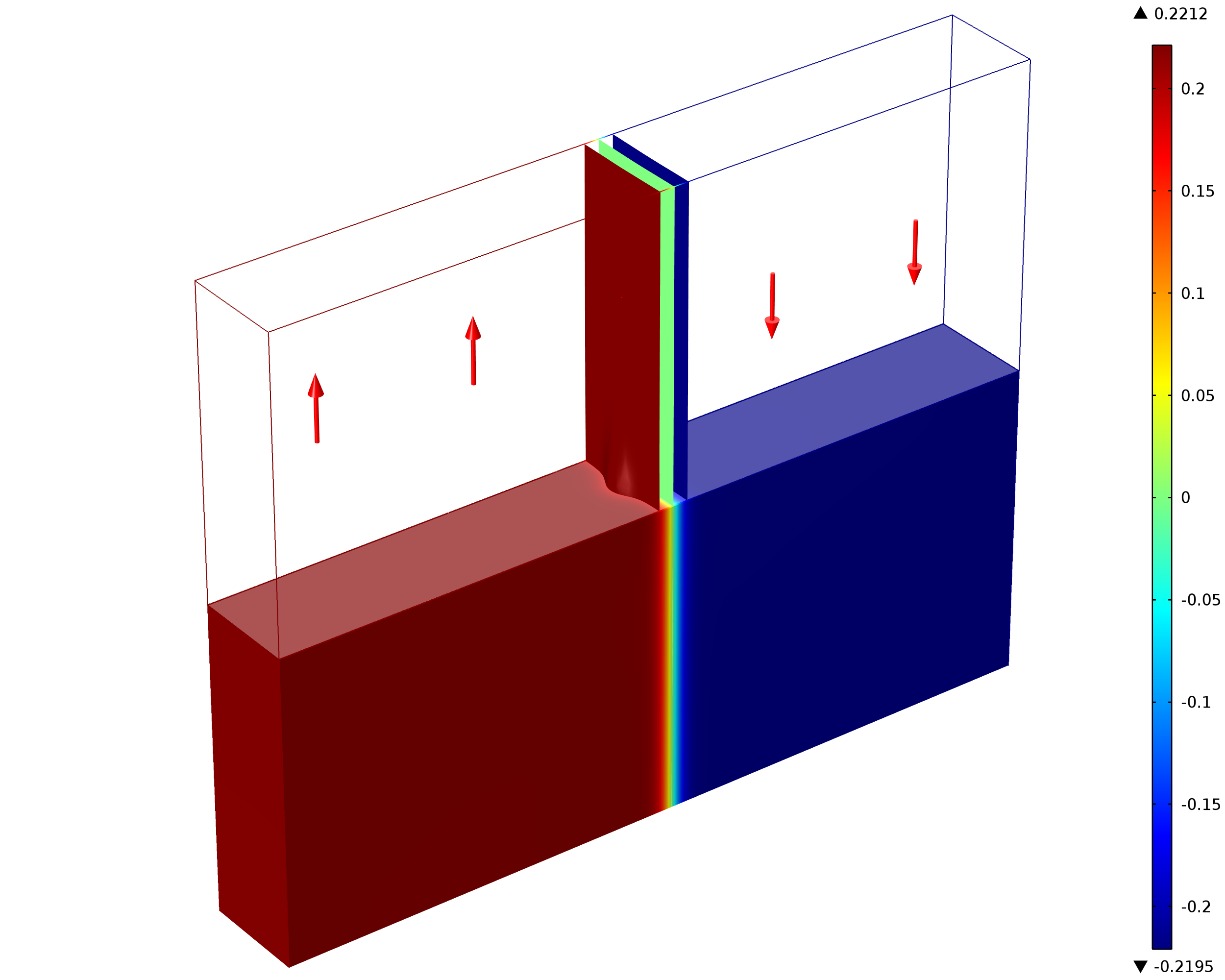}
		\label{fig:PFMres:b}
	}
	\caption{Result of the phase-field model simulation of the \dw{180} domain wall interacting with a pinning center in BaTiO${}_3$ for $r_0=4$\,nm and $r=19$\,nm during aging. Two aging mechanisms were modeled: \subref{fig:PFMres:a} bending mode (BM) aging, where the dielectric response is controlled by bending movements of \dw{180} domain walls, and \subref{fig:PFMres:b} planar mode (PM) aging, where the planar \dw{180} domain walls oscillate around the pinning centers. Magnitude of the external field is $E=180$kV/m. Color on the surface indicates the value of the $P_3$ component of the polarization vector. Blue and red colors indicate the values of $P_3$ equal to -0.221~C/m${}^2$ and 0.221~C/m${}^2$, respectively.}
	\label{fig:PFMres}
\end{figure}
%---- Figure 3
%
Figure~\ref{fig:PFMres} shows results of the numerical simulations of the two considered aging mechanisms in BT considering $r_0=4$\,nm, $r=19$\,nm, and $E=180$\,kV/m. 
It indicates the clear differences in the reversible \dw{180} domain wall motion during PM and BM aging mechanisms. 
Spatially dependent bending displacement of the curved domain wall, which dominates the BM aging is clearly presented in Fig.~\ref{fig:PFMres:a}.
On the other hand, Fig.~\ref{fig:PFMres:b} shows the spatially uniform displacement of the planar \dw{180} domain wall is characteristic for the PM aging.
Numerical results make it possible to estimate the domain wall displacements. The displacement of the domain wall in the PM mode is about $\pin{\delta}=0.45$~nm in BT at 104~$^\circ$C and considering the conditions specified in Fig.~\ref{fig:PFMres}.
The displacement of the apex of the domain wall during the BM mode is about 1.82~nm under the same external conditions. 
This gives the maximum deflection of the bent domain wall $\ben{\delta}^{\star}=1.37$~nm [See Eq.~(\ref{eq:C:DeltaBen-is})]. 
In the PT single crystal, the displacement of the domain wall is 0.15~nm in the PM at external electric field 441~kV/m, temperature 25~$^\circ$C, and $r=25$~nm and $r_0=4$~nm. 
The displacement of the apex of the domain wall during the BM mode is about 0.38~nm under the same external conditions. 
The value of the maximum deflection due to domain wall bending is $\ben{\delta}^{\star}=0.28$~nm. 
It should be noted that the simulations were computed under the conditions of relatively high pinning centers concentrations, which yields to the small domain wall displacements (compared to the lattice constant) at the given magnitude of the external field.

%----

The result of the numerical simulations of aging gives the spatial distribution of polarization $P_i$ at a given external field $E$.
The field dependence of the lattice polarization $P_{\rm up}$ and $P_{\rm dn}$ and the field dependence of the macroscopic polarization $P_f$ in the whole ferroelectric RVE were computed according to the following formulae:
\begin{subequations}
	\label{eq:V:PostProc}
\begin{align}
	\label{eq:V:PostProc:Pdn}
\iftwocolumn\hspace{-5mm}\else \fi 
	P_{\rm dn}(E) &= \frac 1{lr_yr_z}\int_{0}^{l} dx \int_{0}^{r_y} dy \int_{0}^{r_z} P_3(E,\,x,\,y,\,z)\,dz,\\
	\label{eq:V:PostProc:Pup}
\iftwocolumn\hspace{-5mm}\else \fi 
	P_{\rm up}(E) &= \frac 1{lr_yr_z}\int_{a-l}^{a} dx \int_{0}^{r_y} dy \int_{0}^{r_z} P_3(E,\,x,\,y,\,z)\,dz,\\
	\label{eq:V:PostProc:Pf}
\iftwocolumn\hspace{-5mm}\else \fi 
	P_{f}(E) &= \frac 1{ar_yr_z}\int_{0}^{a} dx \int_{0}^{r_y} dy \int_{0}^{r_z} P_3(E,\,x,\,y,\,z)\,dz,
\end{align}
\end{subequations}
where $l$ is the distance at which the averaged dielectric response of the crystal lattice is not affected by the movement of the domain wall. Such a condition is usually satisfied for $l\approx a/4$.
Thickness of the domain wall $a_w$ was numerically estimated from the numerical simulations according to the formula:
\begin{equation}
\label{eq:V:aw-num}
a_w \approx \frac{
		4r_yr_z P_0^2
	}{
		\int_{0}^{a} dx \int_{0}^{r_y} dy \int_{0}^{r_z} 
		\left(
			\partial P_3/\partial x
		\right)^2
		(x,\,y,\,z)\,dz
	}.
\end{equation}

%----

%---- Table 4
{\setlength\tabcolsep{1mm}
\begin{table}
	\caption{Lattice values fitted from the phase field model simulations.}
	\label{tab:DielLatPhaseField}
\iftwocolumn
	\begin{tabular}{lrrrp{15mm}p{15mm}r}
\else
	\begin{tabular}{lrrrrrr}
\fi 
		\hline \hline
		Material & 
		$T$ & 
		$P_0$ & 
		$\varepsilon_c/\varepsilon_0$ &
		$\beta_c$ &
		$\gamma_c$ &
		$a_w$\\
		& 
		[${}^\circ$C] & 
		[C/m${}^2$] & 
		[1] &
[10${}^{-17}\times$
		C/V${}^2$] &
[10${}^{-25}\times$
		m\,C/V${}^3$] &
		[nm]
		\\ \hline 
		PbTiO${}_3$ & 
		25 & 
		0.61 & 
		53.5 &
		1.57 &
		9.5 &
		2.00
		\\
		& 
		50 & 
		0.59 & 
		58.4 &
		1.97 &
		10.5 &
		2.09
		\\
		& 
		75 & 
		0.57 & 
		64.6 &
		2.63 &
		12.0 &
		2.20
		\\
		BaTiO${}_3$ & 
		104 & 
		0.22 & 
		273.1 &
		98.4 &
		404.8 &
		1.44
		\\
		& 
		106 & 
		0.22 & 
		289.3 &
		117.8 &
		486.3 &
		1.48
		\\
		& 
		108 & 
		0.22 & 
		289.3 &
		114.2 &
		450.3 &
		1.48
		\\ \hline\hline
	\end{tabular}
\end{table}
}
%---- Table 4
%
The computed values of $P_{\rm dn}(E)$ and $P_{\rm up}(E)$ were used for the identification of parameters that control the nonlinear dielectric response of the crystal lattice, i.e. $P_0$, $\varepsilon_c$, $\beta_c$, and $\gamma_c$, from Eqs.~(\ref{eq:II:Pup-Pdn}) using the method of least squares. 
The resulting numerical values are presented in Tab.~\ref{tab:DielLatPhaseField}. 

%----

%---- Figure 4
\begin{figure*}[t]
	\centering
	\subfigure[\quad PbTiO${}_3$ (BM aging)]{
		\includegraphics[width=0.48\textwidth]{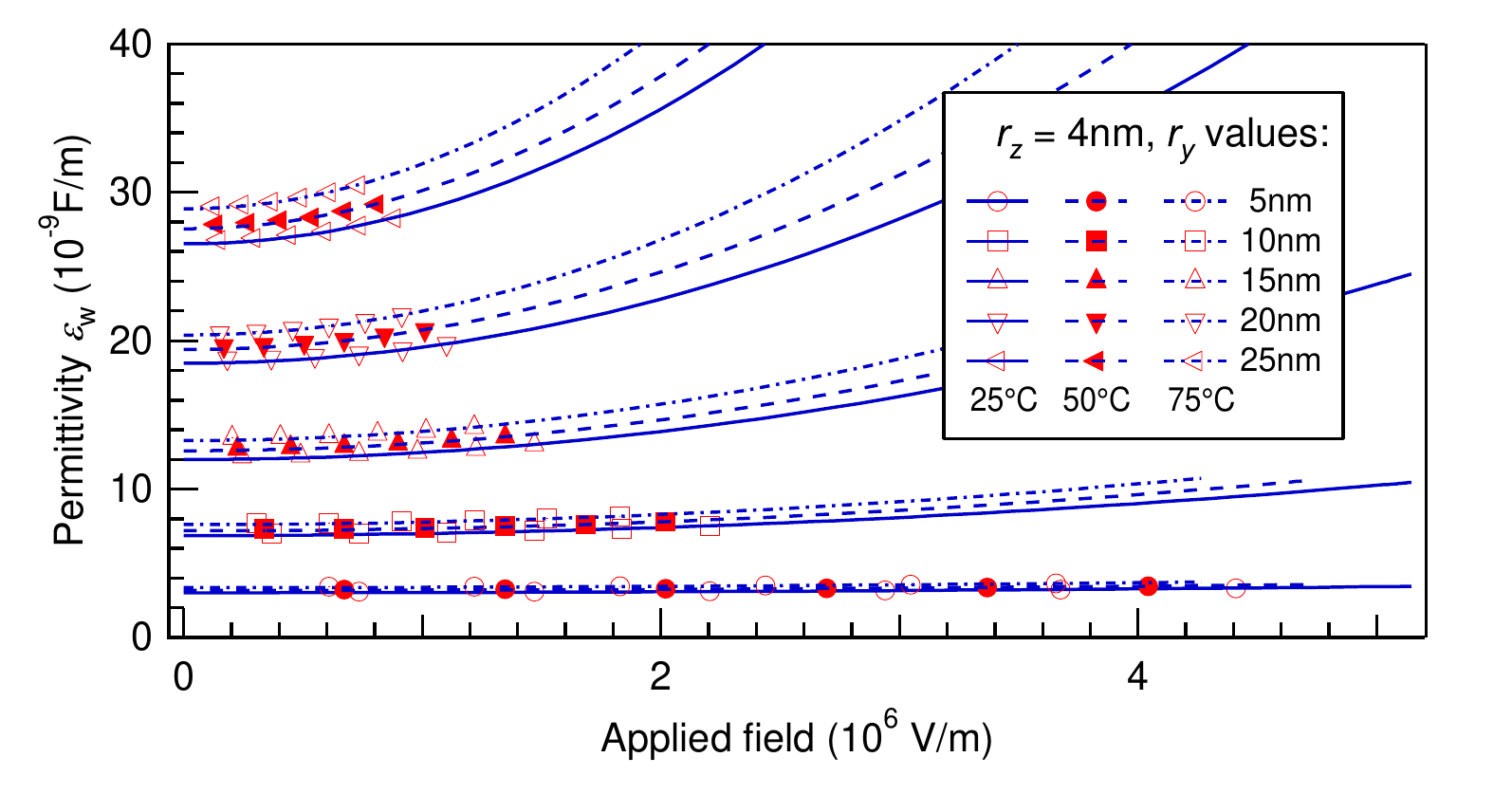}
		\label{fig:EpsW-E:a}
	}
	%\vspace{-3mm}
	\subfigure[\quad PbTiO${}_3$ (PM aging)]{
		\includegraphics[width=0.48\textwidth]{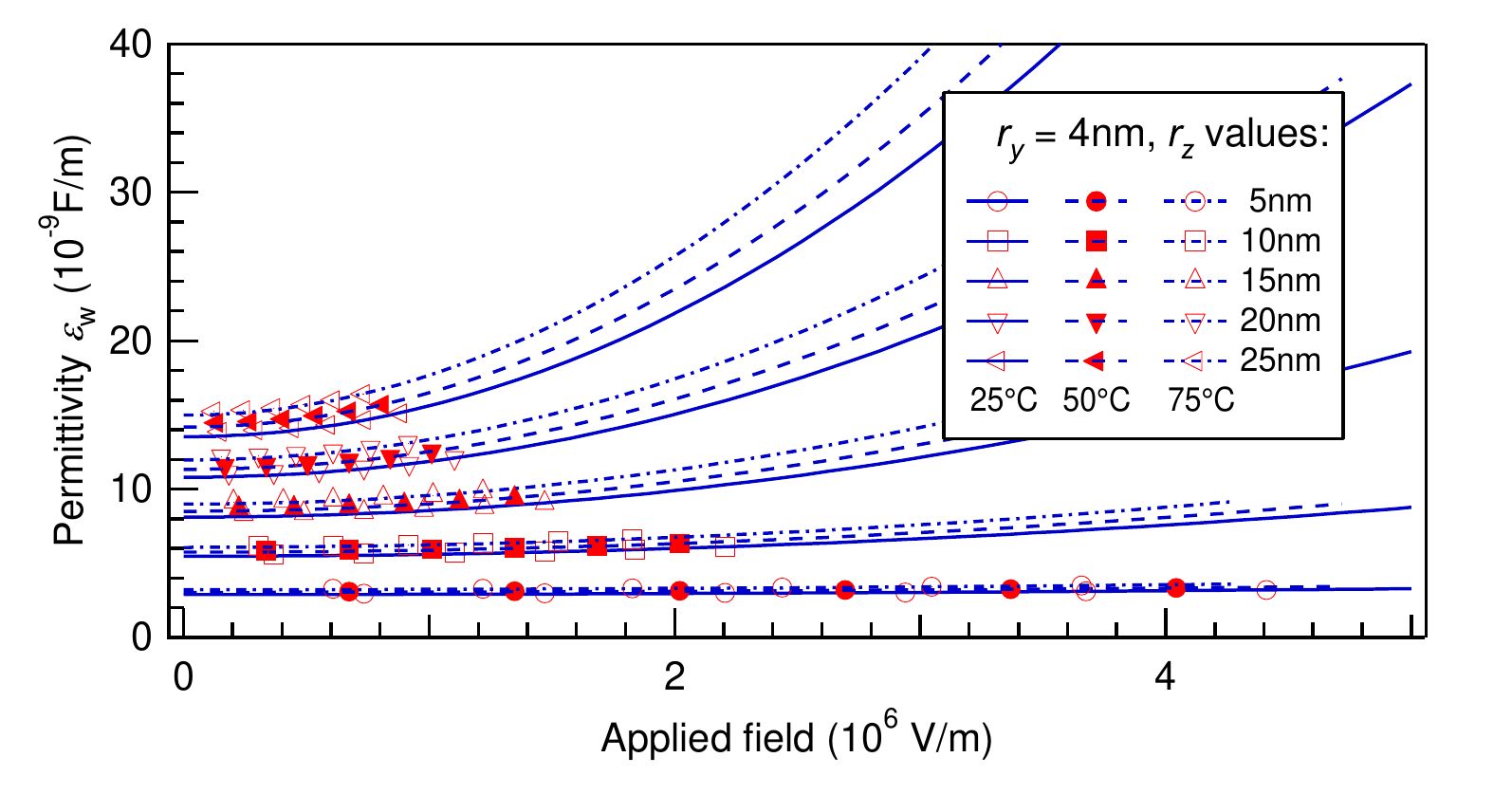}
		\label{fig:EpsW-E:b}
	}
	\subfigure[\quad BaTiO${}_3$ (BM aging)]{
		\includegraphics[width=0.48\textwidth]{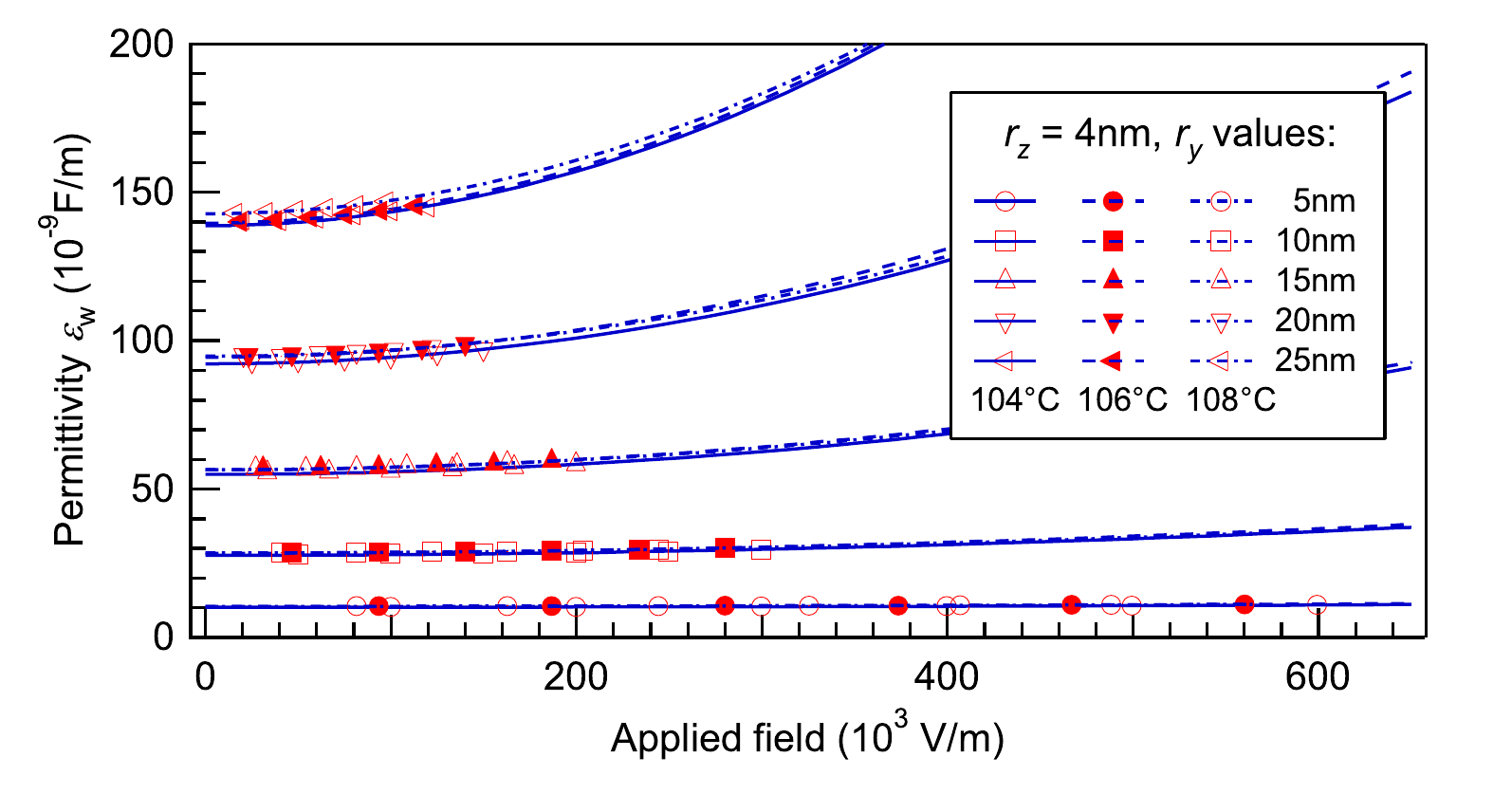}
		\label{fig:EpsW-E:c}
	}
	%\vspace{-3mm}
	\subfigure[\quad BaTiO${}_3$ (PM aging)]{
		\includegraphics[width=0.48\textwidth]{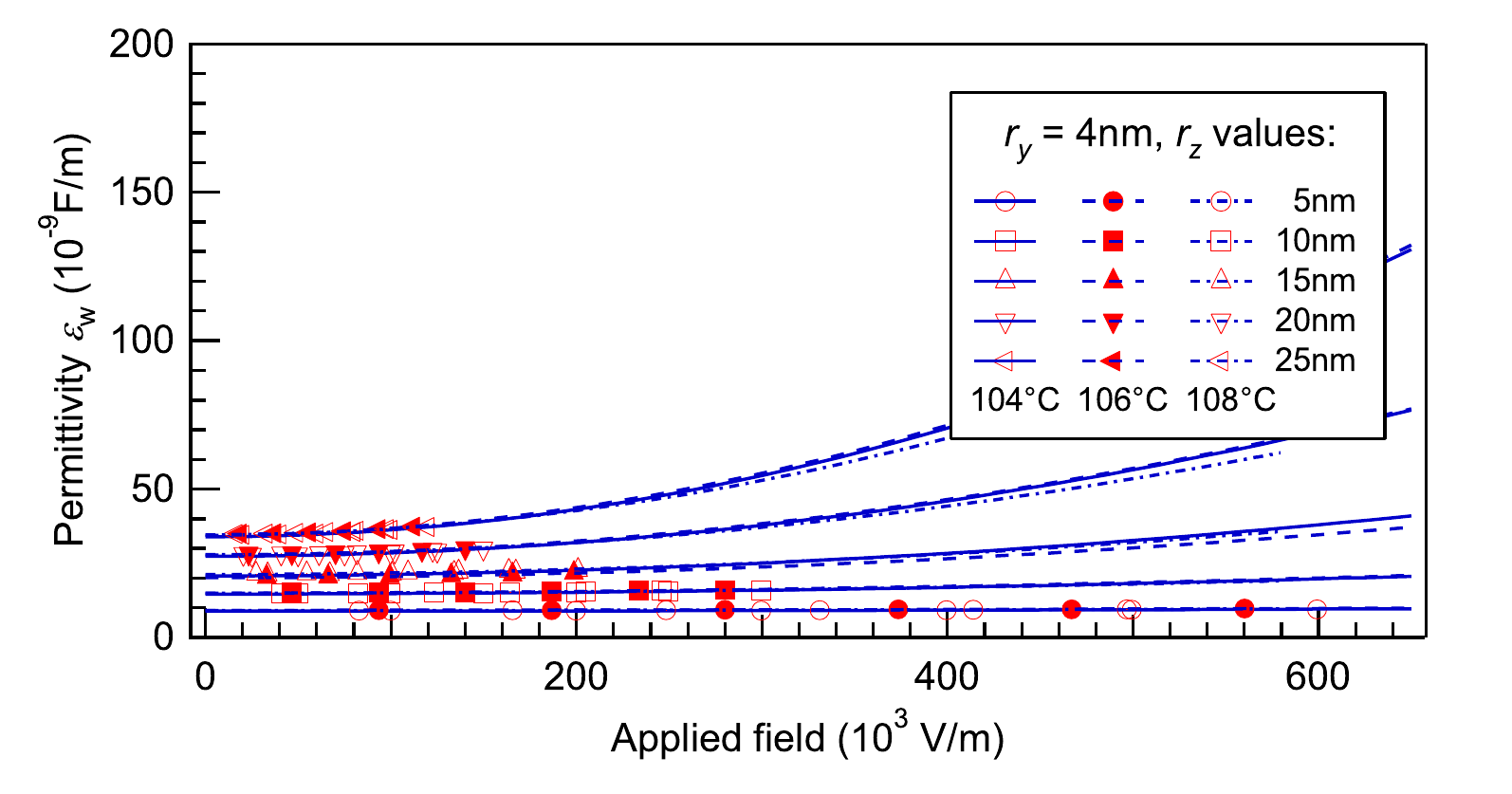}
		\label{fig:EpsW-E:d}
	}
	\caption{
Evolution of the electric field dependence of extrinsic permittivity computed from numerical phase-field model simulations of BM aging \subref{fig:EpsW-E:a} and \subref{fig:EpsW-E:c}; and PM aging \subref{fig:EpsW-E:b} and \subref{fig:EpsW-E:d}. Numerical simulations computed for PbTiO${}_3$ are presented in graphs \subref{fig:EpsW-E:a} and \subref{fig:EpsW-E:b}. Graphs \subref{fig:EpsW-E:c} and \subref{fig:EpsW-E:d} show results for BaTiO${}_3$.
}
	\label{fig:EpsW-E:}
\end{figure*}
%---- Figure 4
%
In the next step, the field dependence of the net spontaneous polarization $P_N(E)$ can be expressed from Eq.~(\ref{eq:II:Pf-is}) in the form:
\begin{equation}
	\label{eq:IV:PNfromPF}
	P_N(E) = \frac{P_f(E) - \varepsilon_c E - \gamma_c\,E^3}{1 - \beta_c E^2/P_0},
\end{equation}
where $P_f(E)$ is the macroscopic polarization at given external field $E$ computed using Eq.~(\ref{eq:V:PostProc:Pf}) from the results of numerical simulations.
It follows from Eq.~(\ref{eq:IV:Epsw-def}) that the field dependence of the extrinsic permittivity $\varepsilon_e(E)$ can be expressed as:
\begin{equation}
	\label{eq:IV:EpsWfromPN}
	\varepsilon_w(E) = \frac{P_N(E) - P_N(0)}{E}.
\end{equation}
The values of $\varepsilon_w(E)$ are then fitted to Eq.~(\ref{eq:IV:Epsw-ser}) and numerical values of the small-signal extrinsic permittivity $\varepsilon_L$ and the nonlinearity constant of the field dependence of extrinsic permittivity $\gamma_w$ at given stage of aging are computed using the method of least squares.  
The evolution of electric field dependence of extrinsic permittivity computed from numerical simulations of the considered PM and BM aging mechanisms is presented in Fig.~\ref{fig:EpsW-E:}.

%----

%---- Figure 5
\begin{figure}[t]
	\centering
	\subfigure[\quad PbTiO${}_3$]{
		\includegraphics[width=0.48\textwidth]{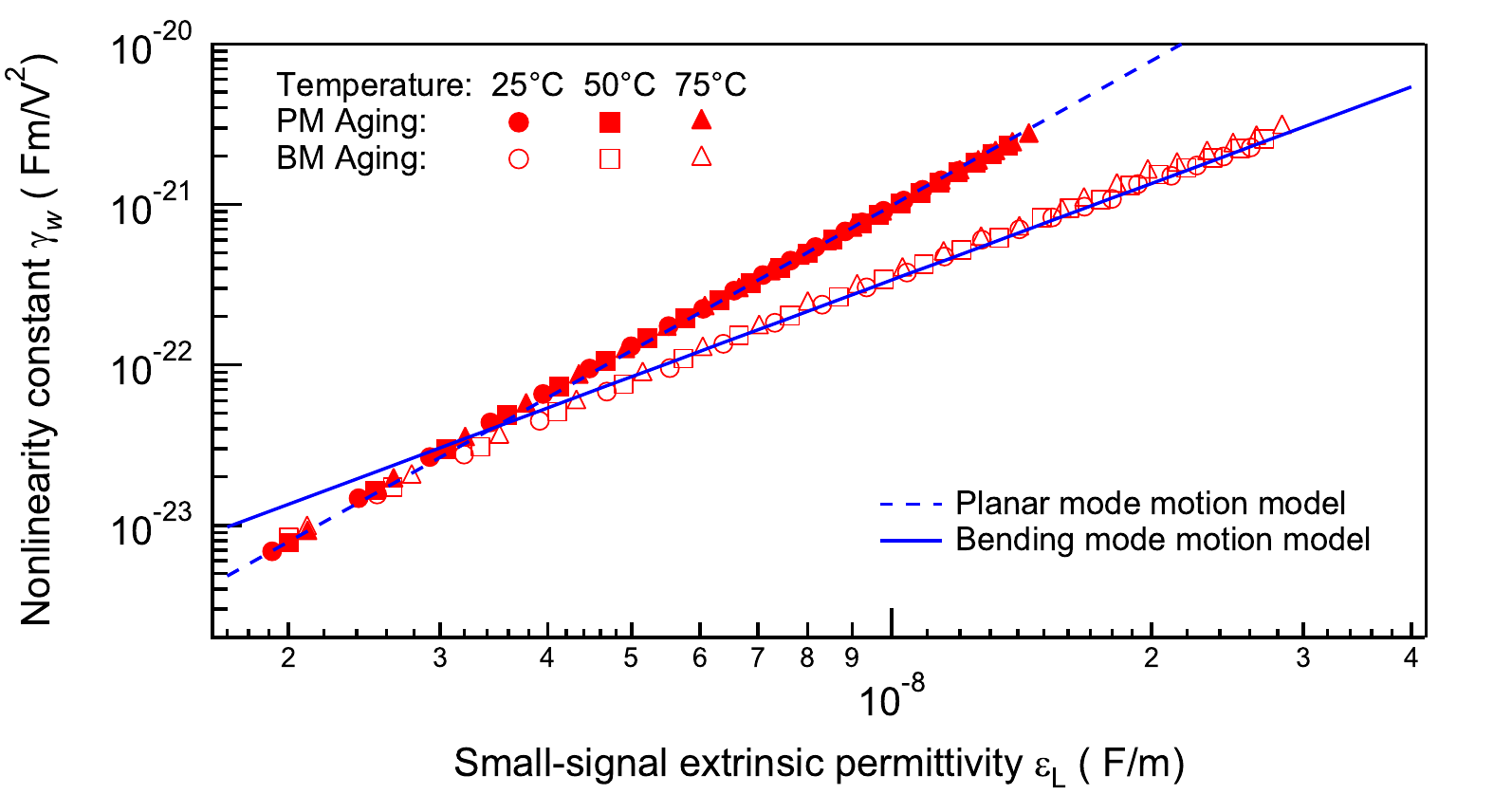}
		\label{fig:GammaW-EpsL:a}
	}
	\subfigure[\quad BaTiO${}_3$]{
		\includegraphics[width=0.48\textwidth]{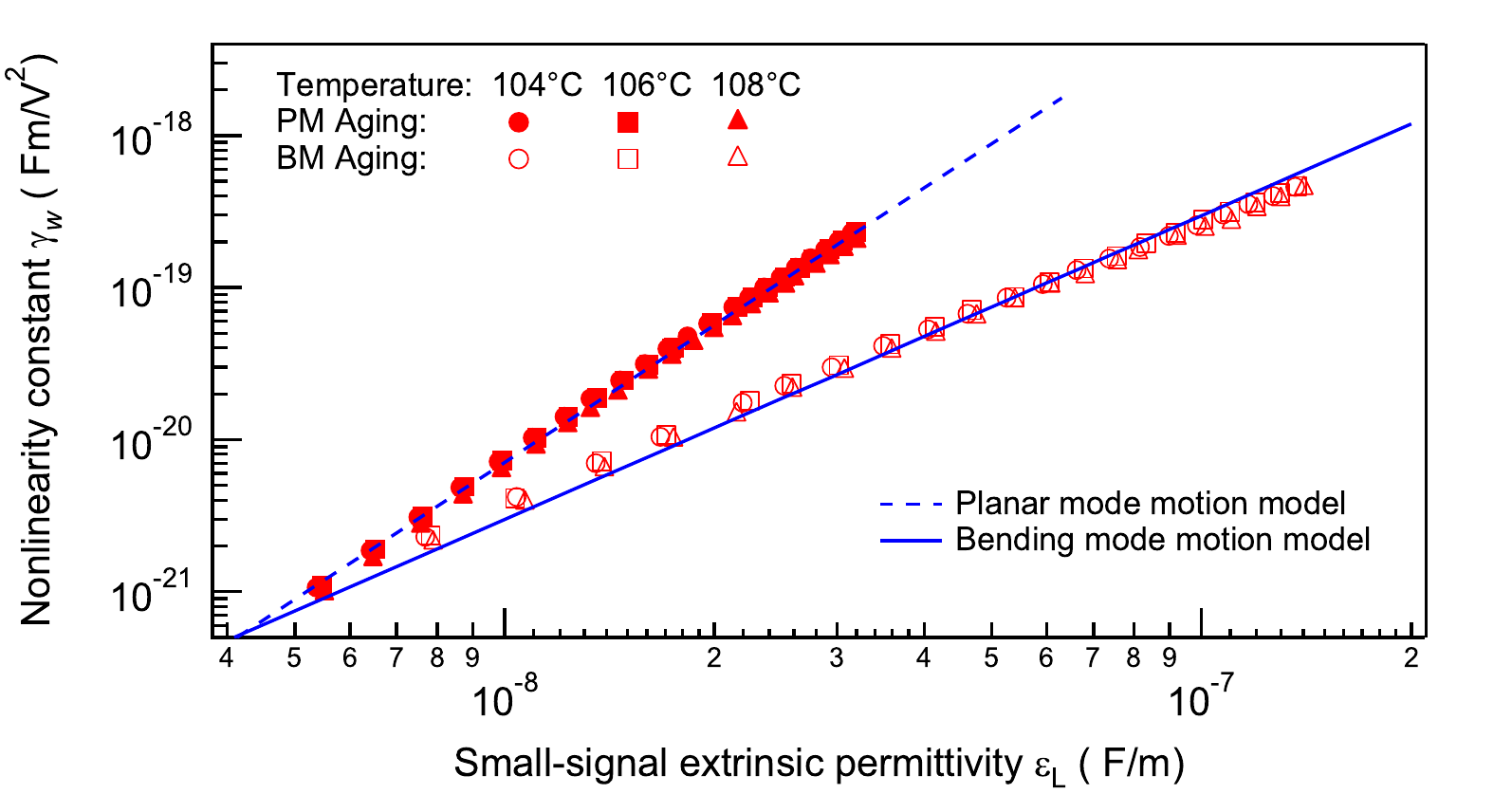}
		\label{fig:GammaW-EpsL:b}
	}
	\caption{Dielectric nonlinearity constant $\gamma_w$ versus small-signal extrinsic permittivity $\varepsilon_{L}$ during PM (filled markers) and BM (open markers) aging. Numerical computations were carried out for PbTiO${}_3$~\subref{fig:GammaW-EpsL:a} and BaTiO${}_3$~\subref{fig:GammaW-EpsL:b}. Dashed and solid lines indicate fitting to the model of PM domain wall motion [see Eq.~(\ref{eq:IV:GammaVsEpsL:pin})] and BM domain wall motion [see Eq.~(\ref{eq:IV:GammaVsEpsL:ben})], respectively.}
	\label{fig:GammaW-EpsL}
\end{figure}
%---- Figure 5
%
Figure~\ref{fig:GammaW-EpsL} shows plots of the dielectric nonlinearity constant $\gamma_w$ versus small-signal extrinsic permittivity $\varepsilon_{L}$ in the logarithmic scales for PT~[Fig.~\ref{fig:GammaW-EpsL:a}] and BT~[Fig.~\ref{fig:GammaW-EpsL:b}], respectively. 
The values of $\varepsilon_L$ and $\gamma_w$ correspond to the best fit of the field dependence of the nonlinear extrinsic permittivity at different stages during the considered PM and BM aging mechanisms.
Filled and empty markers correspond to the PM and BM aging mechanisms, respectively.
Both graphs in Fig.~\ref{fig:GammaW-EpsL} clearly indicate that the curves presented by filed and empty markers differ in their slope.
Curves, which correspond to the PM aging mechanism have the slope equal approximately to 3.
On the other hand, the slope of curves corresponding to the BM aging mechanism equals to 2.
It should be however noted that graphs shown in Fig.~\ref{fig:GammaW-EpsL} were displayed for the parameters of extrinsic nonlinear permittivity, which were extracted using the numerical procedure described above. 
The essential point of this numerical procedure is the knowledge of dielectric response of the crystal lattice, which is possible to compute from the numerical simulations with a high accuracy. 
This is a principal difference from the analysis of experimental data where the dielectric response of the crystal lattice is sometimes difficult to obtain. 

%----

%---- Figure 6
\begin{figure}[t]
	\centering
	\subfigure[\quad PbTiO${}_3$]{
		\includegraphics[width=0.48\textwidth]{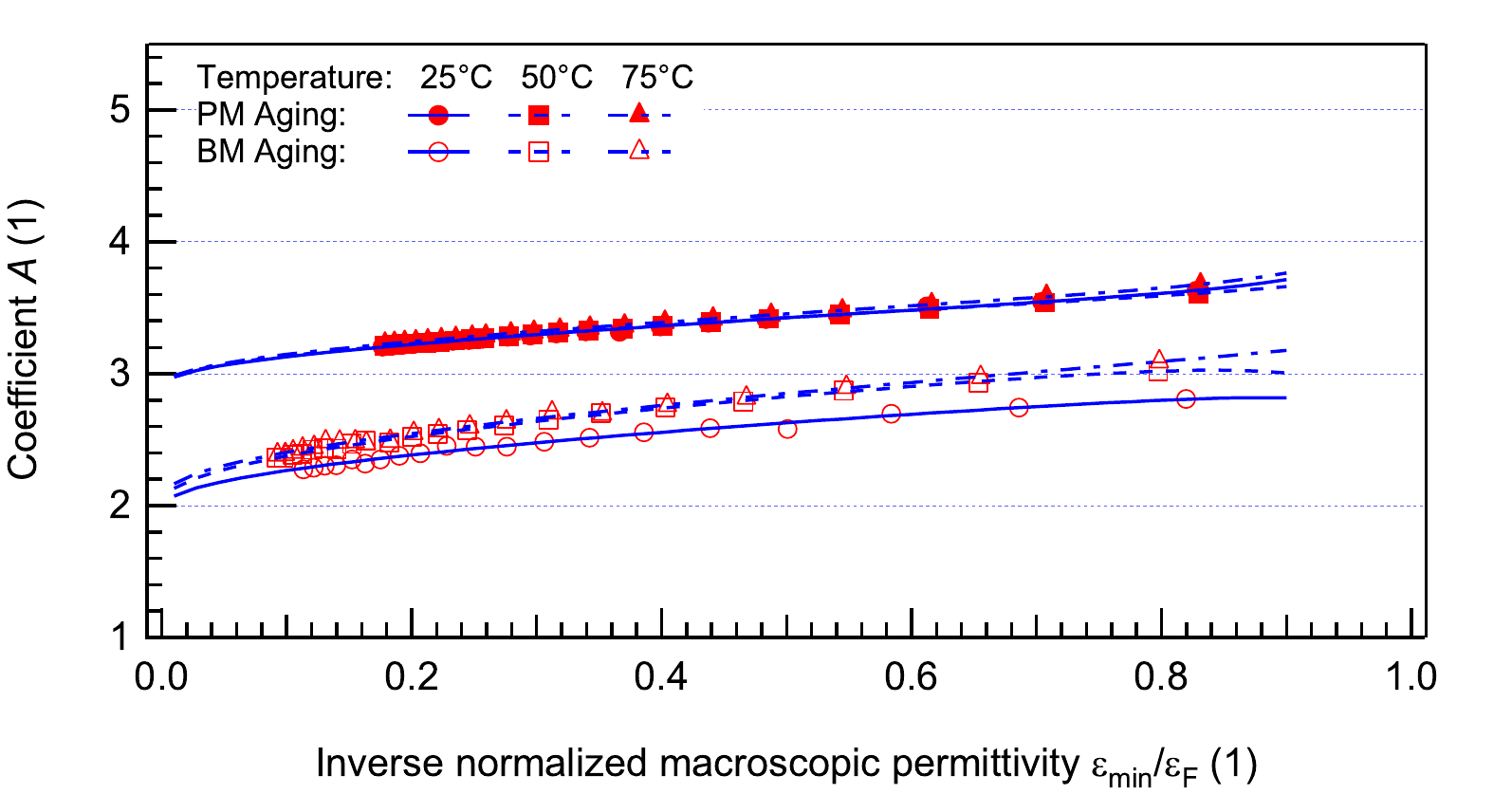}
		\label{fig:CoefA-EpsF:a}
	}
	\subfigure[\quad BaTiO${}_3$]{
		\includegraphics[width=0.48\textwidth]{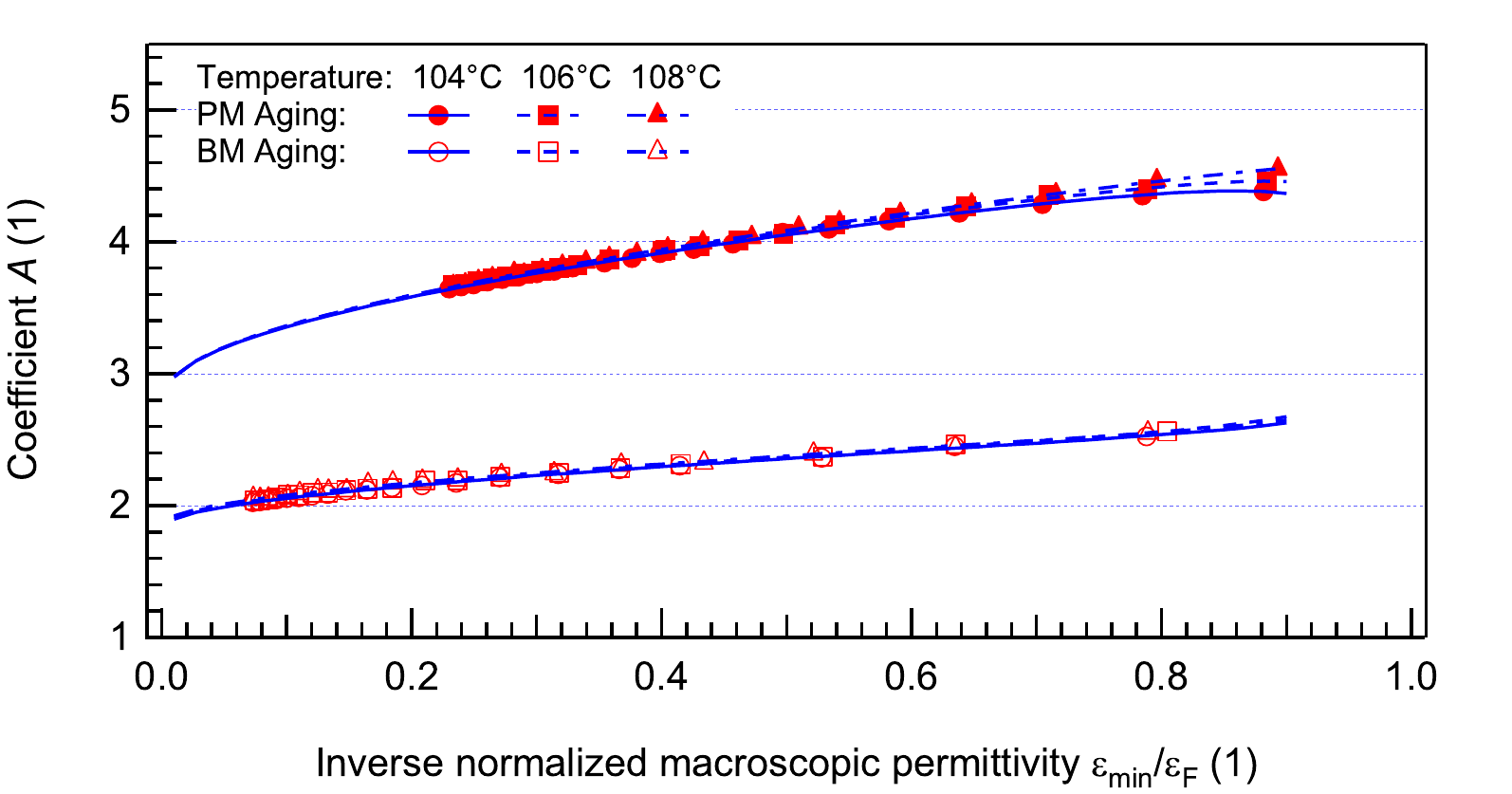}
		\label{fig:CoefA-EpsF:b}
	}
	\caption{Parameter $A$ computed according to Eq.~(\ref{eq:IV:A:def}) as a function of the inverse macroscopic small-signal permittivity $\varepsilon_F$ normalized to its minimal value measured during aging experiment for the PM (filled markers) and BM (open markers) aging mechanisms. It is seen that values of the parameter $A$ can be used for the identification of aging mechanism. Numerical computations were carried out for PbTiO${}_3$~\subref{fig:CoefA-EpsF:a} and BaTiO${}_3$~\subref{fig:CoefA-EpsF:b}.}
	\label{fig:CoefA-EpsF}
\end{figure}
%---- Figure 6
%
In the next step, it is necessary to verify whether it is possible to determine the dominant aging mechanism from measurements of the macroscopic dielectric response, which is given by the superposition of the time-constant intrinsic and time-dependent extrinsic contributions. 
Filled and open markers in Fig.~\ref{fig:CoefA-EpsF} show values of the parameter $A$ computed from the numerical simulations of the PM and BM aging experiments, respectively.
In order to estimate the correlation between the dielectric nonlinearity constant $\gamma$ and small-signal permittivity $\varepsilon_F$, which is expressed by the parameter $p$ [see Eq.~(\ref{eq:IV:Gamma:Is})], the numerically computed values of the parameter $A$ have been fitted to the function given by Eq.~(\ref{eq:IV:ATaylor:Fit}).

%----

%---- Table 4
{\setlength\tabcolsep{1mm}
	\begin{table}
		\caption{Results of the numerical simulation of BM and PM aging mechanisms. The value of parameter $p$ was extrapolated using Eqs.~(\ref{eq:IV:ATaylor:Fit}) and (\ref{eq:IV:AgingExtrapol:n}). It is seen that the extrapolated values of $p$ correspond to the theoretical values 2 and 3 for BM and PM aging mechanisms. The values of lattice permittivity $\varepsilon_c$ were calculated from simulations of aging using Eqs.~(\ref{eq:IV:ATaylor:Fit}) and (\ref{eq:IV:AgingExtrapol:EpsC}). It is seen that the values of $\varepsilon_c$ calculated from aging simulations are in the agreement with values computed using Eqs.~(\ref{eq:II:Pup-Pdn}) and (\ref{eq:V:PostProc:Pdn}) and presented in Tab.~\ref{tab:DielLatPhaseField}.}
		\label{tab:AgingResult}
\iftwocolumn
		\begin{tabular}{lrrrrrr}
\else
			\begin{tabular}{lrrrrrr}
\fi 
				\hline \hline
				Material & 
				$T$ [${}^\circ$C]& 
				$\varepsilon_c/\varepsilon_0$ [1]&
				$p$ [1]&
				$\varepsilon_c/\varepsilon_0$ [1]&
				$p$ [1]&
				$\varepsilon_c/\varepsilon_0$ [1]\\
				& 
				& 
				Tab.~\ref{tab:DielLatPhaseField} &
				\multicolumn{2}{r}{BM Aging} &
				\multicolumn{2}{r}{PM Aging} 
				\\ \hline 
				PbTiO${}_3$ & 
				25 & 
				53.5 &
				1.8 &
				82.6 &
				2.8&
				36.0
				\\
				& 
				50 & 
				58.4 &
				1.8 &
				87.3 &
				2.8&
				38.2
				\\
				& 
				75 & 
				64.6 &
				1.9 &
				88.2 &
				2.8&
				41.8
				\\
				BaTiO${}_3$ & 
				104 & 
				273.1 &
				1.7 &
				218.7 &
				2.5&
				331.6
				\\
				& 
				106 & 
				289.3 &
				1.7 &
				220.4 &
				2.5&
				347.2
				\\
				& 
				108 & 
				289.3 &
				1.7 &
				224.7 &
				2.5&
				347.9
				\\ \hline\hline
			\end{tabular}
		\end{table}
	}
	%---- Table 4
	%

%----

Result of this fitting procedure are presented in Tab.~\ref{tab:AgingResult}.
The value of parameter $p$ was extrapolated using Eqs.~(\ref{eq:IV:ATaylor:Fit}) and (\ref{eq:IV:AgingExtrapol:n}). 
It is seen that the extrapolated values of $p$ correspond to the theoretical values 2 and 3 for BM and PM aging mechanisms, respectively. 
The values of lattice permittivity $\varepsilon_c$ were calculated from the simulations of aging using Eqs.~(\ref{eq:IV:ATaylor:Fit}) and (\ref{eq:IV:AgingExtrapol:EpsC}). 
It is seen that the values of $\varepsilon_c$ calculated from aging simulations are in the good agreement with values computed using Eqs.~(\ref{eq:II:Pup-Pdn}) and (\ref{eq:V:PostProc:Pdn}) and presented in Tab.~\ref{tab:DielLatPhaseField}.

%----

Therefore, it was concluded that the measurement of nonlinear dielectric response may serve as a tool for identification of the mechanism of reversible domain wall motion, which controls the dielectric response during the processes of aging in the polydomain ferroelectric samples.
In addition, it is possible to estimate important material parameters and microstructure of the pinning centers of the ferroelectric system by proper analysis of nonlinear dielectric data.

%====
%
%====
\section{Conclusions}
\label{sec:Conslusions}

In this Article, we presented the main results of our detailed theoretical analysis of the so-called extrinsic contributions to macroscopic dielectric constant due to reversible \dw{180} domain wall movements in polydomain ferroelectric samples.
It was considered that the reversible movement of \dw{180} domain walls is hindered by their interaction with crystal lattice defects, which behave as pinning centers.
It was demonstrated that the general reversible motion of \dw{180} domain walls can be decomposed into two modes: (i) oscillation of planar walls around a pinning center (PM mode) and (ii) bending movements of domain walls between pinning centers (BM mode).
We have expressed simple formulae for the extrinsic contributions to the small-signal permittivity and dielectric nonlinearity constant due to the PM and BM modes as functions of dielectric parameters of the crystal lattice and the parameters of the pinning center distribution on the domain wall.

%----

\begin{table*}[t]
\centering
{\setlength\tabcolsep{2mm}
\renewcommand\arraystretch{1.3}
\caption{The table concludes the main results of this work, i. e. identification of the microscopic mechanism that controls the reversible domain wall motion at small and high concentration of pinning centers in polydomain ferroelectrics. In this work, two microscopic mechanisms responsible for dielectric aging are considered: (i) redistribution of pinning centers along certain directions and (ii) uniform accumulation of pinning centers on the domain wall. Figure~\ref{fig:DWPinning} shows two modes of the reversible domain wall motion considered in this work: (a) uniform motion of the domain wall as a plane (PM) and (b) bending movement of free segments of the domain wall between pinning centers (BM). 
Identification of the mode of the reversible domain wall motion using the measurements of nonlinear permittivity it is demonstrated in Sec.~\ref{sec:Aging}. Table briefly reviews the origins of the PM and BM domain wall motion at small and high pinning centers concentrations.}
\label{tab:AgingConclusion}
\begin{tabular}{p{0.25\textwidth}|p{0.32\textwidth}|p{0.33\textwidth}} 
    \hline\hline
    Domain wall motion& \multicolumn{2}{c}{Pinning centers concentration}  \\ \cline{2-3}
    & ``Small" ($n\ll n_{\rm crit}$) & ``High" ($n\gg n_{\rm crit}$)\\ \hline
    Bending mode \newline ($\gamma \propto \varepsilon^2$) & Aging can originate from the both mechanisms. & Aging originates from redistribution of pinning centers along certain directions inside the domain wall planes. \\  \hline
    Planar mode \newline ($\gamma \propto \varepsilon^3$) & Not available & Aging originates from uniform accumulation of pinning centers at domain walls.\\
    \hline\hline
\end{tabular}
}
\end{table*}
Therefore, it is natural to expect that the extrinsic contributions may change when the distribution of pinning centers is rearranged during dielectric aging of polydomain ferroelectric.
We have find out that the PM mode of domain wall motion is controlled solely by the concentration $n$ of pinning center on the domain wall and that it is insensitive to the anisotropy $\eta$ of the pinning centers distribution on the domain wall. 
On the other hand, the BM mode is controlled by the anisotropy parameter $\eta$ in addition to the concentration $n$.
It can be concluded from the above statements that, when the PM domain wall motion is identified as a dominant contribution to the macroscopic dielectric response during aging, the microscopic evolution of the pinning centers is dominated by the uniform accumulation of the pinning centers on the domain wall and not by the redistribution and alignment of pinning centers along certain directions on the domain wall.
On the contrary, identification of the microscopic detail in the evolution of pinning centers on the domain wall is more difficult when the BM domain wall motion is identified as a dominant in the macroscopic dielectric response.
The reason is that PM mode becomes negligible compared to BM mode in the extrinsic dielectric response in systems with low concentration of pinning centers and the identification of the mechanism of the pinning centers evolution is impossible in this situation. 
However, BM mode may suggest that the evolution of pinning centers is dominated by the alignment of pinning centers along certain directions in systems with a high concentration of the pinning centers.
These results are briefly concluded in Tab.~\ref{tab:AgingConclusion}.

%----

In order to identify the dominant mode of the reversible domain wall motion, we have suggested a method based on the nonlinear analysis of macroscopic dielectric response of polydomain films.
The identification can be obtained by plotting the nonlinearity constant $\gamma$ of the macroscopic dielectric response versus the small-signal macroscopic permittivity $\varepsilon_F$ in the logarithmic scale (see Fig.~\ref{fig:GammaW-EpsL}).
When the condition $\Delta\log\gamma/\Delta\log\varepsilon_F\approx 3$ is valid, it is an indication that the dominant mode is the oscillation of planar \dw{180} domain walls around pinning centers. 
Bending movements of \dw{180} domain walls are indicated by the condition $\Delta\log\gamma/\Delta\log\varepsilon_F\approx 2$.
In addition, we have developed a refinement of this simple method, which is applicable for the analysis of nonlinear dielectric data measured in the intermediate stages of the aging process and in systems with small extrinsic contributions compared to intrinsic ones [see Eq.~(\ref{eq:IV:A:def}), (\ref{eq:IV:ATaylor:Fit}) and (\ref{eq:IV:AgingExtrapol})].

%----

In order to support our statements, we have performed several phase-field model simulations of two aging mechanisms in polydomain ferroelectric: planar mode aging, where the macroscopic dielectric response is controlled by the oscillation of planar \dw{180} domain walls around pinning centers, and bending mode aging, where the dielectric response is controlled by the bending movements of \dw{180} domain walls. 
We have processed numerically computed data according to our suggested method and demonstrated that it is possible to identify which mechanism of the domain wall motion is dominant in a particular aging mechanism using simple macroscopic dielectric measurements and proper analysis of the nonlinear response.
We have further demonstrated that using the proper analysis of nonlinear dielectric data, it is possible to obtain information on the domain pattern microstructure and further material parameters of the ferroelectric polydomain system.

%----

\section*{Acknowledgment}
\label{sec:acknowledgment}

The support from the Czech Science Foundation, Project No. GA14-32228S is acknowledged. The research leading to these results has received also funding from the European Research Council under the EU 7th Framework Program (FP7/2007–2013)/ERC grant agreement no 268058, MOBILE-W.
Authors would like to thank Alexander K. Tagantsev and Dragan Damjanovic of Ceramics Laboratory, Swiss Federal Institute of Technology, Lausanne, Switzerland for reading the manuscript and many useful discussions.

%\bibliographystyle{apsrev}
%\bibliography{bibliography_dw}

\appendix

%====
%
%====
\section{Thermodynamic model of pinned \dw{180} domain wall}
\label{apndx:ThermoModel}

The thermodynamic model of the interaction of \dw{180} domain walls with pinning centers is based on the standard thermodynamic approach~\cite{zhirnov_contribution_1959,cao_theory_1991,hu_three-dimensional_1998}.
Introduction of the model will be carried out in three steps.
The definition of state quantities and thermodynamic potentials is presented in Subsection~\ref{apndx:ThermoModel:Pots}.
Subsection~\ref{apndx:ThermoModel:EoS} presents the equations of state governing the evolution of domain pattern. 
Finally, geometry of the representative volume element (RVE) model of the polydomain ferroelectric with a pinned \dw{180} domain wall and the boundary conditions are specified in Subsection~\ref{apndx:ThermoModel:RVE+BC}.

%----

For the sake of convenience of notation in this Appendix, the $x$, $y$, and $z$ axes of the coordinate system are denoted by symbols $x_1=x$, $x_2=y$, and $x_3=z$, respectively. 
Einstein summation is considered for indexes $i,\,j=1,\,2,\,3$ if not indicated otherwise.

\subsection{State quantities and thermodynamic potentials}
\label{apndx:ThermoModel:Pots}

In the thermodynamic model, we consider seven independent state quantities: scalar electrostatic potential $\varphi$, three components of polarization $P_i$, and three components of mechanical displacement $u_i$. 
For the purpose of clearer definition of thermodynamic potentials, it is convenient to introduce following physical quantities: electric field
\begin{subequations}
	\label{eq:A:SQ}
	\begin{equation}
	\label{eq:A:SQ:Ei}
	E_i = - \varphi_{,i}
	\end{equation}  
	where the symbol $(\cdot)_{,i}=\partial(\cdot)/\partial x_i$ stands for the partial derivative, elastic strain
	\begin{equation}
	\label{eq:A:SQ:eij}
	e_{ij} = \left(1/2\right)\left(u_{i,j}+u_{j,i}\right),
	\end{equation}  
	and the dielectric displacement
	\begin{equation}
	\label{eq:A:SQ:Di}
	D_{i} = \varepsilon_0 \varepsilon_B E_i + P_i,
	\end{equation}  
\end{subequations}
where symbols $\varepsilon_0$ and $\varepsilon_B$ stand for the permittivity of vacuum and the background permittivity, respectively. 

%----

In order to calculate the dielectric response, it is necessary to introduce the Helmholtz free energy (per unite volume), which is given by the sum of following contributions:
\iftwocolumn
\begin{multline}
\label{eq:A:psi-def}
\psi\left(P_i,\,P_{i,j},\,e_{ij},\,D_i\right)=\\
\psi^{(e)}_{\rm bulk} + \psi_{\rm ela} + \psi_{\rm es} + \psi_{\rm wall} + \psi_{\rm ele} + \psi_{\rm pin},
\end{multline}
\else
\begin{equation}
\label{eq:A:psi-def}
\psi\left(P_i,\,P_{i,j},\,e_{ij},\,D_i\right)=
\psi^{(e)}_{\rm bulk} + \psi_{\rm ela} + \psi_{\rm es} + \psi_{\rm wall} + \psi_{\rm ele} + \psi_{\rm pin},
\end{equation}
\fi
where
\begin{subequations}
	\label{eq:A:f}
	\begin{multline}
	\label{eq:A:f:bulk}
	\hspace{-8mm}
	\psi_{\text{bulk}}^{(e)}\left(P_{j}\right)=
	\alpha_{1}\displaystyle\sum_i P_i^2 + 
	\alpha_{11}^{(e)}\displaystyle\sum_iP_i^4 + 
	\alpha_{12}^{(e)}\displaystyle\sum_{i>j}P_i^2 P_j^2+\\
	\hspace{-5mm}
	\alpha_{111}\displaystyle\sum_{i}P_i^6 +
	\alpha_{112}\displaystyle\sum_{i>j}(P_i^4 P_j^2+P_j^4 P_i^2)
	+ 
	\alpha_{123}\displaystyle\prod_{i} P_i^2
	\end{multline}
	is the bulk free energy density, where zero-strain coefficients $\alpha^{(e)}_{ij}$ can be expressed in terms of the stress-free coefficient $\alpha_{ij}$ as follows:
	\iftwocolumn
	\begin{align}
	\nonumber
	\alpha^{(e)}_{11} &= \alpha_{11} + \frac 16\,
	\left[
	\frac{
		2\left(q_{11}-q_{12}\right)^2
	}{
	c_{11}-c_{12}
} +
\frac{
	\left(q_{11}+2q_{12}\right)^2
}{
c_{11}+2c_{12}
}
\right], \\
\nonumber
\alpha^{(e)}_{12} &= \alpha_{12} + 
\\ \nonumber\lefteqn{
	\frac 16\,
	\left[
	\frac{
		2\left(q_{11}+2q_{12}\right)^2
	}{
	c_{11}+2c_{12}
} +
\frac{
	2\left(q_{11}-q_{12}\right)^2
}{
c_{11}-c_{12}
} + 
\frac{3q^2_{44}}{4c_{44}}
\right].
}
\end{align}
\else
\begin{align}
\nonumber
\alpha^{(e)}_{11} &= \alpha_{11} + \frac 16\,
\left[
\frac{
	2\left(q_{11}-q_{12}\right)^2
}{
c_{11}-c_{12}
} +
\frac{
	\left(q_{11}+2q_{12}\right)^2
}{
c_{11}+2c_{12}
}
\right], \\
\nonumber
\alpha^{(e)}_{12} &= \alpha_{12} + 
\frac 16\,
\left[
\frac{
	2\left(q_{11}+2q_{12}\right)^2
}{
c_{11}+2c_{12}
} +
\frac{
	2\left(q_{11}-q_{12}\right)^2
}{
c_{11}-c_{12}
} + 
\frac{3q^2_{44}}{4c_{44}}
\right].
\end{align}
\fi
Symbol
\begin{equation}
\label{eq:A:f:ela}
\psi_{\text{ela}}\left(e_{ij}\right)= 
\frac{c_{11}}{2} \sum_{i} e_{ii}^2 +
\sum_{i>j} \left(
c_{12}\,  e_{ii} e_{jj} +
2 c_{44}\, e_{ij}^2		
\right)
\end{equation}
is the elastic energy density,
\iftwocolumn
\begin{multline}
\label{eq:A:f:wall}
\psi_{\text{wall}}\left(P_{i,j}\right)= 
\frac {G_{11}}2 \sum_{i} P_{i,i}^2 +
G_{12} \sum_{i>j} P_{i,i} P_{j,j} + 
\\
\frac {G_{44}}2 \sum_{i>j} \left(P_{i,j}+P_{j,i}\right)^2
\end{multline}
\else
\begin{equation}
\label{eq:A:f:wall}
\psi_{\text{wall}}\left(P_{i,j}\right)= 
\frac {G_{11}}2 \sum_{i} P_{i,i}^2 +
G_{12} \sum_{i>j} P_{i,i} P_{j,j} + 
\frac {G_{44}}2 \sum_{i>j} \left(P_{i,j}+P_{j,i}\right)^2
\end{equation}
\fi
is the domain wall energy density, 
\iftwocolumn
\begin{multline}
\label{eq:A:f:es}
\psi_{\text{es}}\left(P_i,e_{ij}\right)= 
- q_{11} \sum_{i} e_{ii} P_{i}^2 - 
\\	
q_{12} \sum_{i>j} \left(e_{ii} P_{j}^2 + e_{jj} P_{i}^2\right) -
q_{44} \sum_{i>j}  e_{ij} P_{i} P_{j}
\end{multline}
\else
\begin{equation}
\label{eq:A:f:es}
\psi_{\text{es}}\left(P_i,e_{ij}\right)= 
- q_{11} \sum_{i} e_{ii} P_{i}^2 - 
q_{12} \sum_{i>j} \left(e_{ii} P_{j}^2 + e_{jj} P_{i}^2\right) -
q_{44} \sum_{i>j}  e_{ij} P_{i} P_{j}
\end{equation}
\fi
is the electrostriction energy density,
\begin{equation}
\label{eq:A:f:el}
\psi_{\text{el}}\left(P_i,D_{i}\right)= \frac 1{2\varepsilon_0\varepsilon_B} \left(D_i-P_i\right)^2
\end{equation}
is the electrostatic energy density. 
Symbol $P_{i,j}$ stands for $\partial P_i/\partial x_j$.

%----

Interaction of the domain wall with a pinning center has been introduced by an additional term in the Helmholtz free energy (referred to as ``pinning energy density" hereafter):
\iftwocolumn
\begin{multline}
\label{eq:A:f:pin}
\psi_{\text{pin}}\left(x,\,y,\,z,\,P_i\right)= 
\frac \xi{\pi^{3/2}w^3}\,P_i^2\, \times \\
\exp\left[{-\frac{\left(x-x_d\right)^2 + \left(y-y_d\right)^2 + \left(z-z_d\right)^2}{w^2}}\right],
\end{multline}
\else
\begin{equation}
\label{eq:A:f:pin}
\psi_{\text{pin}}\left(x,\,y,\,z,\,P_i\right)= 
\frac \xi{\pi^{3/2}w^3}\,P_i^2\, 
\exp\left[{-\frac{\left(x-x_d\right)^2 + \left(y-y_d\right)^2 + \left(z-z_d\right)^2}{w^2}}\right],
\end{equation}
\fi
\end{subequations}
where symbols $x_d$, $y_d$, and $z_d$ stand for the $x$-, $y$-, and $z$-coordinate of the pinning center,
symbols $\xi$ and $w$ stand for the strength and radius of domain wall-defect interaction.

\subsection{Equations of state}
\label{apndx:ThermoModel:EoS}

In order to express the equations of state that govern the evolution of the domain pattern at given boundary conditions, it is convenient to introduce the volume density of the electric enthalpy $h$:
\begin{equation}
\label{eq:A:h}
h\left(P_i,\,P_{i,j},\,e_{ij},\,E_i\right)=
\psi\left(P_i,\,P_{i,j},\,e_{ij},\,D_i\right) - D_iE_i,
\end{equation}

%----

Evolution of the domain pattern is then governed by the following equations: 
\begin{subequations}
	\label{eq:A:EoS}
	\begin{align}
	\label{eq:A:EoS:eqmechequilib}
	\left(\frac{\partial h}{\partial e_{ij}}\right)_{,j}&=0,\\
	\label{eq:A:EoS:eqdielgauss}
	\left(\frac{\partial h}{\partial E_{i}}\right)_{,i}&=0,\\
	\label{eq:A:EoS:eqkhalatnikov}
	\frac{1}{\Gamma}\frac{\partial P_i}{\partial t} - 
	\left(\frac{\partial h}{\partial P_{i,j}}\right)_{,j}&=-\frac{\partial h}{\partial P_i},
	\end{align}
\end{subequations}
which express the mechanical equilibrium, the Poisson equation for the electric displacement $D_i$, and the time-dependent Landau-Khalatnikov equation. Symbol $\Gamma$ stands for the kinetic constant.

%---- Table 2
\begin{table}
	\caption{Numerical values of PT~\cite{rossett_thermodynamic_1990} and BT~\cite{hlinka_piezoelectric_2009,marton_domain_2010} material parameters used in the thermodynamic model.}
	\label{tab:PFMMatParams}
\iftwocolumn
	\begin{tabular}{lp{25mm}p{29mm}}
\else
	\begin{tabular}{lrr}
\fi 
		\hline \hline
		Parameter & 
		PbTiO${}_3$ &
		BaTiO${}_3$ \\ \hline 
		$\alpha_{1}$ (J\,m\,C${}^{-2}$) &
		$(T-479)\,3.8\times 10^{5}$ & 
		$(T-381)\,3.34\times 10^{5}$\\
		$\alpha_{11}$	(J\,m${}^{5}$\,C${}^{-4}$) &  
		-7.3$\times 10^{8}$ & 
		$(T-393)\,4.69\times10^{6}-2.02\times10^{8}$ \\
		$\alpha_{12}$	(J\,m${}^{5}$\,C${}^{-4}$) &  
		7.5$\times 10^{8}$ & 
		3.23$\times 10^{8}$\\
		$\alpha_{111}$	(J\,m${}^{9}$\,C${}^{-6}$) &   
		2.6$\times 10^{8}$ & 
		2.76$\times 10^{9}-(T-393)\,5.52\times 10^{7}$\\
		$\alpha_{112}$	(J\,m${}^{9}$\,C${}^{-6}$) &
		6.1$\times 10^{8}$ &
		4.47$\times 10^{9}$\\
		$\alpha_{123}$	(J\,m${}^{9}$\,C${}^{-6}$) &
		-3.7$\times 10^{9}$ &
		4.91$\times 10^{9}$\\
		$c_{11}$ (N\,m${}^{-2}$) &
		1.75$\times 10^{11}$ &
		2.75$\times 10^{11}$\\
		$c_{12}$ (N\,m${}^{-2}$) &
		7.94$\times 10^{10}$ &
		1.79$\times 10^{11}$\\
		$c_{44}$ (N\,m${}^{-2}$) &
		1.11$\times 10^{11}$ &
		5.43$\times 10^{10}$\\
		$G_{11}$ (J\,m${}^{3}$C${}^{-2}$) &
		4.14$\times 10^{-10}$ &
		5.1$\times 10^{-10}$\\
		$G_{12}$ (J\,m${}^{3}$C${}^{-2}$) &
		0 & 
		-2$\times 10^{-11}$\\
		$G_{44}$ (J\,m${}^{3}$C${}^{-2}$) &
		2.07$\times 10^{-10}$ &
		2 $\times 10^{-11}$\\
		$q_{11}$ (J\,m\,C${}^{-2}$) &
		1.14$\times 10^{10}$ &
		1.42$\times 10^{10}$\\
		$q_{12}$ (J\,m\,C${}^{-2}$) &
		4.61$\times 10^{8}$ &
		-7.40$\times 10^{8}$\\
		$q_{44}$ (J\,m\,C${}^{-2}$) &
		7.49$\times 10^{9}$ &
		6.28$\times 10^{9}$\\
		$\varepsilon_B$ (1) &
		10 &
		7.35\\                                               
		$\Gamma$ (C${}^2$\,J${}^{-1}$\,m${}^{-1}$\,s${}^{-1}$) &
		4$\times 10^{4}$ & 
		4$\times 10^{4}$\\  
		\hline\hline
	\end{tabular}
\end{table}
%---- Table 2
%
%
%---- Table 3
{\setlength\tabcolsep{3mm}
\begin{table}
	\caption{Numerical values of the thermodynamic term expressing the interaction of the pinning centers with \dw{180} ferroelectric domain walls [see Eq.(\ref{eq:A:f:pin})] PT and BT material parameters used in the thermodynamic model.}
	\label{tab:PFMPinParams}
	\begin{tabular}{lrr}
		\hline \hline
		Parameter & 
		PbTiO${}_3$ &
		BaTiO${}_3$ \\ \hline 
		$\xi$ (J\,m\,C${}^{-2}$) &
		$6.96\times 10^{-19}$ & 
		$3.21\times 10^{-19}$\\
		$w$	(m) &  
		$5.00\times 10^{-10}$ & 
		$6.32\times 10^{-10}$\\
		\hline\hline
	\end{tabular}
\end{table}
}
%---- Table 3
%
Numerical values of material parameters used in the thermodynamic model are presented in Table~\ref{tab:PFMMatParams}. 
Numerical parameters expressing the interaction of the pinning center with the \dw{180} ferroelectric domain wall according to Eq.~(\ref{eq:A:f:pin}) in PT and BT materials are specified in Tab.~\ref{tab:PFMPinParams}.

\subsection{Representative volume element and boundary conditions}
\label{apndx:ThermoModel:RVE+BC}

%---- Figure 2
\begin{figure}[t]
	\begin{center}
		\includegraphics[width=0.48\textwidth]{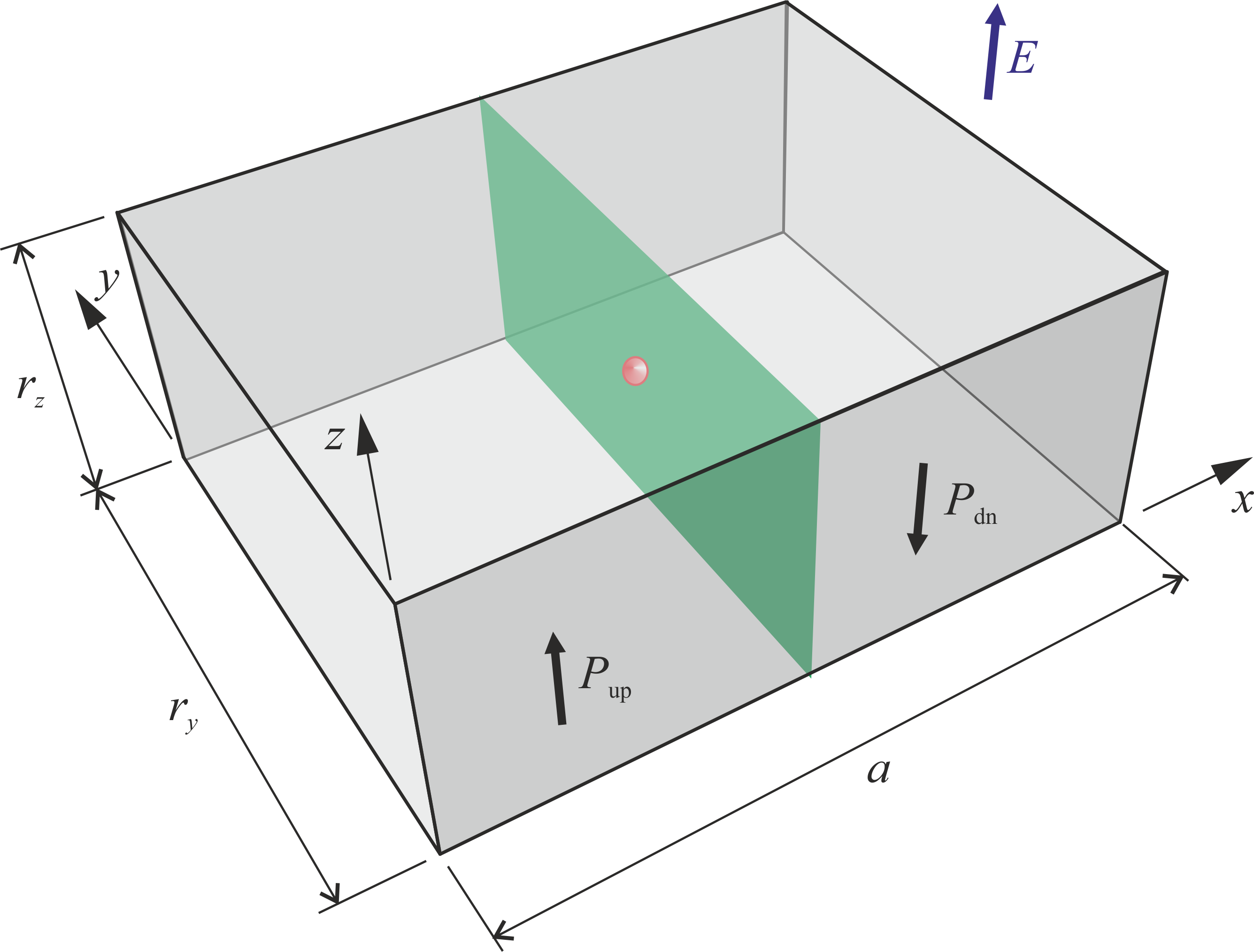}
	\end{center}
	\caption{Geometry of the representative volume element (RVE) in periodic structure of ferroelectric polydomain system with a single \dw{180} domain wall (green rectangle), which interacts with a pinning center (red sphere) in the applied external field.}
	\label{fig:RVE}
\end{figure}
%---- Figure 2
%
Figure~\ref{fig:RVE} shows the geometry of the representative volume element (RVE) in periodic structure of ferroelectric polydomain system with a single \dw{180} domain wall (green rectangle), which interacts with a pinning center (red sphere) in the external field.
We consider that the crystallographic axes of the ferroelectric perovskite material in tetragonal phase are oriented along $x$, $y$ and $z$ axes of the coordinate system.
The RVE is a cuboid with dimensions $a$, $r_y$, and $r_z$ along $x$, $y$ and $z$ directions, respectively. 
Coordinates of the pinning center have, therefore, following values: $x_d=a/2$,  $y_d=r_y/2$, and $z_d=r_z/2$.
The domain wall is perpendicular to the $x$ axis and its position is $x_w=a/2$ in the absence of the external field, i.e. $E=0$.
The domain wall separates anti-parallel domains with the vectors of polarizations $P_{\rm up}$ and $P_{\rm dn}$. 
The vector of external electric field has the magnitude $E$ and orientation along the $z$ axis. 

%----

In order to calculate the average dielectric response of the ferroelectric polydomain system of macroscopic dimensions, periodic boundary conditions for the vectors of polarization $P_i$ are introduced into the thermodynamic model in the directions of $y$ and $z$ axes: 
\begin{subequations}
	\label{eq:A:BC-Pi}
	\begin{align}
	\label{eq:A:BC-Pi:y}
	P_i(x,\,0,\,z)&=P_i(x,\,r_y,\,z),\\
	\label{eq:A:BC-Pi:z}
	P_i(x,\,y,\,0)&=P_i(x,\,y,\,r_z).
	\end{align}
\end{subequations}
Periodic boundary condition for the electrostatic potential $\varphi$ are specified in such a way that it allows the control of the applied electric field $E$ in the RVE:
\begin{subequations}
	\label{eq:A:BC-phi}
	\begin{align}
	\label{eq:A:BC-phi:y}
	\varphi(x,\,0,\,z)&=\varphi(x,\,r_y,\,z),\\
	\label{eq:A:BC-phi:z}
	\varphi(x,\,y,\,0)&=\varphi(x,\,y,\,r_z)-E\,r_z.
	\end{align}
\end{subequations}
where $E$ is the external field magnitude.

%----

Periodic boundary conditions for the vector of elastic displacement $u_i$ are rather difficult to formulate.
The problem is that average values of $u_i$ are given by integrating the nonzero average values of diagonal components of strain tensor $e_{ii}$ and, thus, cannot satisfy the periodic boundary conditions in a finite sample, which are similar to Eqs.~(\ref{eq:A:BC-Pi}).
Another point is that as the $z$ component of polarization, i.e. $P_3(x)$, changes its value from $P_{\rm up}$ through zero to $P_{\rm dn}$, when crossing the domain wall in the $x$-direction, the stress-free strain $e_{33}\propto P_3^2$ is not uniform in the RVE.
Actually, the local value of $e_{33}$ has a finite value within the anti-parallel ferroelectric domains and zero value in the middle of the \dw{180} ferroelectric domain wall in the stress-free sample~\cite{shapovalov_elastic_2014}. 
This results in bulging the surfaces of the stress-free RVE with the \dw{180} domain wall.
This bulging violates the geometric boundary conditions for periodicity of the RVE. 
In order to overcome this difficulty in numerical simulations, we adopted the approach developed by Steiger and Mokrý~\cite{steiger_finite_2015} for numerical simulations of piezoelectrically active composite systems.
In this approach, the periodic boundary conditions are approximated by conditions for the flatness of the RVE surfaces, which are perpendicular to $y$ and $z$ axis, respectively.

%----

In the first step, we introduce the average values of elastic displacement at particular surfaces of the RVE:
\begin{subequations}
	\label{eq:A:BC:uavrg}
	\begin{align}
	\label{eq:A:BC:uavrg:y}
	\avrgs{u_2}(y_0) &= \frac{1}{ar_z} \int_{0}^{a} dx \int_{0}^{r_z} u_2(x, y_0, z) dz, \\
	\label{eq:A:BC:uavrg:z}
	\avrgs{u_3}(z_0) &= \frac{1}{ar_y} \int_{0}^{a} dx \int_{0}^{r_y} u_3(x, y, z_0) dy.
	\end{align}
\end{subequations}

%----

In the second step, we introduce a local force $F_n(x)$, which is normal to the particular surface of the RVE and its magnitude is proportional to the difference between particular components of the local $u_n(x)$ and average $\avrgs{u_n}$ mechanical displacements at particular surface of the RVE, i.e. $F_n(x)\propto\zeta\left( u_n(x)-\avrgs{u_n}\right)$, where $\zeta$ is an arbitrary numerical constant, which is denoted as a surface stiffness.
The physical meaning of the surface stiffness $\zeta$ can be explained as follows. 
The coefficient $\zeta$ represents the proportionality constant between the local surface force $F_n$, which acts in the normal direction at given point on the surface, and the difference between local and average displacements in the normal direction at given point on the surface $u_n(x)-\avrgs{u_n}$. 
Thus, the greater value of $\zeta$, the greater exerted force $F_n$, which acts in such a way that the resulting local displacement in the normal direction $u_n$ is converging to the uniform average $\avrgs{u_n}$ normal displacement on the given surface, i.e. $u_n(x)\rightarrow\avrgs{u_n}$ in the limit $\zeta\rightarrow\infty.$
Our analytical calculations and numerical simulations reveal that the value of $\zeta$ greater than $10^4$ is sufficient to numerically approximate a perfectly flat surface. 

%----

Therefore, the boundary conditions for the flatness of the RVE surfaces have been introduced in the following form:
\begin{subequations}
	\label{eq:A:BC-tau}	
\allowdisplaybreaks
\iftwocolumn
\begin{align}
\label{eq:A:BC-tau-1i-x0}
\hspace{-5mm}
\frac{\partial h}{\partial e_{1i}}(0,\,y,\,z)&=0,\mbox{\qquad for\quad} i=1,2,3, \\
\label{eq:A:BC-tau-1i-a}
\hspace{-5mm}
\frac{\partial h}{\partial e_{1i}}(a,\,y,\,z)&=0,\mbox{\qquad for\quad} i=1,2,3,\\
\label{eq:A:BC-tau-22-y0}
\hspace{-5mm}
\frac{\partial h}{\partial e_{22}}(x,\,0,\,z)&=-\frac{\zeta\,c_{11}}{r_y}\left[u_2(x,\,0,\,z)-\avrgs{u_2}(0)\right],\\
\label{eq:A:BC-tau-2i-y0}
\hspace{-5mm}
\frac{\partial h}{\partial e_{2i}}(x,\,0,\,z)&=0, \mbox{\qquad for\quad} i \ne 2, \\
\label{eq:A:BC-tau-22-ry}
\hspace{-5mm}
\frac{\partial h}{\partial e_{22}}(x,\,r_y,\,z)&=\frac{\zeta\,c_{11}}{r_y}\left[u_2(x,\,r_y,\,z)-\avrgs{u_2}(r_y)\right],\\
\label{eq:A:BC-tau-2i-ry}
\hspace{-5mm}
\frac{\partial h}{\partial e_{2i}}(x,\,r_y,\,z)&=0, \mbox{\qquad for\quad} i \ne 2, \\
\label{eq:A:BC-tau-33-z0}
\hspace{-5mm}
\frac{\partial h}{\partial e_{33}}(x,\,y,\,0)&=-\frac{\zeta\,c_{11}}{r_z}\left[u_3(x,\,y,\,0)-\avrgs{u_3}(0)\right],\\
\label{eq:A:BC-tau-3i-z0}
\hspace{-5mm}
\frac{\partial h}{\partial e_{3i}}(x,\,y,\,0)&=0, \mbox{\qquad for\quad} i \ne 3, \\
\label{eq:A:BC-tau-33-rz}
\hspace{-5mm}
\frac{\partial h}{\partial e_{33}}(x,\,y,\,r_z)&=\frac{\zeta\,c_{11}}{r_z}\left[u_3(x,\,y,\,r_z)-\avrgs{u_3}(r_z)\right],\\
\label{eq:A:BC-tau-3i-rz}
\hspace{-5mm}
\frac{\partial h}{\partial e_{3i}}(x,\,y,\,r_z)&=0, \mbox{\qquad for\quad} i \ne 3.
\end{align}
\else
\begin{align}
	\label{eq:A:BC-tau-1i-x0}
	\hspace{-5mm}
	\frac{\partial h}{\partial e_{1i}}(0,\,y,\,z)&=0,& i &=1,2,3, \\
	\label{eq:A:BC-tau-1i-a}
	\hspace{-5mm}
	\frac{\partial h}{\partial e_{1i}}(a,\,y,\,z)&=0,& i &=1,2,3,\\
	\label{eq:A:BC-tau-22-y0}
	\hspace{-5mm}
	\frac{\partial h}{\partial e_{22}}(x,\,0,\,z)&=-\frac{\zeta\,c_{11}}{r_y}\left[u_2(x,\,0,\,z)-\avrgs{u_2}(0)\right],\\
	\label{eq:A:BC-tau-2i-y0}
	\hspace{-5mm}
	\frac{\partial h}{\partial e_{2i}}(x,\,0,\,z)&=0, & i &\ne 2, \\
	\label{eq:A:BC-tau-22-ry}
	\hspace{-5mm}
	\frac{\partial h}{\partial e_{22}}(x,\,r_y,\,z)&=\frac{\zeta\,c_{11}}{r_y}\left[u_2(x,\,r_y,\,z)-\avrgs{u_2}(r_y)\right],\\
	\label{eq:A:BC-tau-2i-ry}
	\hspace{-5mm}
	\frac{\partial h}{\partial e_{2i}}(x,\,r_y,\,z)&=0, & i &\ne 2, \\
	\label{eq:A:BC-tau-33-z0}
	\hspace{-5mm}
	\frac{\partial h}{\partial e_{33}}(x,\,y,\,0)&=-\frac{\zeta\,c_{11}}{r_z}\left[u_3(x,\,y,\,0)-\avrgs{u_3}(0)\right],\\
	\label{eq:A:BC-tau-3i-z0}
	\hspace{-5mm}
	\frac{\partial h}{\partial e_{3i}}(x,\,y,\,0)&=0, & i &\ne 3,\\
	\label{eq:A:BC-tau-33-rz}
	\hspace{-5mm}
	\frac{\partial h}{\partial e_{33}}(x,\,y,\,r_z)&=\frac{\zeta\,c_{11}}{r_z}\left[u_3(x,\,y,\,r_z)-\avrgs{u_3}(r_z)\right],\\
	\label{eq:A:BC-tau-3i-rz}
	\hspace{-5mm}
	\frac{\partial h}{\partial e_{3i}}(x,\,y,\,r_z)&=0, & i &\ne 3.
\end{align}
\fi
\end{subequations}

%----

The system of partial differential equations given by Eqs.~(\ref{eq:A:psi-def})-(\ref{eq:A:EoS}) with the complementary system of boundary conditions given by Eqs.~(\ref{eq:A:BC-Pi})-(\ref{eq:A:BC-tau}) can be solved analytically or numerically for given values of the external electric field $E$. 

\subsection{Numerical implementation of the phase-field model}

In the numerical simulations, the system of partial differential equations given by Eqs.~(\ref{eq:A:EoS}) complemented with the system of boundary conditions given by Eqs.~(\ref{eq:A:BC-Pi})-(\ref{eq:A:BC-tau}) has been solved numerically for variables $\varphi$, $P_i$, and $u_i$.
The initial conditions were specified as follows: external electric field $E=0$, polarization $P_1=P_2=0$, and the value of $P_3(x,\,y,\,z)$ was set to equal $P_0$ for $x<a/2$ and $-P_0$ for $x>a/2$ considering the numerical values for $P_0$ specified in Tab.~\ref{tab:DielLatPhaseField}. 
We have conducted the numerical simulations for tetragonal PT~\cite{rossett_thermodynamic_1990} and BT~\cite{hlinka_piezoelectric_2009,marton_domain_2010}.

%----

The numerical model was implemented and solved in COMSOL Multiphysics, which is based on the finite element method (FEM).
Since the results of numerical computations were used to obtain the nonlinear dielectric response, the discretization error of the FEM had to be reduced much below the expected magnitude of nonlinear contributions to dielectric response.
It was achieved by severe limitations on the maximum size of the mesh element, which was smaller than approximately 0.1$\times$0.6$\times$1\,nm${}^3$.

%====
%
%====
\section{Extrinsic permittivity due to the uniform displacement of the planar domain wall}
\label{apndx:Model:pin}

In this Section, we calculate the extrinsic permittivity due to the uniform displacement of the planar domain wall.
We follow the approach that has been recently presented by Mokrý \textit{et al.}~\cite{mokry_nonlinear_2013} for the case of ferroelastic \dw{90} domain wall.
The detailed analysis of the equations of state Eqs.~(\ref{eq:A:EoS}) indicates that the presence of pinning center within the domain wall yields a local increase in the domain wall thickness in the vicinity of the pinning center. 
This interaction of \dw{180} domain wall with pinning center and the domain wall widening produces local stray electrostatic fields and elastic stresses.
Thus, the electric enthalpy $h$ has nonzero contributions from $\psi_{\rm el}$, $\psi_{\rm ela}$, and $\psi_{\rm es}$. 
This makes the exact solution of Eqs.~(\ref{eq:A:EoS}) rather difficult. 

%----

In order to proceed all analytical calculations, two additional considerations have been taken into account:
\begin{subequations}
	\label{eq:B:PinApprox}
\begin{enumerate}
\item The radius of the interaction of the \dw{180} domain wall with the pinning center $w$ is much smaller than the representative distances of the RVE:
\begin{equation}
	\label{eq:B:PinApprox:2}
	a\gg w, \quad r_y\gg w, \quad\mbox{and}\quad r_z\gg w.
\end{equation}
\item Spatial distribution of polarization within the domain wall is approximated by Eqs.~(\ref{eq:III:PinApprox:3}) and (\ref{eq:III:Piw}).
\item The constant domain wall thickness is considered is considered in the vicinity of the pinning center.
\end{enumerate}
\end{subequations}

%----

Consider that $\pin{\delta}$ is the uniform displacement of the \dw{180} domain wall from the pinning center at $x_d=a/2$.
Then, the position of the displaced planar wall is $x_{w}=x_{d}+\pin{\delta}$. 
The total electric enthalpy of the system is expressed by integration of electric enthalpy over the volume of the RVE:
\iftwocolumn
\begin{multline}
\label{eq:B:Hpin:def}
\pin{H}=\int_0^a dx \int_0^{r_y} dy \int_0^{r_z} h\left[P_i^{(w)}(x-x_{d}-\pin{\delta}),\,
\right.\\ \left.
P_{i,j}^{(w)}(x-x_{d}-\pin{\delta}),\,e_{ij},\,E_i\right]\,dz,
\end{multline}
\else
\begin{equation}
\label{eq:B:Hpin:def}
\pin{H}=\int_0^a dx \int_0^{r_y} dy \int_0^{r_z} h\left[P_i^{(w)}(x-x_{d}-\pin{\delta}),\,
P_{i,j}^{(w)}(x-x_{d}-\pin{\delta}),\,e_{ij},\,E_i\right]\,dz,
\end{equation}
\fi 
where $E_i$ is the vector of the external uniform field of the magnitude $E$ applied in the direction of $z$ axis and $e_{ij}$ is considered zero for simplicity, since it does not enter the expansion of $\pin{H}$ with respect to $P_N$ at all.

%----

Integrand on the right hand side has the form
\begin{equation}
	h = h_0 + \psi_{\rm pin}^{(w)}(x,\,y,\,z) - E P_3^{(w)}(x-x_{d}-\pin{\delta}), 
\end{equation}
where
\iftwocolumn
\begin{multline}
\psi_{\rm pin}^{(w)}(x,\,y,\,z) = \frac \xi{\pi^{3/2}w^3}\,\left[P_3^{(w)}(x-x_{d}-\pin{\delta})\right]^2\,\times \\ 
\exp\left[{-\frac{\left(x-x_d\right)^2 + \left(y-y_d\right)^2 + \left(z-z_d\right)^2}{w^2}}\right]
\end{multline}
\else
\begin{equation}
	\psi_{\rm pin}^{(w)}(x,\,y,\,z) = \frac \xi{\pi^{3/2}w^3}\,\left[P_3^{(w)}(x-x_{d}-\pin{\delta})\right]^2\, 
	\exp\left[{-\frac{\left(x-x_d\right)^2 + \left(y-y_d\right)^2 + \left(z-z_d\right)^2}{w^2}}\right]
\end{equation}
\fi
and $h_0$ is the contribution to the volume density of electric enthalpy $h$, which is independent on the uniform displacement of the domain wall $\pin{\delta}$.
With the use of Eq.~(\ref{eq:II:PN-def}) and considering $\pin{P}/P_0=2\pin{\delta}/a$, function $\pin{H}$ given by Eq.~(\ref{eq:B:Hpin:def}) can be expressed in the form
\begin{equation}
	\pin{H} = \pinex{H}{,0} + ar_yr_z\,\left[
		\pin{h}\left(\pin{P}\right) - E P_N
	\right],
\end{equation}
where
\iftwocolumn
\begin{multline}
	\pin{h}\left(\pin{P}\right) =
	\frac{\xi
		\erf\left(\frac{r_y}{2w}\right)
		\erf\left(\frac{r_z}{2w}\right)
	}{ar_yr_zw\sqrt{\pi}}
	\times \\
	\int_{0}^{a}
	e^{-\frac{\left(x-a/2\right)^2}{w^2}}
	\left\{P_3^{(w)}\left[x-\frac{a}{2}\left(1+\frac{\pin{P}}{P_0}\right)\right]\right\}^2\,
	dx
\end{multline}
\else
\begin{equation}
\pin{h}\left(\pin{P}\right) =
\frac{\xi
	\erf\left(\frac{r_y}{2w}\right)
	\erf\left(\frac{r_z}{2w}\right)
}{ar_yr_zw\sqrt{\pi}}
\int_{0}^{a}
e^{-\frac{\left(x-a/2\right)^2}{w^2}}
\left\{P_3^{(w)}\left[x-\frac{a}{2}\left(1+\frac{\pin{P}}{P_0}\right)\right]\right\}^2\,
dx
\end{equation}
\fi
and $\pinex{H}{,0}$ is the contribution to the electric enthalpy $H$, which is independent on $\pin{P}$.
When we consider assumptions given by Eq.~(\ref{eq:B:PinApprox:2}), the function $\pin{h}$ can be expressed in the following integral form:
\iftwocolumn
\begin{align}
\pin{h}\left(\pin{P}\right) &\approx
\frac{4P_0^2\xi}{ar_yr_za_w^2}
\int_{0}^{\frac{a \pin{P}}{2 P_0}}
\\ \nonumber\lefteqn{\hspace{-15mm}
\left\{
\delta\,
\left[
\erf\left(
\frac{a_w+2\delta}{2w}
\right)
- 
\erf\left(
\frac{a_w-2\delta}{2w}
\right)
\right]
- \right. }\\ \nonumber\lefteqn{\hspace{-15mm} \left. -
\frac{1}{\sqrt{\pi}}
\exp\left[-\frac{(\tau +2 \delta)^2}{4w^2}\right]
\left[\exp\left({\frac{2 a_w \delta}{w^2}}\right)-1\right]
\right\}\,d\delta.}
\end{align}
\else
\begin{multline}
	\pin{h}\left(\pin{P}\right) \approx	
	\frac{4P_0^2\xi}{ar_yr_za_w^2}
\int_{0}^{\frac{a \pin{P}}{2 P_0}}
\left\{
\delta\,
\left[
\erf\left(
\frac{a_w+2\delta}{2w}
\right)
- 
\erf\left(
\frac{a_w-2\delta}{2w}
\right)
\right]
- \right. \\ \left. -
\frac{1}{\sqrt{\pi}}
	\exp\left[-\frac{(\tau +2 \delta)^2}{4w^2}\right]
	\left[\exp\left({\frac{2 a_w \delta}{w^2}}\right)-1\right]
\right\}\,d\delta.
\end{multline}
\fi 

%----

It is a straightforward task to show that the thermodynamic function $\pin{h}$ can be expressed in the form of Taylor series:
\begin{equation}
\label{eq:B:Hpin:ser}
\pin{h}(\pin{P}) = \pinsup{\alpha_{1}}\, \pin{P}^2 + \pinsup{\alpha_{11}}\, \pin{P}^4,
\end{equation}
where
\begin{subequations}
	\label{eq:B:AlphaPin}
	\begin{align}
	\label{eq:B:AlphaPin:1}
	\pinsup{\alpha_{1}} &= \frac{a\xi}{r_yr_zw^2}\, \Phi\left(a_w/w\right),\\
	\label{eq:B:AlphaPin:11}
	\pinsup{\alpha_{11}} &= 
	-\frac{a^3 \xi}{48P_0^3r_yr_zw^4}\, \Psi\left(a_w/w\right),
	\end{align}
\end{subequations}
and
\begin{subequations}
	\label{eq:B:PinNumFunc}
	\begin{align}
	\label{eq:B:PinNumFunc:Phi}
	\Phi\left(\tau\right) &= \frac {1}{\tau^2}\left[
	{\rm erf}\left(\tau/2\right)-\frac{\tau\,\exp\left(-\tau^2/4\right)}{\sqrt{\pi}}
	\right],\\
	\label{eq:B:PinNumFunc:Psi}
	\Psi\left(\tau\right) &= 
	\frac{\tau}{\sqrt{\pi}} \exp\left(-\tau^2/4\right)
	\end{align}
\end{subequations}
are numerical functions that depend on the spatial distribution of polarization within the domain wall, domain wall thickness and the interaction radius of the pinning center. 

%----

The field dependence of extrinsic polarization response $\pin{P}$ is calculated from the equation of state Eq.~(\ref{eq:IV:EoS:hpin}). 
If we consider the Taylor expansion given by Eq.~(\ref{eq:B:Hpin:ser}), the solution of the equation of state for $\pin{P}$ has the form:
\begin{equation}
\label{eq:IV:PNw:Taylor}
\pin{P}(E)\approx \frac{1}{2}\left(\pinsup{\alpha_{1}}\right)^{-1} E - \frac{\pinsup{\alpha_{11}}}{4}\left(\pinsup{\alpha_{1}}\right)^{-4} E^3 + \cdots.
\end{equation}
It immediately follows from Eqs.~(\ref{eq:B:AlphaPin}) that
\begin{subequations}
	\label{eq:B:PNPinTaylor}
	\begin{align}
	\label{eq:B:PNPinTaylor:EpsL}
	\pin{\varepsilon} &= \frac{r_y r_z w^2}{2 a \xi\, \Phi\left(a_w/w\right)},\\
	\label{eq:B:PNPinTaylor:Gamma}
	\pin{\gamma} &= \frac{
		r_y^3 r_z^3 w^4\,\Psi\left(a_w/w\right)
	}{
	192\, a P_0^2 \xi^3\, \Phi^4\left(a_w/w\right)
}.
\end{align}
\end{subequations}

%====
%
%====
\section{Extrinsic permittivity due to domain wall bending}
\label{apndx:Model:ben}

In this Section, we calculate the extrinsic permittivity due to domain wall bending.
We follow the approach that has been recently presented by Mokrý \textit{et al.}~\cite{mokry_evidence_2009}.
However, we consider a more general case of anisotropic distribution of pinning centers in this Article.

%----

We consider that $\ben{\delta}(y,\,z)$ is a spatially dependent displacement of the domain wall. 
Deflection of the domain wall, which is defined for $y$ running from $-r_y/2$ to $r_y/2$ and for $z$ running from  $-r_z/2$ to $r_z/2$ can be approximated by the function
\begin{equation}
\label{eq:C:DeltaBen-is}
\ben{\delta}(y,\,z) = 4\,\ben{\delta}^{\star}\, 
	\left(	
		\frac{1}{4}
		-
		\frac{
			y^2+\kappa\,z^2
		}{
			r_y^2+\kappa\,r_z^2
		}
	\right),
\end{equation}
where $\ben{\delta}^{\star}$ is the maximum deflection of the bent domain wall at the apex point, which lies on the $x$-axis, and $\kappa$ is the ratio of the radii of curvature at the apex of the bent domain wall, i.e. $\kappa=(\partial^2\ben{\delta}/\partial z^2)/(\partial^2\ben{\delta}/\partial y^2)$, at the apex point of the domain wall due to the strong depolarizing effect in the direction of the ferroelectric $z$-axis. 

%----

Then the $x$-coordinate of the bent domain wall is $x_{w}(y,\,z)=x_{d}+\ben{\delta}(y,\,z)$. 
The total electric enthalpy of the system is expressed by the integration of the electric enthalpy over the volume of the RVE:
\begin{multline}
\label{eq:C:Hben}
\iftwocolumn\hspace{-5mm}\else \fi 
	\ben{H}=
	\int_0^a\!\! dx 
	\int_{-\frac{r_y}2}^{\frac{r_y}2}\!\!  dy 
	\int_{-\frac{r_z}2}^{\frac{r_z}2}\!\!
	h\left[
		P_i^{(w)}\left(x-x_{d}-\ben{\delta}(y,\,z)\right),
		\right.
		\\
		\left.
		P_{i,j}^{(w)}\left(x-x_{d}-\ben{\delta}(y,\,z)\right),\,e_{ij},\,E_i^{(ext)}
	\right]\,dz,
\end{multline}
where again $E_i$ is the vector of the external uniform field of the magnitude $E$ applied in the direction of $z$ axis and $e_{ij}$ is considered zero otherwise the problem is not solvable analytically.

%----

Detailed analysis of this problem has been recently presented by Mokrý \textit{et al.}\cite{mokry_evidence_2009}.
In accordance with this analysis, following additional considerations have been taken into account:
\begin{subequations}
	\label{eq:C:BenApprox}
\begin{enumerate}
\item Maximum deflection of the bent \dw{180} domain wall $\ben{\delta}^{\star}$ is much smaller than the representative distances between the pinning centers $r_y$ and $r_z$, i.e.:
\begin{equation}
	\label{eq:C:BenApprox:1}
	\ben{\delta}^{\star}\ll r_y, \qquad \ben{\delta}^{\star}\ll r_z.
\end{equation}	
\item Thickness of the \dw{180} domain wall is much smaller than the representative distances between the pinning centers $r_y$ and $r_z$, i.e.:
\begin{equation}
\label{eq:C:BenApprox:2}
a_w\ll r_y, \qquad a_w\ll r_z.
\end{equation}	
\item Total electric enthalpy $\ben{H}$ of the RVE that controls the net spontaneous polarization response to the electric field $P_N(E)$ is dominated by the lowest (quadratic)	term, i.e.:
\begin{equation}
	\label{eq:C:BenApprox:3}
	\left|\frac{\partial^2\ben{H}}{\partial P_N^2}\right| \gg \left|\frac{\partial^4\ben{H}}{\partial P_N^4}\right|\cdot\left|P_N^2\right|.
\end{equation}	
	\end{enumerate}
\end{subequations}

%----

When the \dw{180} domain wall is bent due to the action of external electric field, the uncompensated bound charge $\rho_b=P_{i,i}$ appears within the volume of the domain wall. 
The volume density of bound charge $\rho_b$ is equal to the divergence of polarization $P_i^{(w)}\left(x-x_{d}-\ben{\delta}(y,\,z)\right)$, which yields:
\begin{equation}
	\label{eq:C:RhoB}
	\rho_b\left(x,\,y,\,z\right) = 
	\begin{dcases*}
		-\frac{2\,P_0}{a_w}\,\delta_{{\rm ben},z}(z) \hspace{-15mm} & \\
		& for $\left|x-\ben{\delta}(y,\,z)\right|<\frac{a_w}{2}$,\\
		0 & 
		otherwise.
	\end{dcases*}
\end{equation}
The uncompensated bound charges $\rho_b$ produces a local depolarizing field $-\varphi_{d,i}$, where $\varphi_d$ is the electrostatic potential of the depolarizing field.
The electrostatic potential $\varphi_d$ satisfies the boundary condition $\varphi_d(x,\,y,\,-r_z/2)=\varphi_d(x,\,y,\,r_z/2)=0$.
In accordance with this boundary conditions for $\varphi_d$, the total electrostatic potential $\varphi$ in the RVE can be expressed in the form:
\begin{equation}
	\label{eq:C:varphi}
	\varphi(x,\,y,\,z) = \varphi_b(x,\,y,\,z) - E\,z,
\end{equation}
where $E$ is the magnitude of the external electric field applied in the direction along the $z$ axis. 
Then, the main contributions to the volume density of the electric enthalpy, which represents the integrand in Eq.~(\ref{eq:C:Hben}), are as follows
\iftwocolumn
\begin{multline}
\label{eq:C:hBen}
	h = h_1 + \left(\psi_{\rm bulk}^{(w)} + \psi_{\rm wall}^{(w)}\right)(x,\,y,\,z) + \psi_{\rm el}^{(w)}(x,\,y,\,z) - \\
	- EP_3^{\rm (w)}\left(x- x_d-\ben{\delta}(y,\,z)\right),
\end{multline}
\else
\begin{equation}
	\label{eq:C:hBen}
	h = h_1 + \left(\psi_{\rm bulk}^{(w)} + \psi_{\rm wall}^{(w)}\right)(x,\,y,\,z) + \psi_{\rm el}^{(w)}(x,\,y,\,z) - EP_3^{\rm (w)}\left(x- x_d-\ben{\delta}(y,\,z)\right),
\end{equation}
\fi
where the symbols $\left(\psi_{\rm bulk}^{(w)} + \psi_{\rm wall}^{(w)}\right)$ and $\psi_{\rm el}^{(w)}(x,\,y,\,z)$ stand for the contributions to electric enthalpy due to the energy stored in the domain wall and due to the electrostatic energy of the depolarizing field.
Symbol $h_1$ stand for the contribution to the volume density of electric enthalpy, which is independent on the displacement due to the domain wall bending $\ben{\delta}$.

%----

The volume density of energy stored in the domain wall stems from the bulk free energy $\psi_{\rm bulk}$ and from the wall energy $\psi_{\rm wall}$. 
Its value strongly depends on the domain wall thickness $a_w$ and reaches minimum in the real systems.
Calculation of minimum value of this energy, which corresponds to the equilibrium value of the domain wall thickness, represents a task explained in many classical textbooks on ferroelectricity~\cite{tagantsev_domains_2010}.
Thus, it is possible to express the volume density of free energy stored in the domain wall in the form:
\iftwocolumn
\begin{multline}
	\label{eq:C:psi:wall}
	\left(\psi_{\rm bulk}^{(w)} + \psi_{\rm wall}^{(w)}\right)(x,\,y,\,z) = \\
	= \begin{dcases*}
	\frac{\sigma_w}{a_w}
	\left(
	1+\frac{G_{11}}{G_{44}}\,\benex{\delta}{,y}^2+\benex{\delta}{,z}^2
	\right) \hspace{-35mm}& \\
	& for $\left|x-\ben{\delta}(y,\,z)\right|<\frac{a_w}{2}$,\\
	0 & 
	otherwise,
	\end{dcases*}
\end{multline}
\else
\begin{equation}
	\label{eq:C:psi:wall}
	\left(\psi_{\rm bulk}^{(w)} + \psi_{\rm wall}^{(w)}\right)(x,\,y,\,z) = 
	\begin{dcases*}
		\frac{\sigma_w}{a_w}
		\left(
			1+\frac{G_{11}}{G_{44}}\,\benex{\delta}{,y}^2+\benex{\delta}{,z}^2
		\right) & 
		for $\left|x-\ben{\delta}(y,\,z)\right|<\frac{a_w}{2}$,\\
		0 & 
		otherwise,
	\end{dcases*}
\end{equation}
\fi
where symbols
\begin{align}
	\label{eq:C:SigmaW}
	\sigma_w&\approx \frac{a_wP_0^2}{6\,\varepsilon_c},\\
	\label{eq:C:aW}
	a_w&\approx \sqrt{30\, \varepsilon_c G_{44}}
\end{align}
stand for the surface energy of the domain wall and the equilibrium thickness of the domain wall, respectively.
Finally, the volume energy density due to depolarizing field can be expressed in the following form
\begin{equation}
\label{eq:C:psi:el}
\psi_{\rm el}^{(w)}(x,\,y,\,z) = \frac 12\,
\rho_b\,\varphi_d.
\end{equation}

%----

With the use of Eq.~(\ref{eq:II:PN-def}) and considering $\ben{P}/P_0=(4/3)\,(\ben{\delta}^\star/a)$, function $\ben{H}$ given by Eq.~(\ref{eq:C:Hben}) can be expressed in the form
\begin{equation}
	\label{eq:C:Hben:sim}
	\ben{H} = \benex{H}{,1} + ar_yr_z\,\left[
		\ben{h}\left(\ben{P}\right) - E P_N
	\right],
\end{equation}
where it is convenient to express the function $\ben{h}$ in the form of a surface integral over the planar segment of the non-displaced domain wall at $x=a/2$:
\iftwocolumn
\begin{multline}
	\ben{h}(\ben{P}) = \frac{1}{ar_yr_z}
	\int_{-\frac{r_y}2}^{\frac{r_y}2}dy 
	\int_{-\frac{r_z}2}^{\frac{r_z}2}
	\left(
	\sigma_w + \frac{1}{2} \sigma_b\varphi_d
	\right) \times \\
	\sqrt{1 + \benex{\delta}{,y}^2 + \benex{\delta}{,z}^2}\,dz.
\end{multline}
\else
\begin{equation}
	\ben{h}(\ben{P}) = \frac{1}{ar_yr_z}
		\int_{-\frac{r_y}2}^{\frac{r_y}2}dy 
		\int_{-\frac{r_z}2}^{\frac{r_z}2}
		\left(
			\sigma_w + \frac{1}{2} \sigma_b\varphi_d
		\right) 
		\sqrt{1 + \benex{\delta}{,y}^2 + \benex{\delta}{,z}^2}\,dz.
\end{equation}
\fi
Symbol $\benex{H}{,1}$ is the contribution to the electric enthalpy $H$, which is independent on $\ben{P}$.
In the above equation, the symbol $\sigma_b$ stands for the surface density of uncompensated bound charge, which would be located on the non-displaced domain wall, i.e. at $x=a/2$.
It can be expressed from the volume density of bound charge $\rho_b$ given by Eq.~(\ref{eq:C:RhoB}) using the equality:
\begin{equation}
	\label{eq:C:SigmaB}
	\sigma_b(y,\,z) = 
	\int_{0}^{a}	
	\rho_b\left(x,\,y,\,z\right) \,dx = 
	-\frac{
		2\,P_0\, \benex{\delta}{,z}(z)
	}{
		\sqrt{1 + \benex{\delta}{,y}^2 + \benex{\delta}{,z}^2}
	}.
\end{equation}

%----

In accordance with the discussion above, we consider that the volume density of electric enthalpy $\ben{h}$ due to bending of the \dw{180} domain wall has two contributions:
\begin{equation}
	\label{eq:C:hben:def}
	\ben{h}(\ben{P}) = h_{\rm wall}(\ben{P}) + h_{\rm el}(\ben{P}).
\end{equation}
The first contribution $h_{\rm wall}$ is due to the energy stored in the domain wall:
\iftwocolumn
\begin{multline}
	h_{\rm wall}(\ben{P}) = \frac{1}{ar_yr_z}	
	\int_{-\frac{r_y}2}^{\frac{r_y}2}dy 
	\int_{-\frac{r_z}2}^{\frac{r_z}2}
	\sigma_w\,\times \\
	\sqrt{
		1 + 
		\frac{
			36\, a^2 
			\left(
			y^2+\kappa^2 z^2
			\right)
		}{\left(r_y^2+\kappa r_z^2\right)^2}
		\left(\frac{\ben{P}}{P_0}\right)^2
	}\,dz.
\end{multline}
\else
\begin{equation}
	h_{\rm wall}(\ben{P}) = \frac{1}{ar_yr_z}	
		\int_{-\frac{r_y}2}^{\frac{r_y}2}dy 
		\int_{-\frac{r_z}2}^{\frac{r_z}2}
		\sigma_w 
		\sqrt{
			1 + 
			\frac{
				36\, a^2 
				\left(
					y^2+\kappa^2 z^2
				\right)
			}{\left(r_y^2+\kappa r_z^2\right)^2}
			\left(\frac{\ben{P}}{P_0}\right)^2
		}\,dz.
\end{equation}
\fi
Considering assumptions given by Eqs.~(\ref{eq:C:BenApprox:1}) and (\ref{eq:C:BenApprox:2}), it possible to approximate the integrand in the above equation by the Taylor series with respect to $\ben{P}$.
After performing the integration, the function $h_{\rm wall}$ can be expressed in the form:
\iftwocolumn
\begin{multline}
	h_{\rm wall}(\ben{P}) \approx h_{\rm wall,0} + 
	\frac{
		3 a \sigma_w \left(r_y^2+\alpha ^2
		r_z^2\right)
	}{
	4
	P_0^2 \left(r_y^2+\alpha  r_z^2\right)^2
}\, \ben{P}^2
- \\
-\frac{
	9 a^3 \sigma_w \left(9 r_y^4+10 \alpha ^2 r_y^2
	r_z^2+9 \alpha ^4 r_z^4\right)
}{
80
P_0^4 \left(r_y^2+\alpha 
r_z^2\right)^4
}\,\ben{P}^4.
\end{multline}
\else
\begin{equation}
	h_{\rm wall}(\ben{P}) \approx h_{\rm wall,0} + 
		\frac{
			3 a \sigma_w \left(r_y^2+\alpha ^2
			r_z^2\right)
		}{
			4
			P_0^2 \left(r_y^2+\alpha  r_z^2\right)^2
		}\, \ben{P}^2
		-\frac{
			9 a^3 \sigma_w \left(9 r_y^4+10 \alpha ^2 r_y^2
			r_z^2+9 \alpha ^4 r_z^4\right)
		}{
			80
			P_0^4 \left(r_y^2+\alpha 
			r_z^2\right)^4
		}\,\ben{P}^4.
\end{equation}
\fi

%----

The second contribution $h_{\rm el}$ is due to the depolarizing field:
\iftwocolumn
\begin{multline}
\label{eq:C:hel:def}
h_{\rm el}(\ben{P}) = \frac{1}{ar_yr_z}
\int_{-\frac{r_y}2}^{\frac{r_y}2}dy 
\int_{-\frac{r_z}2}^{\frac{r_z}2}
\frac{1}{2} \sigma_b\varphi_d\,\times \\
\sqrt{1 + \benex{\delta}{,y}^2 + \benex{\delta}{,z}^2}\,dz.
\end{multline}
\else
\begin{equation}
\label{eq:C:hel:def}
h_{\rm el}(\ben{P}) = \frac{1}{ar_yr_z}
\int_{-\frac{r_y}2}^{\frac{r_y}2}dy 
\int_{-\frac{r_z}2}^{\frac{r_z}2}
\frac{1}{2} \sigma_b\varphi_d 
\sqrt{1 + \benex{\delta}{,y}^2 + \benex{\delta}{,z}^2}\,dz.
\end{equation}
\fi
The formula for $h_{\rm el}$ cannot be obtained by a straightforward integration as it has been done for $\pin{h}$ and $h_{\rm wall}$.
The first factor in the integrand in Eq.~(\ref{eq:C:hel:def}), i.e. the bound charge density, is given by Eq.~(\ref{eq:C:SigmaB}) and can be expressed in the form of Taylor series with respect to $\ben{P}$:
\begin{equation}
\sigma_b(y,\,z) \approx 
\frac{12\, a \kappa z}{r_y^2+\kappa r_z^2}\, \ben{P} 
-
\frac{
	216\, a^3 \kappa 
	z \left(y^2+\kappa ^2 z^2\right)
}{
\text{Ps}^2 \left(r_y^2+\kappa 
r_z^2\right)^3
}\, \ben{P}^3.
\end{equation}
The second factor in the integrand on the right-hand side of Eq.~(\ref{eq:C:hel:def}), i.e. the electrotatic potential of the depolarizing field, is given by the solution of equations of state Eqs.~(\ref{eq:A:EoS}). 

%----

Considering the hard ferroelectric approximation and considering the assumptions given by Eqs.~(\ref{eq:C:BenApprox:1}) and (\ref{eq:C:BenApprox:2}), the solution of the equation of motion Eqs.~(\ref{eq:A:EoS}) for electrostatic potential $\varphi_d$ can be approximated by the solution of Laplace equation:
\begin{subequations}
\label{eq:C:Pot}
\begin{equation}
	\label{eq:C:PotLap}
	\varepsilon_a
	\left(
	\frac{\partial^2\varphi_d}{\partial x^2} +
	\frac{\partial^2\varphi_d}{\partial y^2}
	\right) +
	\varepsilon_c\frac{\partial^2\varphi_d}{\partial z^2} 
	= 0,
\end{equation}
where $\varepsilon_a$ is the component of permittivity tensor in the direction of $x$ and $y$ axis, respectively.
The above Laplace equation should be completed with the internal boundary conditions for the continuity of the normal component of electric displacement and for the continuity of the electrostatic potential at the domain wall, i.e., at $x=a/2$,
\begin{align}
	\label{eq:C:PotBoundCond:1}
	\varphi_d^{(+)}
	&=
	\varphi_d^{(-)},\\
	\label{eq:C:PotBoundCond:2}
	\frac{\partial\varphi_d^{(+)}}{\partial x}
	-
	\frac{\partial\varphi_d^{(-)}}{\partial x}
	&=
	\frac{\sigma_b}{\varepsilon_0\varepsilon_a},
\end{align}
\end{subequations}
where the superscripts $(+)$ and $(-)$ denote the electrostatic potential for
$x>a/2$ and $x<a/2$, respectively.
It is possible to show that functions 
\iftwocolumn
\begin{align}
\label{eq:C:PotFourierGen}
\varphi_d^{(\pm)}(x,\,y,\,z)
&=
\sum_{n=1}^\infty
\phi_{n0}
\sin \left(\frac{2n\pi z}{r_z}\right)\,\times \\ \nonumber
\lefteqn{
e^{
	\mp\frac{2n\pi \left(x-a/2\right)}{r_z}\,
	\sqrt{\varepsilon_c/\varepsilon_a}
} +
}\\ \nonumber
\lefteqn{\hspace{-15mm}+
\sum_{n,m=1}^\infty
\phi_{nm}
\cos \left(\frac{2m\pi y}{r_y}\right)
\sin \left(\frac{2n\pi z}{r_z}\right)\,\times }\\ \nonumber 
\lefteqn{e^{
	\mp \frac{2\pi \left(x-a/2\right)}{r_yr_z}\,
	\sqrt{m^2 r_z^2 + n^2 r_y^2 \,\varepsilon_c/\varepsilon_a}
}
}
\end{align}
\else
\begin{multline}
\label{eq:C:PotFourierGen}
	\varphi_d^{(\pm)}(x,\,y,\,z)
	=
	\sum_{n=1}^\infty
	\phi_{n0}
		\sin \left(\frac{2n\pi z}{r_z}\right)
		e^{
			\mp\frac{2n\pi \left(x-a/2\right)}{r_z}\,
			\sqrt{\varepsilon_c/\varepsilon_a}
		} +
\\ 
	+
	\sum_{n,m=1}^\infty
	\phi_{nm}
	\cos \left(\frac{2m\pi y}{r_y}\right)
	\sin \left(\frac{2n\pi z}{r_z}\right)
	e^{
		\mp \frac{2\pi \left(x-a/2\right)}{r_yr_z}\,
		\sqrt{m^2 r_z^2 + n^2 r_y^2 \,\varepsilon_c/\varepsilon_a}
	}
\end{multline}
\fi
satisfy the Laplace Eq.~(\ref{eq:C:PotLap}) and the internal boundary condition for the continuity of electrostatic potential Eq.~(\ref{eq:C:PotBoundCond:1}). 
The unknown coefficients $\phi_{n0}$ and $\phi_{nm}$ can be obtained by substituting Eqs.~(\ref{eq:C:PotFourierGen}) into the internal boundary for the continuity of the electric displacement Eq.~(\ref{eq:C:PotBoundCond:2}),
\iftwocolumn
\begin{align}
	\phi_{n0}& = 
	-\frac{
		3\, (-1)^n\, a \kappa r_z^2 
	}{
	n^2 \pi^2   
	\sqrt{\varepsilon_a\varepsilon_c} 
	\left( r_y^2+\kappa r_z^2 \right)
}\times \\ \nonumber\lefteqn{\hspace{-5mm}
\left\{
\ben{P}
-
\frac{
	3\, a^2 
	\left[
	n^2 \pi ^2 r_y^2
	+
	3 \kappa ^2 \left(\pi ^2 n^2-6\right) r_z^2
	\right]
}{
2 
n^2 \pi ^2 P_0^2 
\left(r_y^2+\kappa  r_z^2\right)^2
}\,
\ben{P}^3
\right\},} \\
\phi_{nm}& = 
\frac{
	54\,(-1)^{m+n}\, a^3 \kappa r_y^3 r_z^2 \, \ben{P}^3
}{
m^2 n \pi ^4 P_0^2
\left(r_y^2+\kappa  r_z^2\right)^3
\sqrt{m^2 r_z^2 \varepsilon_a^2 + n^2 r_y^2 \varepsilon_c \varepsilon_a}
}. \nonumber \\
\end{align}
\else
\begin{align}
	\phi_{n0}& = 
		-\frac{
				3\, (-1)^n\, a \kappa r_z^2 
		}{
				n^2 \pi^2   
				\sqrt{\varepsilon_a\varepsilon_c} 
				\left( r_y^2+\kappa r_z^2 \right)
		}
		\left\{
			\ben{P}
			-
			\frac{
					3\, a^2 
					\left[
						n^2 \pi ^2 r_y^2
						+
						3 \kappa ^2 \left(\pi ^2 n^2-6\right) r_z^2
					\right]
			}{
				2 
				n^2 \pi ^2 P_0^2 
				\left(r_y^2+\kappa  r_z^2\right)^2
			}\,
			\ben{P}^3
		\right\}, \\
	\phi_{nm}& = 
		\frac{
			54\,(-1)^{m+n}\, a^3 \kappa r_y^3 r_z^2 
		}{
			m^2 n \pi ^4 P_0^2
			\left(r_y^2+\kappa  r_z^2\right)^3
			\sqrt{m^2 r_z^2 \varepsilon_a^2 + n^2 r_y^2 \varepsilon_c \varepsilon_a}
		}\, \ben{P}^3.
\end{align}
\fi

%----

When Eq.~(\ref{eq:C:SigmaB}) is substituted into Eq. (\ref{eq:C:hel:def}), the function $h_{\rm el}$ can be with use of Eq.~(\ref{eq:C:RhoB}) expressed in the form:
\iftwocolumn
\begin{align}
	\label{eq:C:hel:gen}
	h_{\rm el}(\ben{P}) &= \\ \nonumber \lefteqn{\hspace{-10mm}
	\frac{P_0}{ar_yr_z}
	\int_{-\frac{r_y}2}^{\frac{r_y}2}dy 
	\int_{-\frac{r_z}2}^{\frac{r_z}2}
	\benex{\delta}{,z}(z)\,
	\varphi_d \left(a/2,\,y,\,z\right)
	\,dz}\\
	\label{eq:C:hel:sum}
	{}&=	-\sum_{n=1}^\infty
	\frac{6\,(-1)^n\,\kappa r_z \,\ben{P}}{n\pi (r_y^2+\kappa r_z^2)}\,\phi_{n0}.		
\end{align}
\else
\begin{align}
	\label{eq:C:hel:gen}
	h_{\rm el}(\ben{P}) &= 
		\frac{P_0}{ar_yr_z}
		\int_{-\frac{r_y}2}^{\frac{r_y}2}dy 
		\int_{-\frac{r_z}2}^{\frac{r_z}2}
		\benex{\delta}{,z}(z)\,
		\varphi_d \left(a/2,\,y,\,z\right)
		\,dz\\
	\label{eq:C:hel:sum}
	{}&=	-\sum_{n=1}^\infty
		\frac{6\,(-1)^n\,\kappa r_z \,\ben{P}}{n\pi (r_y^2+\kappa r_z^2)}\,\phi_{n0}.		
\end{align}
\fi
A note should be given that coefficients $\phi_{nm}$ do not enter the formula  for $h_{\rm el}$.
The reason is that the first factor of the integrand in Eq.~(\ref{eq:C:hel:gen}) is a function of $z$ only and its integration of all terms with $\cos(2m\pi y/r_y)$ with respect to $dy$ over its period $r_y$ gives zero.
After performing the summation in Eq.~(\ref{eq:C:hel:sum}), the function $h_{\rm el}$ has the form:
\iftwocolumn
\begin{multline}
	h_{\rm el}(\ben{P}) = 
	\frac{
		0.17\, a \kappa ^2 r_z^3 
	}{
	\sqrt{\varepsilon_a \varepsilon_c} 
	\left(r_y^2+\kappa  r_z^2\right)^2
}\times \\
\left[
\ben{P}^2
-
\frac{
	a^2 \left(3\, r_y^2 + 4.28\, \kappa ^2 r_z^2\right)
}{
2\,P_0^2 \left(r_y^2+\kappa  r_z^2\right)^2
}\,
\ben{P}^4 
\right].
\end{multline}
\else
\begin{equation}
	h_{\rm el}(\ben{P}) = 
		\frac{
			0.17\, a \kappa ^2 r_z^3 
		}{
			\sqrt{\varepsilon_a \varepsilon_c} 
			\left(r_y^2+\kappa  r_z^2\right)^2
		}
		\left[
			\ben{P}^2
			-
			\frac{
				a^2 \left(3\, r_y^2 + 4.28\, \kappa ^2 r_z^2\right)
			}{
				2\,P_0^2 \left(r_y^2+\kappa  r_z^2\right)^2
			}\,
			\ben{P}^4 
		\right].
\end{equation}
\fi

%----

The function $\ben{h}$ given by Eq.~(\ref{eq:C:hben:def}) can be with use of Eq.~(\ref{eq:C:SigmaW}) expressed in the form:
\begin{equation}
	\ben{h}(\ben{P}) = \bensupex{\alpha_{1}}{,\kappa}\, P_N^2 + \bensupex{\alpha_{11}}{,\kappa}\, P_N^4,
\end{equation}
where
\begin{subequations}
\label{eq:C:AlphaBenKappa}	
\iftwocolumn
\begin{align}
\label{eq:C:AlphaBenKappa:1}	
\bensupex{\alpha_{1}}{,\kappa}(\kappa) &= 
\frac{
	a 
}{
\left(r_y^2+\kappa  r_z^2\right)^2
} \times \\ \nonumber \lefteqn{
\left[
\frac{
	a_w  \left(r_y^2+\kappa ^2 r_z^2\right)
}{
8\, \varepsilon_c
}
+
\frac{
	0.174\, \kappa ^2 r_z^3
}{
\sqrt{\varepsilon_a\,\varepsilon_c}
}
\right],}
\\
\label{eq:C:AlphaBenKappa:11}	
\bensupex{\alpha_{11}}{,\kappa}(\kappa) &=
-
\frac{
	9\, a^3 
}{
80\, P_0^2
\left(r_y^2+\kappa  r_z^2\right)^4
} \times \\ \nonumber \lefteqn{
\left[
\frac{
	a_w  \left(9\, r_y^4+10\, \kappa ^2 r_y^2 r_z^2+9\, \kappa ^4 r_z^4\right)
}{
6\, \varepsilon_c
}
+
\right.} \\ \nonumber \lefteqn{ \left.
	\frac{
		\kappa ^2 r_z^3
		\left(2.33\, r_y^2+3.32\, \kappa ^2 r_z^2\right)
	}{
	\sqrt{\varepsilon_a\,\varepsilon_c}
}
\right],
}
\end{align}
\else
\begin{align}
\label{eq:C:AlphaBenKappa:1}	
	\bensupex{\alpha_{1}}{,\kappa}(\kappa) &= 
		\frac{
			a 
		}{
			\left(r_y^2+\kappa  r_z^2\right)^2
		}
		\left[
			\frac{
				a_w  \left(r_y^2+\kappa ^2 r_z^2\right)
			}{
				8\, \varepsilon_c
			}
			+
			\frac{
				0.174\, \kappa ^2 r_z^3
			}{
				 \sqrt{\varepsilon_a\,\varepsilon_c}
			}
		\right],
	\\
\label{eq:C:AlphaBenKappa:11}	
	\bensupex{\alpha_{11}}{,\kappa}(\kappa) &=
		-
		\frac{
			9\, a^3 
		}{
			80\, P_0^2
			\left(r_y^2+\kappa  r_z^2\right)^4
		}
		\left[
			\frac{
				a_w  \left(9\, r_y^4+10\, \kappa ^2 r_y^2 r_z^2+9\, \kappa ^4 r_z^4\right)
			}{
				6\, \varepsilon_c
			}
			+
\right. \\ \nonumber \lefteqn{ \left.
			\frac{
				\kappa ^2 r_z^3
				\left(2.33\, r_y^2+3.32\, \kappa ^2 r_z^2\right)
			}{
				 \sqrt{\varepsilon_a\,\varepsilon_c}
			}
		\right],
}
\end{align}
\fi
\end{subequations}
are coefficients dependent on the parameter $\kappa$, which expresses the anisotropy of the curvature of the bent domain wall.
In the final step, it is necessary to determine the parameter $\kappa$ from the condition for the minimum of free energy.
If we consider the condition given by Eq.~(\ref{eq:C:BenApprox:3}), the condition for minimum free energy of the bent domain wall $\ben{h}$ is equivalent to the condition for the minimum value of the coefficient of leading term with respect to $\ben{P}$, i.e.:
\begin{equation}
	\label{eq:C:Alpha:CondKappa}
	\partial \bensupex{\alpha_{1}}{,\kappa}(\kappa)/\partial \kappa = 0.
\end{equation}
Solution of the above equation for $\kappa$ gives:
\begin{equation}
	\label{eq:C:Kappa:Is}
	\kappa=\left(1 + \frac{1.4\, r_z}{a_w}\sqrt{\frac{\varepsilon_c}{\varepsilon_a}}\right)^{-1}
\end{equation}
Since it is considered that $a_w\ll r_z$, the above equation can be approximated by 
\begin{equation}
	\label{eq:C:Kappa:Approx}
	\kappa\approx \frac{0.717\, a_w}{r_z}\sqrt{\frac{\varepsilon_a}{\varepsilon_c}}.
\end{equation}
After substitution of Eq.~(\ref{eq:C:Kappa:Approx}) into Eqs.~(\ref{eq:C:AlphaBenKappa}) and considering $a_w\ll r_z$ and $a_w\ll r_y$, the leading terms of $\bensup{\alpha_{1}}$ and  $\bensup{\alpha_{11}}$ with respect to $a_w$, $r_y$ and $r_z$ are as follows:
\begin{subequations}
	\label{eq:C:AlphaBen}
	\begin{align}
	\label{eq:C:AlphaBen:1}
	\bensup{\alpha_{1}} &= \frac{a a_w }{8\,\varepsilon_c r_y^2},\\
	\label{eq:C:AlphaBen:11}
    \bensup{\alpha_{11}} &= 
	-\frac{0.17\, a^3 a_w}{\varepsilon_c P_0^2 r_y^4 }.
	\end{align}
\end{subequations}
It immediately follows from Eqs.~(\ref{eq:IV:PNw:Taylor}) that
\begin{subequations}
	\label{eq:C:PNBenTaylor}
	\begin{align}
	\label{eq:C:PNBenTaylor:EpsL}
	\ben{\varepsilon} &= \frac{4\, \varepsilon_c r_y^2}{a a_w},\\
	\label{eq:C:PNBenTaylor:Gamma}
	\ben{\gamma} &= \frac{
		172.8\, \varepsilon_c^3 r_y^4
	}{
	a a_w^3 P_0^2
}.
\end{align}
\end{subequations}

%====
%
%====
\section{Numerical refinement of analytical models}
\label{apndx:Refinement}

Our results presented in Sec.~\ref{sec:Aging} indicated that it is possible to obtain quantitative estimates on the material parameters of the ferroelectric system, which undergoes an evolution during aging process.
In order to provide tools for these quantitative estimates, it is necessary to improve accuracy of the analytical results presented in Sec.~\ref{sec:ReversibleModel}.
The necessary information for such a refinement of analytical models can be obtained from the performed numerical simulations presented in Sec.~\ref{sec:PFM}.

%----

The concept of numerical simulations of the two considered aging mechanisms makes it possible to separate the extrinsic contributions to permittivity controlled by the two mechanisms of reversible \dw{180} domain wall motions and to precise predictions of the analytical models.
In accordance with the implementation of numerical simulations of aging due to SB and EP, it is convenient to introduce following symbols for numerical values of small-signal extrinsic permittivity fitted from numerical simulations:
\begin{subequations}
	\label{eq:V:EpsSBEP}
	\begin{align}
	\label{eq:V:EpsSBEP:Flat}
	\bensup{\varepsilon_L} &= 
	    \bensup{\pin{\varepsilon}} + 
	    \bensup{\ben{\varepsilon}}
	    \mbox{\quad for } r_y=r,\, r_z=r_0,\\
	\label{eq:V:EpsSBEP:Tall}
	\pinsup{\varepsilon_L} &= 
	    \pinsup{\pin{\varepsilon}} + 
	    \pinsup{\ben{\varepsilon}}
	    \mbox{\quad for } r_y=r_0,\, r_z=r.
	\end{align}
\end{subequations}
Applying the straightforward algebraic manipulations to Eqs.~(\ref{eq:V:EpsSBEP}), (\ref{eq:B:PNPinTaylor:EpsL}), and (\ref{eq:C:PNBenTaylor:EpsL}), it is possible to express the separated contributions to small-signal extrinsic permittivity due to the two considered mechanisms of reversible \dw{180} domain wall motion:
\begin{subequations}
	\label{eq:V:EpsPinBen}
	\begin{align}
	\label{eq:V:EpsPinBen:Ben}
	\bensup{\ben{\varepsilon}} &= 
	    \frac{
	        r^2
	        \left(
	            \bensup{\varepsilon_L} - 
	            \pinsup{\varepsilon_L}
	        \right)
	    }{r^2-r_0^2},\\
	\label{eq:V:EpsPinBen:Pin}
	\pinsup{\pin{\varepsilon}} &= 
	    \frac{
	        r^2\, \pinsup{\varepsilon_L} - 
	        r_0^2\, \bensup{\varepsilon_L}
	    }{r^2-r_0^2}.
	\end{align}
\end{subequations}

%----

%---- Figure 6
\begin{figure}[t]
	%\centering
	\subfigure[]{
		\includegraphics[width=0.48\textwidth]{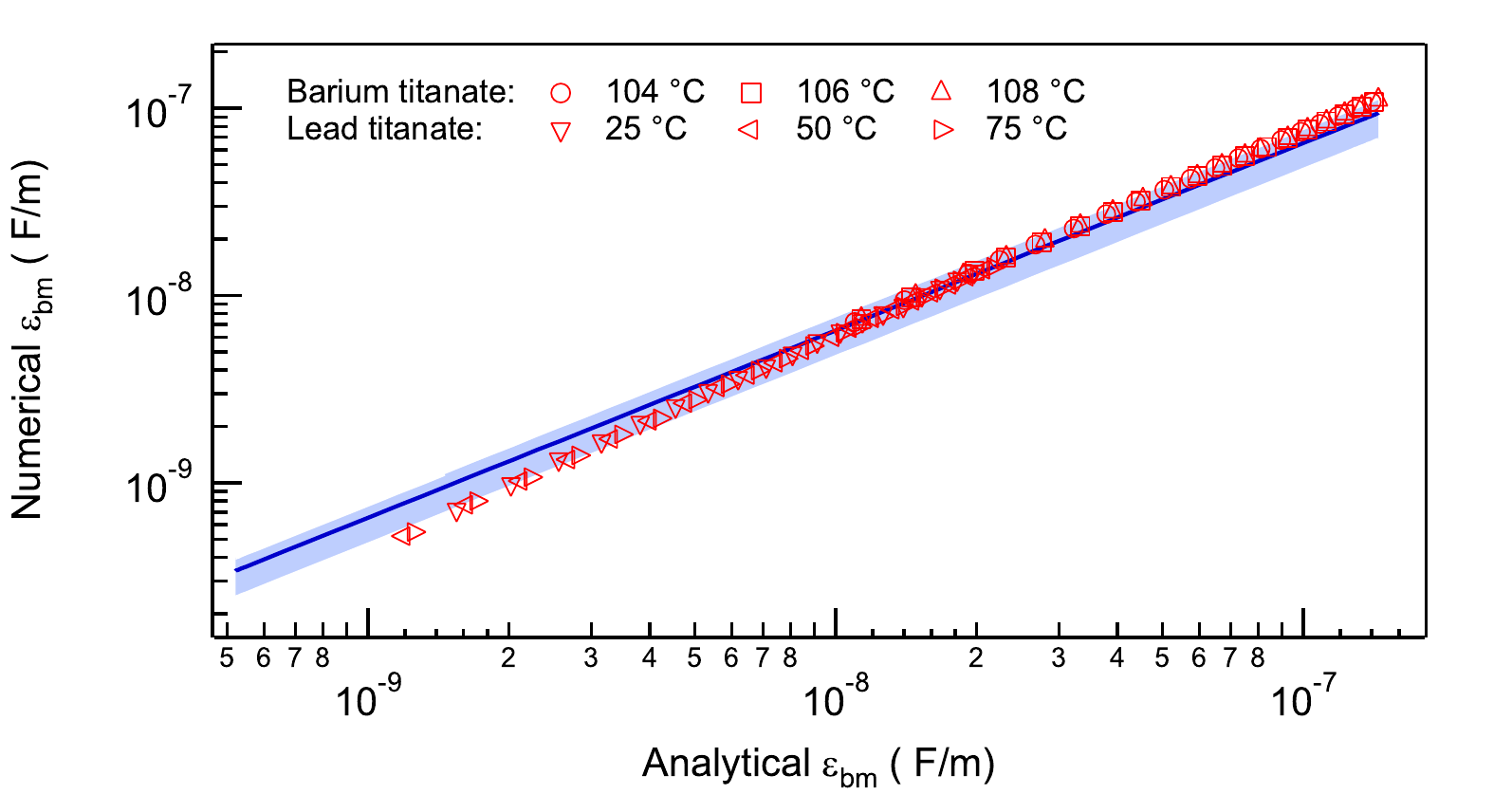}
		\label{fig:EpsL:a}
	}
	\subfigure[]{
		\includegraphics[width=0.48\textwidth]{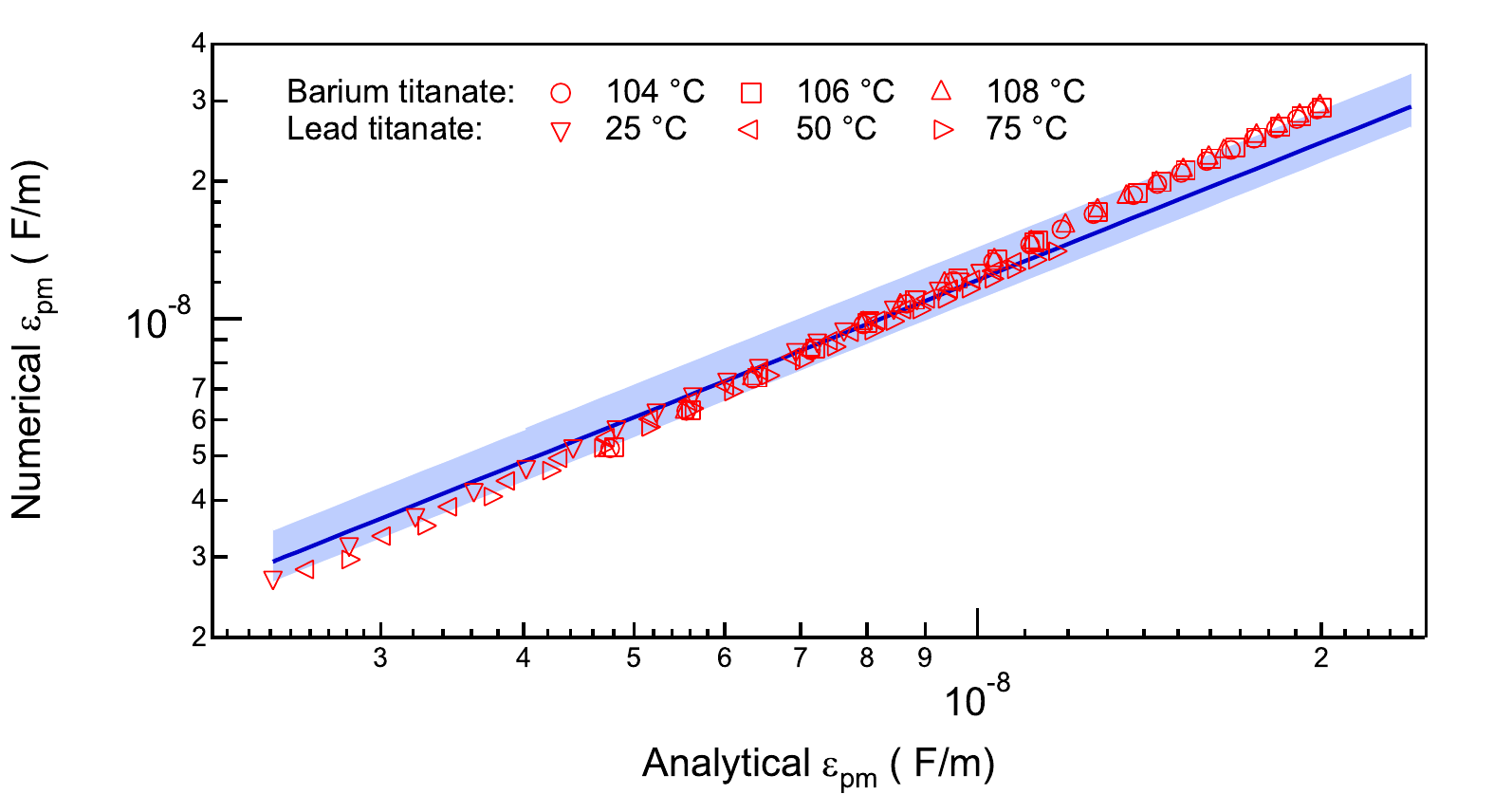}
		\label{fig:EpsL:b}
	}
	\caption{
		Comparison of the numerically computed values [see Eqs.~(\ref{eq:V:EpsPinBen})] versus predictions of analytical formulae [see Eqs.~(\ref{eq:B:PNPinTaylor:EpsL}) and (\ref{eq:C:PNBenTaylor:EpsL})] of small-signal permittivity controlled by bending movements~\subref{fig:EpsL:a} and uniform planar movement~\subref{fig:EpsL:b} of \dw{180} domain walls, respectively. 
	}
	\label{fig:EpsL}
\end{figure}
%---- Figure 6
%
Figure~\ref{fig:EpsL} shows the comparison of the numerically computed values (see Eqs.~\ref{eq:V:EpsPinBen}) versus predictions of analytical formulae [see Eqs.~(\ref{eq:B:PNPinTaylor:EpsL}) and (\ref{eq:C:PNBenTaylor:EpsL})] of small-signal permittivity controlled by bending movements Fig.~\ref{fig:EpsL:a} and uniform planar movement Fig.~\ref{fig:EpsL:b} of \dw{180} domain walls, respectively.

%----

Using the identical approach, it is convenient to introduce following symbols for numerical values of nonlinearity constant of the extrinsic permittivity fitted from numerical simulations:
\begin{subequations}
	\label{eq:V:GamamFlatTall}
	\begin{align}
	\label{eq:V:GamamFlatTall:Flat}
	\bensup{\gamma_w} &= 
	    \bensup{\pin{\gamma}} + 
	    \bensup{\ben{\gamma}},
	    \mbox{\quad for } r_y=r,\, r_z=r_0,\\
	\label{eq:V:GamamFlatTall:Tall}
	\pinsup{\gamma_w} &= 
	    \pinsup{\pin{\gamma}} + 
	    \pinsup{\ben{\gamma}}
	    \mbox{\quad for } r_y=r_0,\, r_z=r.
	\end{align}
\end{subequations}
Applying the straightforward algebraic manipulations to Eqs.~(\ref{eq:V:GamamFlatTall}), (\ref{eq:B:PNPinTaylor:Gamma}), and (\ref{eq:C:PNBenTaylor:Gamma}), it is possible to express the nonlinearity constant of the extrinsic permittivity due to the two considered mechanisms of reversible \dw{180} domain wall motion:
\begin{subequations}
	\label{eq:V:GamamPinBen}
	\begin{align}
	\label{eq:V:GamamPinBen:Ben}
	\bensup{\ben{\gamma}} &= 
	    \frac{
	        r^4 
	        \left(
	            \bensup{\gamma_w} - 
	            \pinsup{\gamma_w}
	        \right)
	    }{r^4-r_0^4},\\
	\label{eq:V:GamamPinBen:Pin}
	\pinsup{\pin{\gamma}} &= 
	    \frac{
	        r^4\, \pinsup{\gamma_w} - 
	        r_0^4\, \bensup{\gamma_w}
	    }{r^4-r_0^4}.
	\end{align}
\end{subequations}

%----

%---- Figure 7
\begin{figure}[t]
	\centering
	\subfigure[]{
		\includegraphics[width=0.48\textwidth]{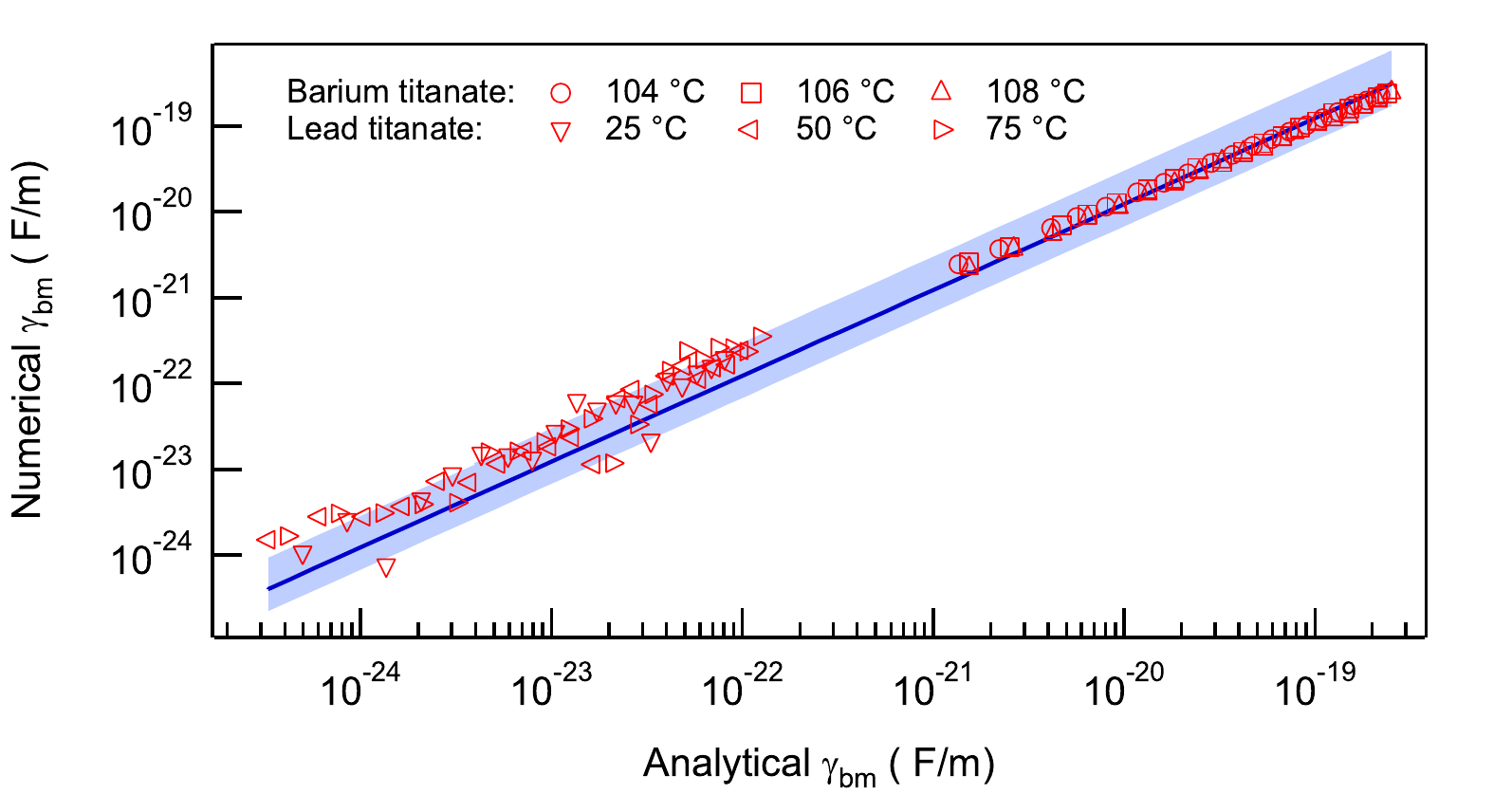}
		\label{fig:GammaW:a}
	}
	\subfigure[]{
		\includegraphics[width=0.48\textwidth]{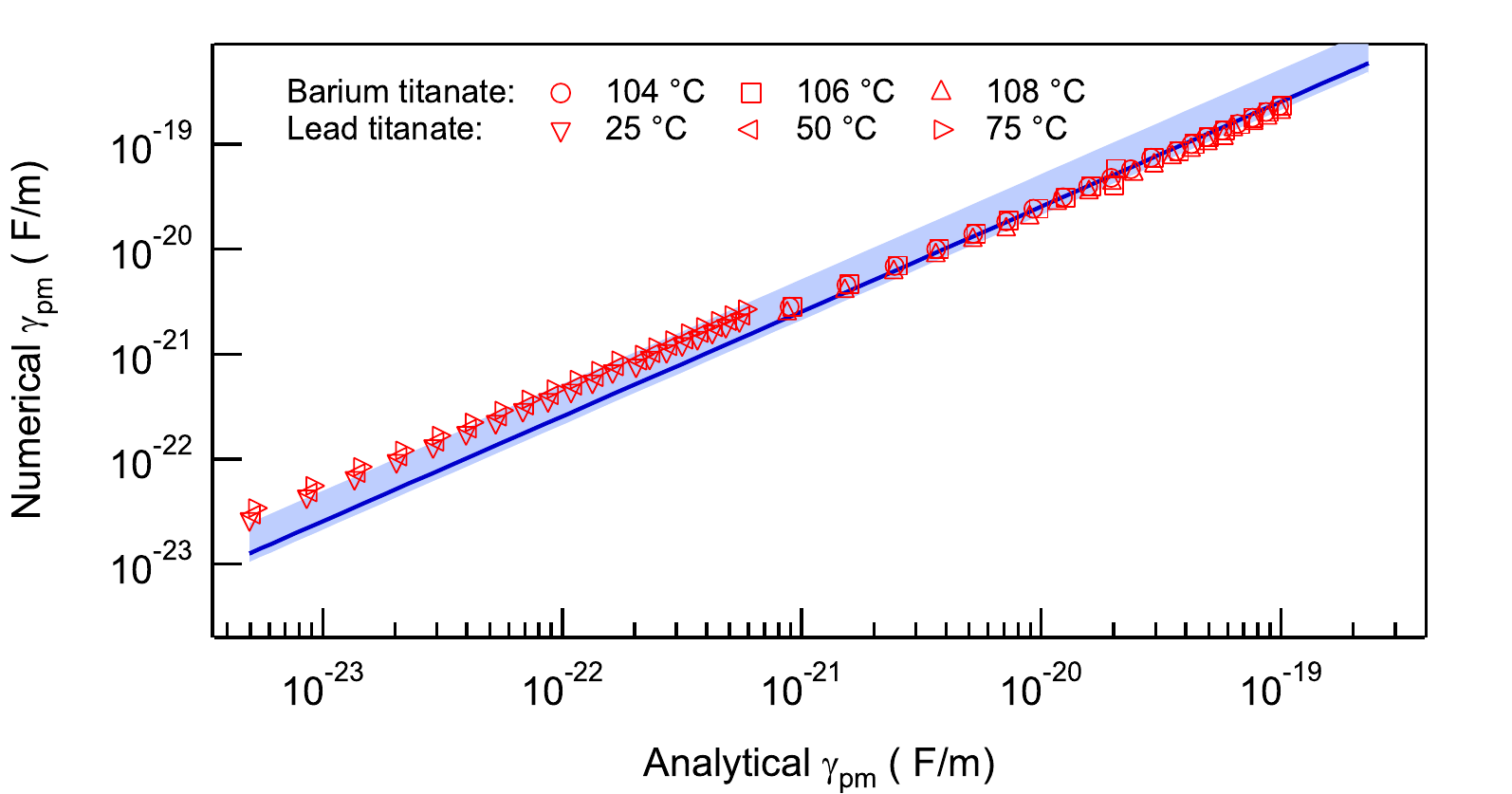}
		\label{fig:GammaW:b}
	}
	\caption{
		Comparison of the numerically computed values [see Eqs.~(\ref{eq:V:GamamPinBen})] versus predictions of analytical formulae [see Eqs.~(\ref{eq:B:PNPinTaylor:Gamma}) and (\ref{eq:C:PNBenTaylor:Gamma})] of nonlinearity constant of extrinsic permittivity controlled by bending movements~\subref{fig:GammaW:a} and uniform planar movement~\subref{fig:GammaW:b} of \dw{180} domain walls, respectively. 
	}
	\label{fig:GammaW}
\end{figure}
%---- Figure 7
%
Figure~\ref{fig:GammaW} shows the comparison of the numerically computed values [see Eqs.~(\ref{eq:V:GamamPinBen})] versus predictions of analytical formulae [see Eqs.~(\ref{eq:B:PNPinTaylor:Gamma}) and (\ref{eq:C:PNBenTaylor:Gamma})] of nonlinearity constant of extrinsic permittivity controlled by bending movements Fig.~\ref{fig:GammaW:a} and uniform planar movement Fig.~\ref{fig:GammaW:b} of \dw{180} domain walls, respectively. 

%----

It has been already mentioned that the analytical solution of the developed thermodynamic model has been subjected to several strong simplifications. 
These simplification have an effect on the numerical values of dimensionless constants in Eqs.~(\ref{eq:B:PNPinTaylor}) and (\ref{eq:C:PNBenTaylor}).
Comparisons presented in Figs.~\ref{fig:EpsL} and \ref{fig:GammaW} allow to compute the numerical corrections factors to analytical formulae.
Hence, we introduce the numerical dimensionless constants:
\begin{subequations}
	\label{eq:V:CorrFacs}
	\begin{align}
	\label{eq:V:CorrFacs:MuPin}
	\pin{\mu} &= \pinsup{\pin{\varepsilon}}/\pin{\varepsilon},\\
	\label{eq:V:CorrFacs:MuBen}
	\ben{\mu} &= \bensup{\ben{\varepsilon}}/\ben{\varepsilon},\\
	\label{eq:V:CorrFacs:GammaPin}
	\pin{\nu} &= \pinsup{\pin{\gamma}}/\pin{\gamma},\\
	\label{eq:V:CorrFacs:GammaBen}
	\ben{\nu} &= \bensup{\ben{\gamma}}/\ben{\gamma}.
	\end{align}
\end{subequations}

%----

%---- Table 5
\begin{table}
	\caption{Numerical values of correction factors $\mu$ and $\nu$ to analytical formulae given by Eqs.~(\ref{eq:B:PNPinTaylor}) and (\ref{eq:C:PNBenTaylor}).}
	\label{tab:V:CorrFacs}
	\begin{tabular}{lrrrrrr}
		\hline \hline
		Parameter & 
		0.05-quantile & 
		Median& 
		0.95-quantile\\ \hline		
		$\pin{\mu}$& 1.1& 1.2& 1.4\\
		$\ben{\mu}$& 0.47& 0.65& 0.76\\
		$\pin{\nu}$& 2.0& 2.5& 5.2\\
		$\ben{\nu}$& 0.6& 1.2& 3.0\\
		\hline\hline
	\end{tabular}
\end{table}
%---- Table 5
%
Table~\ref{tab:V:CorrFacs} shows that the computed median values of all correction factors falls within the interval between 0.5 and 2.5.
This seems to be acceptable considering the simplifications presented Appendices~\ref{apndx:Model:pin} and \ref{apndx:Model:ben}.
It can be, therefore, concluded that analytical solutions give very reasonable qualitative predictions for the extrinsic permittivity due to the reversible \dw{180} domain wall movements. 

%----

Although the total number of numerical simulations was reasonable to use the method of least squares, the combination of the parameters computed for two different materials has introduced a number of outlier and leverage points that may affect the result of the least-square method. 
In order to avoid the incorrect statistical treatment of the results of numerical simulations, we have performed a quantile regression.
The results presented in Table~\ref{tab:V:CorrFacs} indicate the ranges of numerical values that cover the 90\% of the numerically computed phase field values.

\end{document}